\documentclass[12pt]{article}
\usepackage{jheppub}
\pdfoutput=1
\usepackage{epstopdf}

\usepackage{jheppub}
\usepackage{mathrsfs}
\usepackage{psfrag}
\usepackage{color}

\usepackage[table]{xcolor}

\usepackage{mathrsfs,amsmath,bbm,array,amsfonts,graphicx,wrapfig,arydshln,lscape,float,multirow,longtable,rotating,subfigure,makecell}
\usepackage{url}
\usepackage[utf8]{inputenc}

\newcommand{\be}{\begin{equation}}
\newcommand{\ee}{\end{equation}}
\newcommand{\beq}{\begin{equation}}
\newcommand{\beql}[1]{\begin{equation}\label{#1}}
\newcommand{\eeq}{\end{equation}}
\newcommand{\ba}{\begin{array}}
\newcommand{\ea}{\end{array}}
\newcommand{\bea}{\begin{eqnarray}}
\newcommand{\beal}[1]{\begin{eqnarray}\label{#1}}
\newcommand{\eea}{\end{eqnarray}}
\newcommand{\ben}{\begin{enumerate}}
\newcommand{\een}{\end{enumerate}}
\newcommand{\bean}{\begin{eqnarray*}}
\newcommand{\eean}{\end{eqnarray*}}
\newcommand{\eref}[1]{(\ref{#1})}
\newcommand{\sref}[1]{\S\ref{#1}}
\newcommand{\tref}[1]{Table~\ref{#1}}

\newcommand{\fref}[1]{Figure \ref{#1}}
\newcommand{\btab}[1]{\begin{tabular}{#1}}
\newcommand{\etab}{\end{tabular}}

\newcommand{\comment}[1]{}

\newcommand{\qed}{\nobreak \ifvmode \relax \else
      \ifdim\lastskip<1.5em \hskip-\lastskip
      \hskip1.5em plus0em minus0.5em \fi \nobreak
      \vrule height0.75em width0.5em depth0.25em\fi}

\newcolumntype{C}[1]{>{\centering\arraybackslash}m{#1}}

\newcommand{\pl}{Pl\"ucker }

\title{Positivity Sectors and the Amplituhedron}

\author{Daniele Galloni}

\affiliation{INFN, Sezione di Torino \\
Via Pietro Giuria 1, 10125 Torino, Italy
}

\emailAdd{daniele.galloni@to.infn.it}

\abstract{We initiate a detailed investigation into the assembly of simple amplituhedron-like building blocks to obtain spaces of physical interest. In particular, we describe the geometric process through which the building blocks, which we call positivity sectors, glue together to form the desired geometries. Positivity sectors are seen to naturally segment the space describing the $L^{th}$ power of the one-loop amplitude. In this way, we obtain a good understanding of how the geometric complexity of the building blocks can be washed out in the formation of larger spaces. Conversely, the tools we develop allow us to form spaces of ever greater complexity, a process which is crucial to the construction of the amplituhedron from its triangulations, which remains an important open question. We present the full boundary structure of all positivity sectors related to the three-loop amplituhedron. We also construct a practical algorithm that achieves the desired geometric assembly of positivity sectors, and make available supporting Mathematica files containing the full boundary structure of all positivity sectors at three loops.
}

\preprint{
}

\begin{document}

\maketitle

\section{Introduction} 
\label{sec:intro}

Scattering amplitudes constitute one of the main tools with which we study and understand gauge theories. Not only do they closely connect theory and experiment, but their structure allows us to glean powerful insights into the formulation with which to best describe the gauge theory. Unsurprisingly, this has led to explosive progress in the development of tools and techniques with which to compute scattering amplitudes \cite{Bern:1994zx,Bern:1994cg,Cachazo:2004kj,Britto:2004nc,Britto:2004ap,Britto:2005fq,Bern:2005iz} (for reviews, see e.g.\ \cite{Dixon:1996wi,Beisert:2010jr,Drummond:2011ic,Elvang:2013cua}). This progress is particularly pronounced in planar 4-dimensional $\mathcal{N}=4$ Super-Yang-Mills (SYM), where computational results exist even for very high loop orders \cite{Bern:2006ew,Bern:2007ct,ArkaniHamed:2010kv,ArkaniHamed:2010gh,Dixon:2011nj,Bourjaily:2011hi,Dixon:2013eka,Dixon:2014xca}. 

To a great extent, this progress has been achieved through an effort to make the full symmetry group of the theory manifest. In fact, planar 4-dimensional $\mathcal{N}=4$ SYM possesses an infinite-dimensional symmetry known as the \textit{Yangian} \cite{Drummond:2009fd}, which arises from the existence of a conformal and dual conformal symmetry \cite{Drummond:2006rz,Alday:2007hr}. The discovery of the Yangian has triggered an interesting interplay between perturbative diagrammatic approaches and integrable methods for computing the amplitude \cite{Beisert:2003yb,Beisert:2006ez}. It has also indirectly forced a shift in the perturbative approaches, away from the usual Feynman-diagrammatic expansion, to the construction of terms which are individually Yangian-invariant. Such a construction has been explicitly formulated for non-supersymmetric tree-level amplitudes with the celebrated BCFW recursion relations \cite{Britto:2004ap,Britto:2005fq}, which form the foundation of very fruitful unitary methods \cite{Bern:1994zx,Bern:1994cg}. For the loop-level integrand of $\mathcal{N}=4$ SYM there exist similar recursions relations \cite{ArkaniHamed:2010kv}, which generate completely Yangian-invariant terms with which to construct the amplitude.  

The understanding of these recursion relations was greatly augmented by the discovery of a dual, more mathematical formulation of the amplitude, which heavily relies on the Grassmannian \cite{ArkaniHamed:2012nw} (whose formulation is rooted in preceding discoveries made in \cite{ArkaniHamed:2009dn,Mason:2009qx,ArkaniHamed:2009vw,ArkaniHamed:2010kv,Kaplan:2009mh,ArkaniHamed:2009dg} and has undergone considerable progress \cite{Bourjaily:2012gy,Amariti:2013ija,Franco:2013nwa,Du:2014jwa,Franco:2014nca,Olson:2014pfa,Bork:2015fla,Frassek:2015rka,Benincasa:2015zna}).\footnote{This thread of research has also generated interest in a deformed version of the method \cite{Ferro:2012xw,Ferro:2013dga,Beisert:2014qba,Broedel:2014hca,Bargheer:2014mxa,Ferro:2014gca}, where the helicities of the external particles may be deformed in order to regularize the amplitude.} Such a profoundly different expression of scattering amplitudes has the potential to make discoveries that were previously inaccessible through more traditional methods of computing the amplitude. For example, it has already made the $dlog$ form of the amplitude completely explicit, giving a handle with which to tackle non-planar amplitudes, where there is strong evidence suggesting a similar $dlog$ structure \cite{Bern:1997nh,Bern:2007hh,Bern:2010tq,Carrasco:2011mn,Bern:2012uc,Arkani-Hamed:2014via,Arkani-Hamed:2014bca,Bern:2014kca,Franco:2015rma,Chen:2015bnt,Badger:2015lda,Bern:2015ple}. 

The Grassmannian formulation has culminated in the discovery of the \textit{amplituhedron} \cite{Arkani-Hamed:2013jha,Arkani-Hamed:2013kca}, a highly geometric mathematical object whose geometric structure completely mimics the singularity structure of the amplitude. The amplituhedron is conjectured to yield the amplitude for a given process, through the computation of its volume with a specific volume form. The amplituhedron answers the question of how to combine Yangian-invariant building blocks to form the amplitude describing a process, and has been the subject of recent investigation \cite{Enciso:2014cta,Bai:2014cna,Franco:2014csa,Lam:2014jda,Arkani-Hamed:2014dca,Bai:2015qoa,Ferro:2015grk,Bern:2015ple}.\footnote{For an interesting step towards a formulation of an amplituhedron for ABJM amplitudes see \cite{Elvang:2014fja}.} Locality and unitarity, obfuscated by the previously known recursion relations, are seen to be properties derived from the geometry of the amplituhedron. While the amplituhedron remains conjectural, it has passed many non-trivial direct tests \cite{Arkani-Hamed:2013jha,Arkani-Hamed:2013kca,Franco:2014csa}.

The amplituhedron, as the first object which single-handedly encapsulates the entire amplitude, has enormous potential for studying global properties of scattering amplitudes. However, a clear understanding of the geometric tools and techniques required for utilizing its power most efficiently is still lacking. An important goal is the creation of tools with which it is possible to subdivide the amplituhedron into simpler components, in order to construct the full amplitude through geometric means. Such a process requires a detailed understanding of the process of triangulating the amplituhedron, and the inverse process of assembling the triangulations into the full object representing the amplitude. This article aims to shed light on this process. We will gain a geometric understanding of the assembly process and generate a practical algorithm with which geometric components can be assembled to create the spaces of interest. 

As an immediate practical implementation of our results, we illustrate the assembly of the cube of the one-loop amplitude through components which we call positivity sectors. Our techniques are expected to be more generally valid in the construction of the $L^{th}$ power of the one-loop amplituhedron. Furthermore, we use the positivity sectors to construct the geometry associated to terms in the three-loop log of the amplitude. The exploration of ideas through the careful understanding of examples has been a rewarding strategy in recent developments in this area; we contribute in this spirit with very explicit elaborations of examples in order to facilitate understanding and aid future developments.

This article is organized as follows. \sref{sec:amplintro} reviews background on the amplituhedron and its boundary structure. Positivity sectors related to the amplituhedron are defined in \sref{sec:possectors}, where we also show how they may be used to form spaces of physical interest, such as the cube of the one-loop amplitude and the second-order contribution to the log of the amplitude. \sref{sec:3loopsectors} presents the boundary structure of each of the positivity sectors related to the three-loop amplituhedron, highlighting new features seen at three loops. This section presents the number of boundaries of various dimensionality and computes the Euler number of each positivity sector. \sref{sec:assembling} describes in detail how to assemble sectors in order to obtain the spaces of interest, focusing on describing the change in boundary structure through a geometric description of the assembly process; this allows us to present a practical algorithm for implementing the assembly of positivity sectors in \sref{sec:assemblingalgorithm}. Finally, the algorithm is implemented in \sref{sec:formingallspaces} to construct the cube of the one-loop amplitude, as well as terms contributing to the three-loop log of the amplitude. Here we comment on the significance of the remarkable simplification of the Euler number, through a geometric interpretation of our results, highlighting intermediate stages of the full assembly process. We conclude and summarize our results in \sref{sec:conclusion}. Four appendices are provided, which contain supporting material to the article and an explanation of how to use the supporting Mathematica files that may be downloaded with the article.

\section{The Amplituhedron and its Stratification} 
\label{sec:amplintro}

We shall begin by briefly introducing the amplituhedron, and the currently known methods for finding its boundary structure, i.e.\ its \textit{stratification}. We refer the reader to \cite{Arkani-Hamed:2013jha,Arkani-Hamed:2013kca} for details on the introduction to the amplituhedron and to \cite{Franco:2014csa} for details on obtaining its stratification.

The amplituhedron is a generalization of the positive Grassmannian, conjectured to describe all planar scattering amplitudes in $\mathcal{N}=4$ SYM. The space inhabited by the amplituhedron is denoted $G(k,k+4;L)$, where $k$ indicates the $\text{N}^{k}\text{MHV}$ degree of the amplitude, and $L$ is the loop level of the amplitude under consideration. Points in the amplituhedron $\mathcal{Y} \in G(k,k+4;L)$ represent $k$-dimensional planes in $k+4$ dimensions, alongside $L$ 2-dimensional planes extending in the 4-dimensional space transverse to the $k$-plane.\footnote{As is clear from what $\mathcal{Y}$ represents, calling each $\mathcal{Y}$ a \textit{point} in the amplituhedron is of course a misnomer: we should rather say that each $\mathcal{Y}$ is an internal \textit{plane} in the amplituhedron.} The amplituhedron is the collection of all possible $\mathcal{Y}$ satisfying specific positivity conditions, which we will define more precisely below.

Each $\mathcal{Y}$ can be expressed as the product of a matrix $Z \in M_+(n,k+4)$ encoding the external kinematic data, where $n$ is the number of external particles, with a $(k+2L)\times n$ matrix $\mathcal{C}$. Explicitly, this product is
\begin{equation} \label{eq:defamplsmall}
\mathcal{Y} = \mathcal{C} \cdot Z \; ,
\end{equation} 
where each row in $Z$ is the bosonized super-momentum-twistor of an external particle, and the rows have been arranged in such a way  that all $(k+4) \times (k+4)$ minors of $Z$ are positive.\footnote{Here and in what follows, by \textit{positive} we usually mean \textit{non-negative}: boundaries of the spaces we consider are obtained by setting positive quantities to zero. We hope the reader will not be confused by this slight abuse of terminology.} For the amplitude to be non-zero, $k$ must take on values between $k=0$, corresponding to MHV amplitudes, and $k=n-4$, corresponding to $\overline{\text{MHV}}$ amplitudes.

Each matrix $\mathcal{C}$ is formed by stacking $2 \times n$ matrices $D_{(i)}$, where $i=1, \ldots , L$, onto a matrix $C$ belonging to the positive Grassmannian $C \in G_+ (k,n)$, with the condition that the $D_{(i)}$ are transverse to $C$, i.e.\ that $D_{(i)} \in C^{\perp}$. Finally, there are positivity conditions on the matrix $\mathcal{C} \in G_+ (k,n;L)$ as a whole: the matrices
\begin{equation} \label{eq:Cconditions}
\begin{pmatrix}
C
\end{pmatrix} ,
\begin{pmatrix}
D_{(1)}\\
\hline
C
\end{pmatrix} , \cdots,
\begin{pmatrix}
D_{(L)}\\
\hline
C
\end{pmatrix} ,
\begin{pmatrix}
D_{(1)}\\
D_{(2)}\\
\hline
C
\end{pmatrix} , \cdots
\end{equation}
must all have positive maximal minors. In other words, $C$ with any $D_{(i)}$s stacked over it must be positive, where we don't consider stacks of matrices that have more rows than columns.

Using this notation we can write \eref{eq:defamplsmall} more transparently as  
\begin{equation} \label{eq:defamplbig}
\begin{pmatrix}
\mathcal{L}_{(1)} \\
\mathcal{L}_{(2)} \\
\vdots \\
\mathcal{L}_{(L)} \\
\hline
Y
\end{pmatrix} =  \begin{pmatrix}
D_{(1)} \\
D_{(2)} \\
\vdots \\
D_{(L)} \\
\hline
C
\end{pmatrix} \cdot Z \quad \; .
\end{equation}
$Y$ is a $k$-plane in $(k+4)$ dimensions and resides in the \textit{tree-level amplituhedron};\footnote{The tree-level amplituhedron is the amplituhedron corresponding to tree-level processes.} each $\mathcal{L}_{(i)}$ is a 2-plane in the 4-dimensional space transverse to $Y$ and describes the loop-level part of the amplitude. The full $L$-loop amplitude for $n$ particles of helicity $k$ is given by all possible $\mathcal{Y}$, subject to the conditions \eref{eq:Cconditions} and the condition of positivity of $Z$. To explore the interior of the amplituhedron we scan over the various $\mathcal{Y}$, by scanning over all possible $\mathcal{C}$ and kinematic configurations $Z$.

\subsection{Boundaries of the Amplituhedron} 
\label{sec:amplboundaries}

The amplitude is given by integrating over the \textit{volume} of the amplituhedron, with a specific volume form that has logarithmic singularities on its boundaries. This volume form is the integrand of the amplitude. Given the relation between boundaries and singularities, it is clear that planar $\mathcal{N}=4$ SYM amplitudes can only have logarithmic singularities. 

In this framework, the problem of understanding the full singularity structure of the scattering amplitude translates to a geometric problem of understanding the full boundary structure of the amplituhedron. Understanding the singularity structure is extremely desirable. This understanding is what originally allowed for the construction of the celebrated BCFW recursion relations; moreover, all $\mathcal{N}=4$ SYM amplitudes are completely determined by their singularity structure. 

We shall now review the methods introduced in \cite{Franco:2014csa} to obtain the stratification of the amplituhedron, and hence the singularity structure of the amplitude. Before we begin, we point out that $Z$ is a square matrix when $n=4+k$. In these cases we may perform a change of basis such that $Z$ becomes equal to the identity matrix, thus rendering it trivial. Understanding the stratification of the amplituhedron then amounts to understanding the stratification of $\mathcal{C} \in G_+(k,n;L)$. 

Following \cite{Franco:2014csa}, we shall restrict ourselves to these cases, and consider the simplest option where $k=0$, $n=4$ and $L$ is free, i.e.\ we shall outline how to obtain the singularity structure of 4-particle MHV amplitudes at arbitrary loop order. Here $\mathcal{C}$ is simply given by a stack of $L$ matrices $D_{(i)}$, and the positivity conditions \eref{eq:Cconditions} amount to demanding the positivity of each $D_{(i)}$, as well as the positivity of all $4 \times 4$ matrices $\begin{pmatrix} D_{(i)}\\ D_{(j)} \end{pmatrix}$ formed by stacking pairs of such matrices. 

\paragraph{\pl Coordinates.}
The degrees of freedom of $D_{(i)} \in G(2,n)$ are best expressed through their $2 \times 2$ minors, known as their \pl coordinates.\footnote{We note that $D_{(i)} \in G(2,n)$ is generally \textit{not} restricted to the positive Grassmannian. This restriction exists in the special case of $k=0$, where \eref{eq:Cconditions} forces $D_{(i)} \in G_+(2,n)$.} Each \pl coordinate $\Delta^{(i)}_I$ bears an index $I$ which specifies which $k$ columns were used to form the minor, and an index $i$ which indicates that this \pl coordinate is a minor of the matrix $D_{(i)}$. The \pl coordinates are not independent: they are subject to \pl relations; for $G_+(2,4)$ there is only one relation
\begin{equation} \label{eq:pluckrel}
\Delta^{(i)}_{12} \Delta^{(i)}_{34} + \Delta^{(i)}_{14} \Delta^{(i)}_{23} = \Delta^{(i)}_{13} \Delta^{(i)}_{24} \; \; .
\end{equation}
We may also express the $4 \times 4$ minor $\Delta^{(i,j)}_{1234}$ of $\begin{pmatrix} D_{(i)}\\ D_{(j)} \end{pmatrix}$ using \pl coordinates:
\begin{equation} \label{eq:4x4minor}
\Delta^{(i,j)}_{1234} = \Delta^{(i)}_{12} \Delta^{(j)}_{34} + \Delta^{(i)}_{34} \Delta^{(j)}_{12} + \Delta^{(i)}_{14} \Delta^{(j)}_{23} + \Delta^{(i)}_{23} \Delta^{(j)}_{14} - \Delta^{(i)}_{13} \Delta^{(j)}_{24} - \Delta^{(i)}_{24} \Delta^{(j)}_{13} \; \; .
\end{equation}
Hence, for $k=0$, $n=4$ and arbitrary $L$, we may express the positivity conditions \eref{eq:Cconditions} as $\Delta^{(i)}_I >0$ and $\Delta^{(i,j)} > 0$: these are the positive degrees of freedom we need to turn off in order to access boundaries of the amplituhedron.\footnote{When $n=4$, which are the cases we are restricting ourselves to, the $4 \times 4$ minor can only have $I = 1234$ as subindex, so in the interest of simplicity we shall omit this subindex from our expressions.}

\paragraph{The Stratification.}
We may now use $\Delta^{(i)}_I$ and $\Delta^{(i,j)}$ to construct the stratification of the amplituhedron. This procedure is naturally divided in two steps:
\begin{enumerate}
\item We begin by shutting off combinations of $\Delta^{(i)}_I$s in all possible ways compatible with the positivity of $\Delta^{(j,k)} > 0$. The structure that ensues is called $\Gamma_0$. Each element in $\Gamma_0$ constitutes a boundary of the amplituhedron, and is specified by which \pl coordinates are vanishing.\footnote{There is a caveat to this statement, which we shall return to in \sref{sec:regionsintro}.} These elements may be arranged into a poset, ordered by the dimensionality of each site. Some sites in the poset, i.e.\ some elements of $\Gamma_0$, may turn off all terms that contribute to some $\Delta^{(j,k)}$s and hence force them to become zero trivially. Those $\Delta^{(j,k)}$s that do not vanish in this way are free to take on non-zero values.
\item In each site in $\Gamma_0$, we now turn off combinations of $\Delta^{(j,k)}$ in all possible ways compatible with the positivity of the remaining $\Delta^{(l,m)}$s. Each combination corresponds to a separate boundary of the amplituhedron. The structure emanating from each $\Gamma_0$ site is called $\Gamma_1$. Generally the $\Delta^{(j,k)}$s are not all independent, and $\Gamma_1$ may be very non-trivial. An example of the variety of possible $\Gamma_1$s that can occur at 3 loops is shown in Figures \ref{fig:1Delta2Delta} and \ref{fig:3Ltrees}.
\end{enumerate}
There are powerful techniques for obtaining $\Gamma_0$ and $\Gamma_1$. Details on these techniques are found in \cite{Franco:2014csa}. We shall make heavy use of the $\Gamma_0$ and $\Gamma_1$ structures throughout this article.

\paragraph{The Euler Number.}
We may now take all the boundaries found in the above way and count how many there are at each dimension
\begin{equation}
d = D - N_C
\end{equation}
where $D$ is the dimension of the amplituhedron, and $N_C$ is the number of independent conditions imposed by shutting off a given selection of $\Delta^{(i)}_I$s and $\Delta^{(i,j)}$s. For $k=0$ and $n=4$ we have $D=4L$. We denote the total number of $d$-dimensional boundaries $\mathfrak{N}^{(d)}$. Computing the alternating-sign sum of this number gives a quantity known as the Euler number, 
\begin{equation} \label{eq:euler}
\mathcal{E} = \sum_{d=0}^{D} (-1)^d \mathfrak{N}^{(d)} \; \; .
\end{equation}
The Euler number is of great geometric interest. For example, convex polytopes all have $\mathcal{E}=1$, independently of how many vertices they have or how high their dimensionality is. This requires \textit{huge} cancellations between the number of boundaries of various dimensions; the statement that this cancellation happens is essentially rooted in the statement that convex polytopes are ``simple'' geometric objects. Expressed differently, their boundaries of various dimensionality must link up in a combinatorial way so as to construct the polytope, and this ensures that the Euler number is equal to 1.

An arbitrary stratification, however, is \textit{not} expected to have a particularly small Euler number. In fact, naively one might expect the Euler number to be comparable in size to the largest $\mathfrak{N}^{(d)}$. The fact that the singularity structure of scattering amplitudes has an amazingly small Euler number is a strong hint that there should exist a geometric description of this structure. The amplituhedron is conjectured to be that description. Indeed, we shall see that the Euler number of the amplituhedron is many orders of magnitude smaller than the largest $\mathfrak{N}^{(d)}$.

We can heuristically see a small Euler number as indicative of geometric simplicity---convex polytopes being the simplest of all. As we shall see, there are closely related objects to the amplituhedron which exhibit a similar simplicity, and the amplituhedron constitutes a sector within this simpler space. In this article we shall study in detail the process by which various sectors, each with an Euler number $\mathcal{E} \neq 1$, glue together to form the simpler space, with $\mathcal{E}=1$.

\subsection{On-Shell Diagrams in the Amplituhedron} 
\label{sec:onshellinside}

Understanding how components glue together to form more complicated objects is central to constructing the amplituhedron from its triangulations. This process crucially relies on understanding how boundaries glue together, and how the gluing modifies the boundary structure of the assembled object.

Triangulations of the amplituhedron form a promising avenue to obtain the integrand for very complicated processes. The integrand contribution associated to each triangle\footnote{By triangle we mean the fundamental building block of the triangulation. These building blocks do not need to have the structure of a simplex.} is particularly simple: it is simply $\frac{1}{\langle \cdots \rangle \, \cdots \, \langle \cdots \rangle}$, where each $\langle \cdots \rangle$ denotes a determinant of $(k+4)$ bosonized momentum twistors. Tuning a bracket $\langle \cdots \rangle$ to zero corresponds to going to a boundary of the triangle.

There are generally many ways in which the amplituhedron can be triangulated. The BFCW recursion relations constitute one such triangulation: each term in the recursion relations corresponds to a separate triangle, and the statement of which BFCW terms are required to build up the amplitude is a statement of which triangles are necessary to assemble the entire amplituhedron. In this context, the advantage of the amplituhedron over BCFW is that it does not give preferential treatment to any given triangulation, and allows for the important possibility of using significantly more efficient triangulations of the amplitude.\footnote{A full understanding of the various possible triangulations of the amplituhedron is missing and is currently the subject of investigation.} Each triangulation will have boundaries which are internal in the amplituhedron. These boundaries are formed by specifying a choice of sign for some quantity which, while necessary for the triangulation, shouldn't have been specified from the perspective of the entire amplituhedron. Hence, these ``fake'' boundaries end up inside the amplituhedron and ultimately play no role in the full scattering amplitude. 

From a given triangulation, understanding the assembly process whereby triangles glue together to form the amplitude is an important step in the construction of the amplitude integrand. A precise understanding of this geometric process is still lacking however; while the relatively simple case of two loops was treated in \cite{Franco:2014csa}, more general scenarios are not understood. One of the consequences of this work is a better understanding of this process.

\subsection{Regions} 
\label{sec:regionsintro}

When stratifying the amplituhedron, it often happens that specifying which $\Delta^{(i)}_I$s and $\Delta^{(i,j)}$s have been shut off is insufficient information to completely characterize a given boundary. This is due to the fact that the vanishing of those minors has multiple solutions, i.e.\ these conditions are satisfied on disjoint domains of the remaining \pl coordinates. We call each of these domains a \textit{region}. 

The counting of all regions is what is referred to in \cite{Franco:2014csa} as the \textit{full} stratification; merely counting $\{ \Delta^{(i)}_I, \Delta^{(i,j)} \}$ labels, which denote the set of vanishing minors, is referred to as the \textit{mini} stratification. The mini stratification is insensitive to the multiplicity of independent solutions; in other words, if multiple regions have the same set of vanishing minors they are not counted multiple times in the mini stratification.

\begin{figure}[h]
\begin{center}
\includegraphics[scale=0.5]{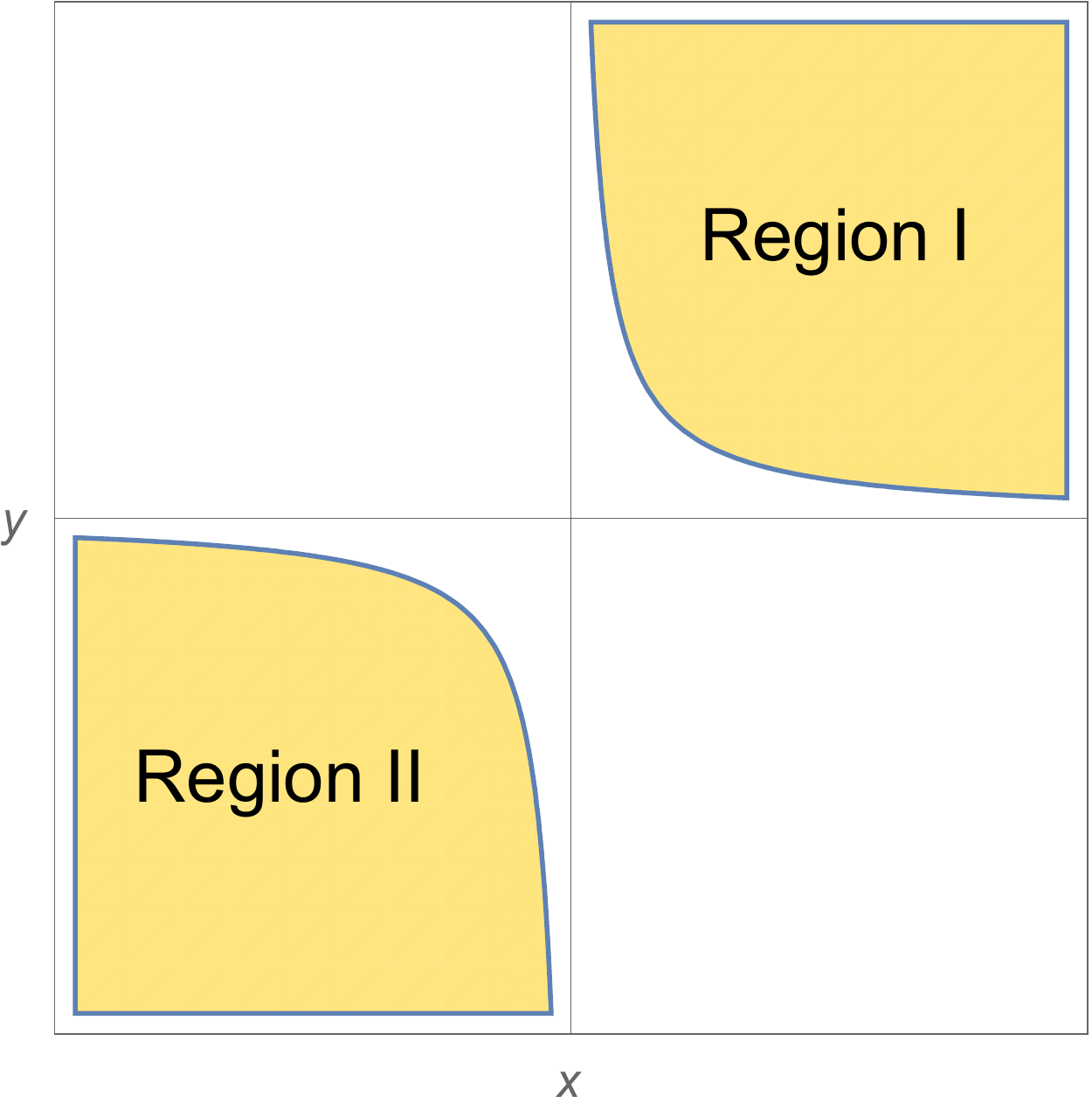}
\caption{For a given set of equalities and inequalities on minors $\Delta^{(i)}_I$s and $\Delta^{(j,k)}$s, there may be multiple regions. The yellow regions both satisfy $\Delta^{(1,2)}>0$, where $\Delta^{(j,k)}$ is given in \eref{eq:regionsminor}.}
\label{fig:regions1and2}
\end{center}
\end{figure}

As an example of the difference between the full and the mini stratification, let us look at a specific boundary in the $k=0$, $n=4$, $L=2$ amplituhedron. We begin by rewriting $\Delta^{(1,2)}$ from \eref{eq:4x4minor} by solving for the \pl relations \eref{eq:pluckrel}. $\Delta^{(1,2)}$ can then be recast into the convenient form
\begin{eqnarray} \label{eq:4x4minorconvenient}
\Delta^{(1,2)}  & = & \frac{\left(\Delta^{(1)}_{12} \Delta^{(2)}_{13} - \Delta^{(1)}_{13} \Delta^{(2)}_{12} \right) \Big(\Delta^{(1)}_{13} \Delta^{(2)}_{34} - \Delta^{(1)}_{34} \Delta^{(2)}_{13} \Big)}{\Delta^{(1)}_{13} \Delta^{(2)}_{13}} \nonumber \\
& + & \frac{\Big(\Delta^{(1)}_{23} \Delta^{(2)}_{13} - \Delta^{(1)}_{13} \Delta^{(2)}_{23} \Big) \Big(\Delta^{(1)}_{13} \Delta^{(2)}_{14} - \Delta^{(1)}_{14} \Delta^{(2)}_{13} \Big)}{\Delta^{(1)}_{13} \Delta^{(2)}_{13}} \, \, .
\end{eqnarray}
If we now go to the boundary obtained by shutting off $\Delta_{23}^{(1)}=\Delta_{14}^{(1)}=0$, $\Delta^{(1,2)}$ takes the schematic form
\begin{equation} \label{eq:regionsminor}
\Delta^{(1,2)}= \frac{1}{k_1} \Big[  \left( \ldots \right)\left( \ldots \right)  -k_2 \Big]
\end{equation}
where $k_1$ and $k_2$ are products of \pl coordinates and hence positive. The round brackets, however, are independently free to be positive or negative. Denoting the first bracket as $x$ and the second as $y$, we see that $\Delta^{(1,2)} > 0$ is equivalent to $x y - k_2 >0$, which is satisfied on disjoint domains of the $x$-$y$ plane, as illustrated in \fref{fig:regions1and2}. Each of these domains is a separate region. 

It may be possible that the additional information provided by the full stratification is unimportant to the task of constructing the amplitude integrand, and that for all practical purposes the mini stratification is sufficient. Such speculations, however, fall beyond the scope of this article. Here we shall primarily be interested in the mini stratification, since as we argue in \sref{sec:assemblyRegions}, the process of gluing boundaries to assemble geometrically simple objects from complicated building blocks may be insensitive to which stratification we're dealing with.

\section{Positivity Sectors} 
\label{sec:possectors}

As described above, the amplituhedron is defined as the region on which it is possible to simultaneously satisfy all positivity conditions \eref{eq:Cconditions}. There are however closely related objects to the amplituhedron which have a simpler definition and a considerably simpler structure; an example of this is the \textit{deformed amplituhedron}, which was discovered and discussed in \cite{Franco:2014csa}. 

A different and particularly simple example is the $L^{th}$ power of the one-loop amplituhedron, which for $n=4$ is simply given by $G_+ (0,4;1)^L$. This space is entirely characterized by the positivity of the $6 L$ \pl coordinates $\Delta^{(i)}_I$, which arise from the $L$ different $D_{(i)}$ in \eref{eq:defamplbig}. In particular, all the $4 \times 4$ minors $\Delta^{(j,k)}$ formed by taking pairs $D_{(j)}$ and $D_{(k)}$, are \textit{not constrained} to the positive domain.

It is a beautiful result, which follows from \cite{2005arXiv09129W}, that $G_+ (0,4;1)^L$ has Euler number $\mathcal{E} =1$ for all $L$. This simple Euler number is a remarkable statement on the geometry of $G_+ (0,4;1)^L$, and can only be satisfied if its $33^L$ boundaries of different dimensionality are finely balanced such that their alternating sum is equal to 1. 

\subsection{Sectors and Their Labels} 
\label{sec:sectorlabels}
The amplituhedron constitutes a subregion of $G_+ (0,4;1)^L$, formed by specifying the positivity of the $4 \times 4$ minors $\Delta^{(i,j)}$. We shall call this subregion a \textit{sector}: a sector in $G_+ (0,4;1)^L$ is given by additionally specifying the sign requirements on the various $\Delta^{(i,j)}$. Under this terminology, the amplituhedron is the sector given by $\Delta^{(i,j)} > 0 $ for all $i$ and $j$.

We see that there are many other sectors, each given by a different choice of signs for the various $\Delta^{(i,j)}$. For $L$ loops, we have $\binom{L}{2}$ different (but generally not independent) $\Delta^{(i,j)}$s. Since each $\Delta^{(i,j)}$ can be constrained to the positive or negative domain, we have $2^{\binom{L}{2}}$ different sectors. 

We shall now construct a convenient labeling for the sectors. The label will be a collection of signs, each one corresponding to the domain of a $\Delta^{(i,j)}$. We shall order this collection according to the lexicographic order of $(i,j)$ in $\Delta^{(i,j)}$; for example, for $L=3$ the first entry corresponds to the sign of $\Delta^{(1,2)}$, the second entry corresponds to the sign of $\Delta^{(1,3)}$ and the third entry corresponds to the sign of $\Delta^{(2,3)}$. Hence, the 3-loop amplituhedron expounded in \cite{Franco:2014csa} is the sector $\{ +, +, +\}$ in $G_+ (0,4;1)^3$; the remaining sectors in $G_+ (0,4;1)^3$ are $\{ +, +, -\}$, $\{ + , - , + \}$, $\{ - , + , + \}$, $\{ + , - , - \}$, $\{ - , + , - \}$, $\{ - , - , + \}$ and $\{ - , - , - \}$.

One of the aims of this paper is to explore the physical significance of the various sectors, and how they glue together to form the simpler object $G_+ (0,4;1)^L$. As mentioned in \sref{sec:amplintro}, our results have the additional benefit of providing a useful handle on the open issue of constructing the amplituhedron from its triangulations.

\subsection{The Log of the Amplitude}
\label{sec:logsectors}

Aside from enabling us to study in generality the important process of gluing boundaries, positivity sectors are interesting in their own right. Indeed, the log of the amplitude, which is what connects the amplitude to the S-matrix through $S \sim \log (A)$, may in certain cases be obtained by gluing certain positivity sectors together. If we write the loop expansion of the amplitude as
\begin{equation}
A = 1 + g A_1 + g^2 A_2 + g^3 A_3 + \ldots 
\end{equation}
where $A_L$ is the $L$-loop contribution to the amplitude, we see that the second-order contribution to the log of the amplitude is $\frac{A_1^2}{2}-A_2$. Neglecting the $\frac{1}{2}$ which is related to the symmetrization of variables, this contribution amounts to the difference between the two-loop amplitude and the square of the one-loop amplitude, i.e.\ between $G_+ (0,4;1)^2$ and $G_+ (0,4;2)$. Following the discussion in the previous section, $G_+ (0,4;2)$ is nothing other than the $\{ + \}$ sector of $G_+ (0,4;1)^2$. Hence, their difference is precisely the $\{ -\}$ sector of $G_+ (0,4;1)^2$, i.e.\ the one given by imposing $\Delta^{(1,2)} < 0$.

It is more difficult, however, to make a similar statement about the third-order contribution to the log of the amplitude. Following the analysis in \cite{Arkani-Hamed:2013kca}, we see that the relevant positivity sectors are $\{ + , - , - \}$, $\{ - , + , - \}$, $\{ - , - , + \}$ and $\{ - , - , - \}$. However, the final positivity sector contributes with a factor of 2, and it is unknown how to turn such prefactors into geometric statements. Nevertheless, in \sref{sec:formingallspaces} we shall see that it is possible to interpret individual terms in the three-loop log of the amplitude as geometric spaces, which naturally segment $G_+ (0,4;1)^3$.

\section{Three-Loop Positivity Sectors} 
\label{sec:3loopsectors}

We shall now study the positivity sectors of $G_+ (0,4;1)^3$ in detail. In this section we will present the stratification of all 8 sectors. In \sref{sec:assembling} we shall turn our attention to the process of gluing sectors together.

\subsection{The $\{+,+,+\}$ Sector} 
\label{sec:+++}

Let us begin by reviewing the stratification of $G_+ (0,4;3)$, already presented in \cite{Franco:2014csa}. As described in \sref{sec:amplboundaries} and discussed in detail in \cite{Franco:2014csa}, we shall begin by constructing the $\Gamma_0$ boundaries and in each element in $\Gamma_0$  find the appropriate $\Gamma_1$ boundaries.

\paragraph{$\boldsymbol{\Gamma_0}$ Boundaries.}
To construct $\Gamma_0$ we begin by shutting off all combinations of \pl coordinates in a manner consistent with the \pl relations and $\Delta^{(i)}_I>0$. We shall initially ignore the requirements from $\Delta^{(j,k)}>0$, and take them into account in a second step. We note that there is already an interesting non-trivial structure here, since many combinations of \pl coordinates cannot be simultaneously turned off. This has multiple causes:
\begin{itemize}
\item The \pl relations may become violated. As an example, we may not turn off both $\Delta^{(i)}_{12}=\Delta^{(i)}_{14}=0$ on the non-zero domain of the remaining \pl coordinates, because the \pl relation \eref{eq:pluckrel} would become 
\begin{equation}
0 + 0 = \Delta^{(i)}_{13} \Delta^{(i)}_{24} \nonumber
\end{equation}
which cannot possibly be satisfied if $\Delta^{(i)}_{13},\Delta^{(i)}_{24} \neq 0$. This statement is simply a statement of the fact that the \pl coordinates are not all independent, and must be shut off consistently.
\item The positivity of the \pl coordinates may become violated. As an example, if we attempt to turn off exclusively $\Delta^{(i)}_{13}=0$, we see that the \pl relation becomes
\begin{equation}
\Delta^{(i)}_{12}\Delta^{(i)}_{34}+\Delta^{(i)}_{14}\Delta^{(i)}_{23}=0
\end{equation}
which is not in itself inconsistent, but may not be satisfied on the positive domain $\Delta^{(i)}_{I}>0$. Here we see that it is the requirement of positivity that removes ``by hand'' certain potential boundaries.
\end{itemize}
In \sref{sec:formingallspaces} we shall see that the structure that ensues from taking the above into consideration is the stratification of $G_+ (0,4;1)^3$, i.e.\ the stratification of all 8 positivity sectors assembled together. The number of $d$-dimensional boundaries of $G_+ (0,4;1)^3$ is denoted $\mathbb{N}^{(d)}$ and presented in \tref{tab:oneLcubestrat}, where we find the Euler number to be equal to 1.

\begin{table}
\begin{center}
\bigskip 
{\small
\begin{tabular}{|c|c|}
\hline
\textbf{Dim} & $\boldsymbol{\mathbb{N}}$ \\
\hline
\textbf{12} &  1 \\
\hline
\textbf{11} & 12 \\
\hline
\textbf{10} & 78 \\
\hline
\textbf{9} & 340 \\
\hline
\textbf{8} & 1\,086 \\
\hline
\textbf{7} & 2\,640 \\
\hline
\textbf{6} & 4\,960 \\
\hline
\textbf{5} & 7\,200 \\
\hline
\textbf{4} & 7\,956 \\
\hline
\textbf{3} & 6\,480 \\
\hline
\textbf{2} & 3\,672 \\
\hline
\textbf{1} & 1\,296 \\
\hline
\textbf{0} & 216 \\
\hline
\end{tabular}
}
\bigskip
\caption{Number of boundaries of the cube of the 1-loop stratification of the amplituhedron. The Euler number is easily computed and found to be $\mathcal{E} = 216 - 1\,296 + \ldots - 12 + 1 = 1 $.\label{tab:oneLcubestrat}}
\end{center}
\end{table}

We shall now study the effects of imposing the positivity of $\Delta^{(j,k)}>0$. This requirement is what forces us into the $\{+,+,+\}$ sector of $G_+ (0,4;1)^3$. There are again two sources of potential obstructions to reaching a boundary by shutting off a set of \pl coordinates:
\begin{itemize}
\item Each $4 \times 4$ minor is composed of positive terms and negative terms, as seen in \eref{eq:4x4minor}. It is possible to turn off a set of \pl coordinates that does not violate the \pl relations or the positivity of the $\Delta^{(i)}_I$s, but that turns off all positive terms in a $\Delta^{(i,j)}$. An example of this is turning off $\Delta^{(i)}_{12}=\Delta^{(i)}_{34}=\Delta^{(i)}_{14}=\Delta^{(i)}_{23}=\Delta^{(i)}_{13}=0$, which forces
\begin{equation}
\Delta^{(i,j)} = 0 + 0 + 0 + 0 - 0 - \Delta^{(i)}_{24} \Delta^{(j)}_{13} \; .
\end{equation}
Here it is clearly impossible for $\Delta^{(i,j)}>0$ on the domain $\Delta^{(i)}_{I},\Delta^{(j)}_{J} >0$. This obstruction is another example of positivity removing boundaries ``by hand'': while there is nothing a priori inconsistent with the requirement that $\Delta^{(i)}_{12}=\Delta^{(i)}_{34}=\Delta^{(i)}_{14}=\Delta^{(i)}_{23}=\Delta^{(i)}_{13}=0$, the extended positivity requirements of $\mathcal{C} \in G_+ (0,4;L)$ oblige us to discard this potential boundary. 
\item The various $4 \times 4$ minors are not always independent: linear relations among them may form on some choices of turned-off \pl coordinates. In such cases, we could envision that the simultaneous positivity of all $\Delta^{(j,k)}$ may become impossible to satisfy. In the $\Gamma_0$ structure, this effect never comes into play in the $\{ + , + , + \}$ sector, but does occur in other sectors. We shall therefore return to this issue in \sref{sec:++-} and \sref{sec:+--and---}. However, when studying the $\Gamma_1$ structure, the linear relations will become extremely important in all sectors and will give rise to an abundance of scenarios, which we shall presently describe.
\end{itemize}
After removing those boundaries in $\mathbb{N}^{(d)}$ that violate the positivity requirements \mbox{$\Delta^{(i,j)}>0$}, we have completed the construction of the $\Gamma_0$ structure for the $\{+,+,+\}$ sector.\footnote{We remind the reader that we are always referring to the mini stratification in this paper, unless otherwise stated.} The number of $d$-dimensional boundaries in $\Gamma_0$ is denoted $\mathcal{N}^{(d)}$ and presented in \tref{tab:+++strat}.

\paragraph{$\boldsymbol{\Gamma_1}$ Boundaries.}
Let us now proceed and construct $\Gamma_1$, formed by shutting off all possible combinations of $\Delta^{(i,j)}$s, without shutting off any \pl coordinates or violating the positivity of those $\Delta^{(k,l)}$s that have not been turned off. If a $\Delta^{(i,j)}$ is trivially zero it cannot of course be shut off, and if a $\Delta^{(i,j)}$ only consists of positive terms, the only way to shut it off is by turning off \pl coordinates, which would take us to a different site in $\Gamma_0$ in which this $\Delta^{(i,j)}$ appears to be zero trivially. Hence, for each site in $\Gamma_0$, i.e.\ for each choice of \pl coordinates tuned to zero, the $\Gamma_1$ structure will depend on how many $\Delta^{(i,j)}$s have both positive and negative terms. We will denote these cases by $N \Delta^{(i,j)}$, where $N$ is the number of such $\Delta^{(i,j)}$s. 

In each $\Gamma_0$ site one of the following scenarios may occur:
\begin{enumerate}
\item[0.] All three minors are trivially zero or manifestly positive, i.e.\ $N=0$. For $\Gamma_0$ sites with $0 \Delta^{(i,j)}$ the $\Gamma_1$ structure is trivial.
\item For $\Gamma_0$ sites with $N=1$, we obtain a new boundary by shutting off the relevant $\Delta^{(i,j)}$. Hence, the $\Gamma_1$ structure as illustrated in \fref{fig:1Delta2Delta}(a).
\item For $\Gamma_0$ sites with $N=2$, we may shut off either minor, or both. This gives a $\Gamma_1$ structure as illustrated in \fref{fig:1Delta2Delta}(b). It is possible to envision that in certain $\Gamma_0$ sites these minors may be linearly related, e.g.\ through $\Delta^{(i,j)} \sim \Delta^{(i,k)}$; this would cause a different $\Gamma_1$ structure, as the two minors would need to either be shut off simultaneously, or not at all. A thorough analysis of this sector shows that this never occurs, and so the only $\Gamma_1$ structure that exists for $2 \Delta^{(i,j)}$ sites in $\Gamma_0$ is the one in \fref{fig:1Delta2Delta}(b).

\begin{figure}[h]
\begin{center}
\includegraphics[width=8cm]{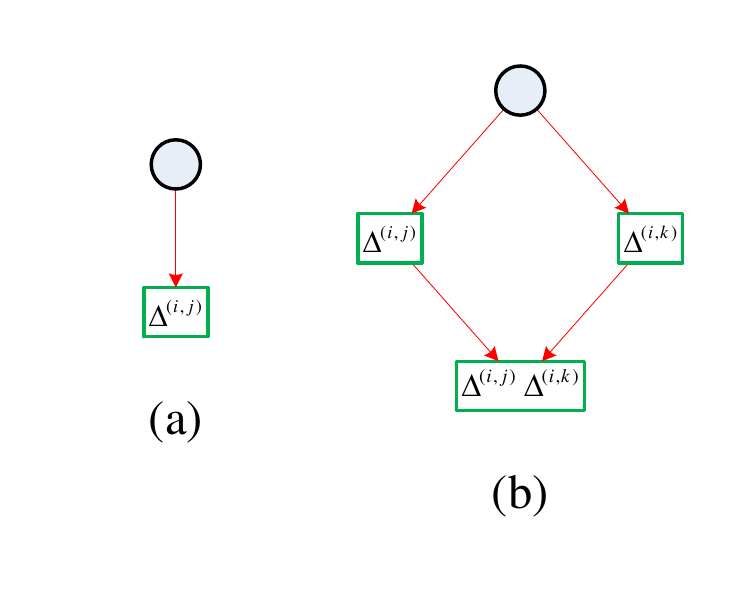}
\vspace{-.6cm}\caption{The general structure of $\Gamma_1$'s emanating from: (a) $1\Delta^{(i,j)}$ and (b) $2\Delta^{(i,j)}$ sites in $\Gamma_0$. The blue circle represents a site in $\Gamma_0$. Each green box is a boundary existing in the $\Gamma_1$ structure, where we have turned off the minors present inside the box. This structure has been ordered by dimensionality, where each vertical downwards step constitutes the loss of one degree of freedom.}
\label{fig:1Delta2Delta}
\end{center}
\end{figure}

\item For $\Gamma_0$ sites with $N=3$, a large number of different $\Gamma_1$ structures arise. These are categorized into Types and are all shown in \fref{fig:3Ltrees}. 
\begin{figure}[ht]
\begin{center}
\includegraphics[width=14cm]{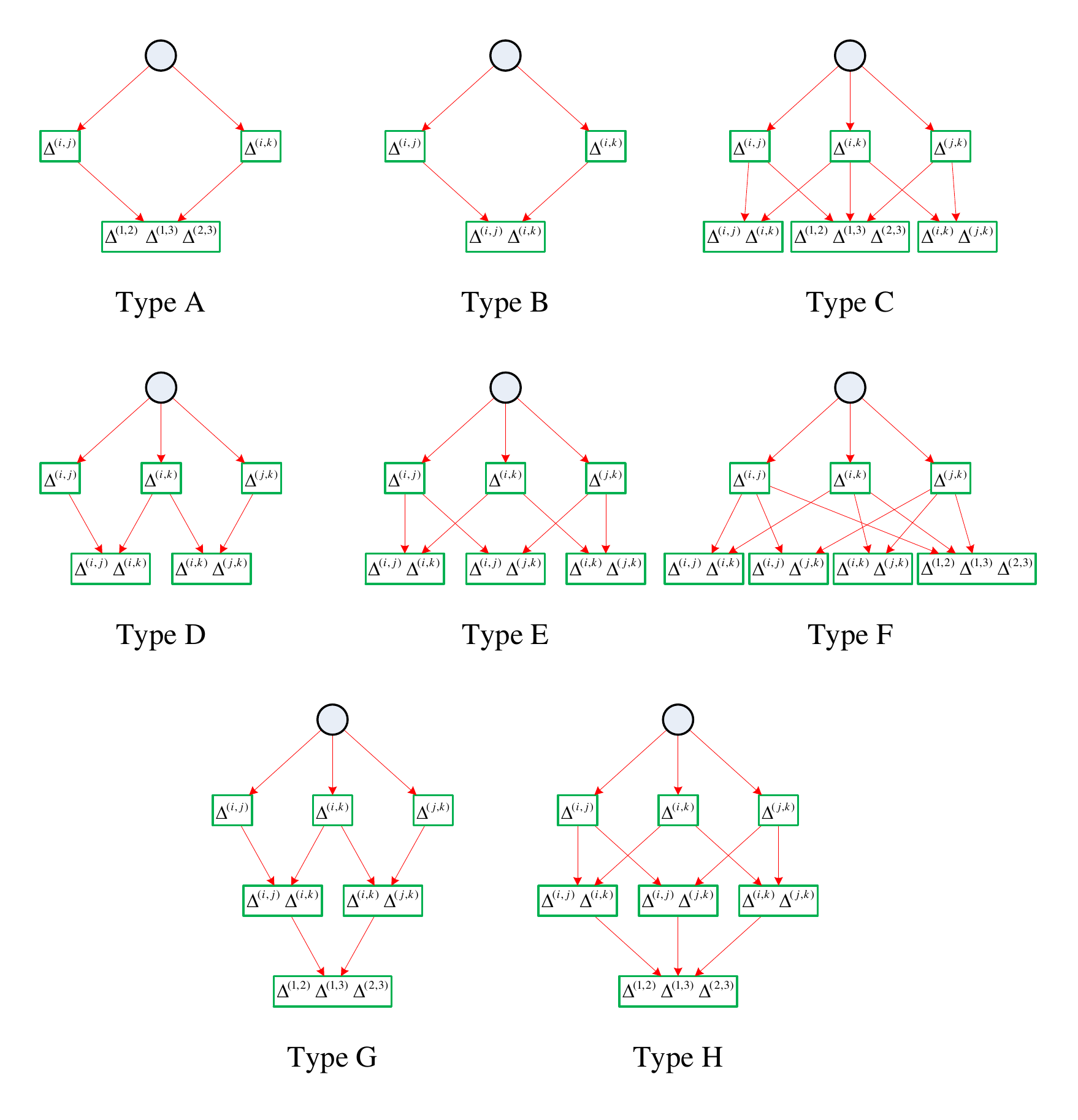}
\caption{Full classification of $\Gamma_1$ boundary structures arising from $3 \Delta^{(i,j)}$ sites in $\Gamma_0$. Each box indicates the minors that have been shut off to reach that boundary, and the blue circle is the relevant site in $\Gamma_0$.}
\label{fig:3Ltrees}
\end{center}
\end{figure}
Let us briefly comment on the origin of the different Types. If $\Delta^{(1,2)}$, $\Delta^{(1,3)}$ and $\Delta^{(2,3)}$ may be shut off completely independently we expect a $\Gamma_1$ structure of Type H, where are all combinations of the various $\Delta^{(i,j)}=0$ appear. If we instead have a situation where for example
\begin{equation} \label{eq:typeAequation}
\Delta^{(1,2)} = k_1 \Delta^{(2,3)} - k_2 \Delta^{(1,3)} \; \; , \quad k_1, k_2 >0
\end{equation}
we see that we are not allowed to exclusively turn off $\Delta^{(2,3)}$, since it would become impossible for $\Delta^{(1,2)}$ and $\Delta^{(1,3)}$ to simultaneously be positive. Additionally, shutting off any pair of $4 \times 4$ minors will automatically force the third to be zero. Hence, the $\Gamma_1$ structure we obtain from this type of situation is that of Type A. The other Types arise from modifications of \eref{eq:typeAequation}, where $k_1$ and $k_2$ are allowed to take on negative values, or be unrestricted, and by the possible presence of another term $k_3$ to the right-hand side of \eref{eq:typeAequation}. An example of a $\Gamma_0$ site which gives rise to a relation of the type \eref{eq:typeAequation} is $\Delta^{(1)} _{14}=\Delta^{(2)} _{12}=\Delta^{(2)} _{14}=\Delta^{(2)} _{24}=\Delta^{(3)} _{13}=\Delta^{(3)} _{14}=\Delta^{(3)} _{34}=0$. Solving for the \pl relations, we see that the $4 \times 4$ minors are related through
\begin{equation}
\Delta^{(1,2)} = \frac{\Delta^{(1)}_{12}}{\Delta^{(3)}_{12}} \Delta^{(2,3)} - \frac{\Delta^{(1)}_{12} \Delta^{(2)}_{13}}{\Delta^{(3)}_{12} \Delta^{(1)}_{13}} \Delta^{(1,3)} \; \; .
\end{equation}
\end{enumerate}

The full information on the number of boundaries for $N \Delta^{(i,j)}$s of various dimensionality is presented in \tref{tab:+++Gamma0Gamma1}.

\begin{table}
\begin{center}
\bigskip
{\footnotesize
\begin{tabular}{|@{\hskip 3pt}c@{\hskip 3pt}|c|c|c|c|c|c|c|c|c|c|c|}
\hline
$\boldsymbol{\{ + , + , + \}}$ & \multicolumn{8}{c|}{\textbf{3} $\boldsymbol{\Delta^{(i,j)}}$} & \multirow{2}{*}{\textbf{2} $\boldsymbol{\Delta^{(i,j)}}$} & \multirow{2}{*}{\textbf{1} $\boldsymbol{\Delta^{(i,j)}}$} & \multirow{2}{*}{\textbf{0} $\boldsymbol{\Delta^{(i,j)}}$} \\
 \cline{1-9}
\multicolumn{1}{|c|}{\textbf{Dim}} & \textbf{A} & \textbf{B} & \textbf{C} & \textbf{D} & \textbf{E} & \textbf{F} & \textbf{G} & \textbf{H} &  & &  \\
\hline
\multicolumn{1}{|c|}{\textbf{12}} & 0 & 0 & 0 & 0 & 0 & 0 & 0 & 1 & 0 & 0 & 0 \\
\hline 
\multicolumn{1}{|c|}{\textbf{11}} & 0 & 0 & 0 & 0 & 0 & 0 & 0 & 12 & 0 & 0 & 0 \\
\hline 
\multicolumn{1}{|c|}{\textbf{10}} & 0 & 0 & 0 & 0 & 0 & 0 & 0 & 78 & 0 & 0 & 0 \\
\hline 
\multicolumn{1}{|c|}{\textbf{9}} & 0 & 0 & 0 & 0 & 0 & 4 & 0 & 324 & 0 & 12 & 0 \\
\hline 
\multicolumn{1}{|c|}{\textbf{8}} & 0 & 12 & 48 & 0 & 0 & 12 & 0 & 726 & 96 & 108 & 0 \\
\hline 
\multicolumn{1}{|c|}{\textbf{7}} & 48 & 96 & 144 & 96 & 48 & 12 & 12 & 600 & 576 & 528 & 0 \\
\hline 
\multicolumn{1}{|c|}{\textbf{6}} & 144 & 120 & 144 & 96 & 0 & 2 & 0 & 144 & 1\,080 & 1\,584 & 176 \\
\hline 
\multicolumn{1}{|c|}{\textbf{5}} & 144 & 0 & 24 & 0 & 0 & 0 & 0 & 0 & 792 & 2\,424 & 1\,056 \\
\hline 
\multicolumn{1}{|c|}{\textbf{4}} & 24 & 0 & 0 & 0 & 0 & 0 & 0 & 0 & 240 & 1\,848 & 2\,544 \\
\hline 
\multicolumn{1}{|c|}{\textbf{3}} & 0 & 0 & 0 & 0 & 0 & 0 & 0 & 0 & 24 & 672 & 3\,264 \\
\hline 
\multicolumn{1}{|c|}{\textbf{2}} & 0 & 0 & 0 & 0 & 0 & 0 & 0 & 0 & 0 & 96 & 2\,424 \\
\hline 
\multicolumn{1}{|c|}{\textbf{1}} & 0 & 0 & 0 & 0 & 0 & 0 & 0 & 0 & 0 & 0 & 1\,008 \\
\hline 
\multicolumn{1}{|c|}{\textbf{0}} & 0 & 0 & 0 & 0 & 0 & 0 & 0 & 0 & 0 & 0 & 186 \\
\hline 
\end{tabular}
}
\bigskip
\caption{Number of boundaries in $\Gamma_0$ with $N \Delta^{(i,j)}$ minors that have both positive and negative terms, and may naively be set to zero. As shown in \fref{fig:3Ltrees} and exemplified in \eref{eq:typeAequation}, for $3 \Delta^{(i,j)}$s there are often cases where some minors may in reality \textit{not} be set to zero, due to linear relations among the various minors.\label{tab:+++Gamma0Gamma1}}
\end{center}
\end{table}

\paragraph{The Total Number of Boundaries.}
We are now ready to present the full boundary structure of the $\{+,+,+\}$ sector of $G_+ (0,4;1)^3$. The number of boundaries in the $\Gamma_0$ structure can be simply computed by summing the number of boundaries in each row of \tref{tab:+++Gamma0Gamma1}. For example, the number of 4-dimensional boundaries is 
\begin{equation}
\mathcal{N}^{(4)} = 24+240+1\,848+2\,544=4\,656 \; .
\end{equation}
All boundaries in $\Gamma_0$ are enumerated in the first column of \tref{tab:+++strat}.

\begin{table}
\begin{center}
\bigskip 
{\small
\begin{tabular}{|c|c|c|c|}
\cline{2-4}
\multicolumn{1}{c|}{} & \multicolumn{2}{c|}{$\boldsymbol{\{ +, +, + \}}$} \\
\hline
\textbf{Dim} & $\boldsymbol{\mathcal{N}}$ & $\boldsymbol{\mathfrak{N}}$ \\
\hline
\textbf{12} &  1 & 1 \\
\hline
\textbf{11} & 12 & 15 \\
\hline
\textbf{10} & 78 & 117 \\
\hline
\textbf{9} & 340 & 611 \\
\hline
\textbf{8} & 1\,002 & 2\,244 \\
\hline
\textbf{7} & 2\,160 & 5\,908 \\
\hline
\textbf{6} & 3\,490 & 10\,996 \\
\hline
\textbf{5} & 4\,440 & 13\,956 \\
\hline
\textbf{4} & 4\,656 & 12\,044 \\
\hline
\textbf{3} & 3\,960 & 7\,488 \\
\hline
\textbf{2} & 2\,520 & 3\,504 \\
\hline
\textbf{1} & 1\,008 & 1\,128 \\
\hline
\textbf{0} & 186 & 186 \\
\hline
\end{tabular}
}
\bigskip
\caption{Number of boundaries $\mathfrak{N}$ of various dimensions of $G_+(0,4;3)$, or equivalently the $\{ +,+,+\}$ sector of $G_+ (0,4;1)^3$. $\mathcal{N}$ is the number of boundaries in the $\Gamma_0$ structure, which does not take into account boundaries arising from $\Delta^{(i,j)}_I \to 0$.\label{tab:+++strat}}
\end{center}
\end{table}

In order to compute the additional boundaries arising from the $\Gamma_1$ structure at the various sites in $\Gamma_0$, we add the number of boundaries shown in Figures \ref{fig:1Delta2Delta} and \ref{fig:3Ltrees} of appropriate codimension, to the boundaries in $\mathcal{N}^{(d)}$. Let us illustrate this computation for the 8-dimensional boundaries $\mathfrak{N}^{(8)}$. Here, 9-dimensional boundaries from $1 \Delta^{(i,j)}$ sites will each contribute by 1 to dimension 8. There are 12 such cases. 9-dimensional boundaries from $3 \Delta^{(i,j)}$ sites of Type F or H will each contribute with 3 additional boundaries of dimension 8. This gives a contribution of $3 \cdot (4+324)=984$. Moreover, 10-dimensional boundaries of Type H which will each contribute an additional 3 boundaries to dimension 8, giving a contribution of $3 \cdot 78 = 234$. Finally, there are 12 boundaries of Type H of dimension 11 that will each give a contribution of 1 to the 8-dimensional boundaries we are counting. Hence, in total we have
\begin{equation}
\mathfrak{N}^{(8)} = \mathcal{N}^{(8)} + 12 + 984 + 234 + 12 = 2\,244 \; \; .
\end{equation}
Performing this computation for $d=0,\ldots,12$ will yield the total number of boundaries of the $\{+,+,+\}$ sector of $G_+ (0,4;1)^3$, which we present in the second column of \tref{tab:+++strat}.

We may now compute the Euler number of this sector:
\begin{equation}
\mathcal{E} = 186 - 1\,128 + \ldots - 15 + 1 = -14 \; \; .
\end{equation}
A few remarks are in order. First, we note that the 3-loop amplituhedron we have just presented has a total of $58\,198$ boundaries of various dimensions, with the largest contribution coming from $\mathfrak{N}^{(5)} = 13\,956$. In order to obtain such a small Euler number, there have to be extraordinary cancellations; the fact that these occur is a reflection of the highly geometric nature of the amplituhedron. A common question regarding the boundaries of the amplituhedron is whether these cancellations are an artefact of an inefficient geometric description of scattering amplitudes, analogously to the inefficient computation of gluon amplitudes through Feynman diagrams. This is however \textit{not} the case: each of the boundaries in \tref{tab:+++strat} represents a \textit{physical} singularity of the scattering amplitude, which cannot possibly depend on the parametrization used to describe it, and can hence not be discarded or reparametrized in a cleverer way. The amplitude really has thousands of distinct soft and collinear singularities.

Secondly, we remark on the fact that while the Euler number is very small, it is still an order of magnitude larger than the Euler number of $G_+ (0,4;1)^3$. In fact, the amplituhedron's Euler number is naively expected to grow for larger $L$. The boundaries of the $\{ +,+,+\}$ sector must glue together with the boundaries of the other sectors in  a rather non-trivial way to produce an Euler number $\mathcal{E}=1$ for all $L$.

\subsection{The $\{+,+,-\}$, $\{+,-,+\}$ and $ \{-,+,+\}$ Sectors} 
\label{sec:++-}

We shall now explore the stratification of the sectors where two $4 \times 4$ minors are positive and one is negative. As for the sector $\{ +,+,+\}$, we begin from the stratification of $G_+ (0,4;1)^3$ and study the effects of imposing the positivity conditions pertaining to the sector under consideration. Without loss of generality we may restrict ourselves to discussing the sector $\{+,+,-\}$, for which $\Delta^{(1,2)} >0$, $\Delta^{(1,3)} >0$ and $\Delta^{(2,3)} <0$; the $\{+,-,+  \}$ and $\{-,+,+ \}$ sectors will have an identical boundary structure, albeit with swapped labels.

\begin{figure}[h]
\begin{center}
\includegraphics[scale=1]{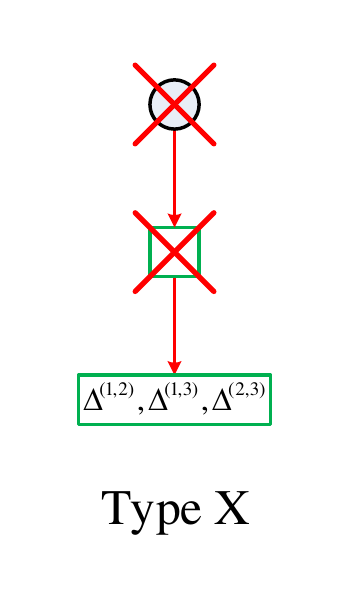}
\vspace{-0.5cm}
\caption{The positivity specifications of the sector under consideration can force certain $\Gamma_0$ sites to disappear, while some parts of their $\Gamma_1$ stratification remain. We have marked with a red cross those parts of the $\Gamma_1$ structure that are absent, also marking the originating site in $\Gamma_0$ following the discussion around \eref{eq:violatedtypeA}.}
\label{fig:typeX}
\end{center}
\end{figure}

\paragraph{$\boldsymbol{\Gamma_0}$ Boundaries.}
The construction of $\Gamma_0$ is obtained from the boundaries in $G_+ (0,4;1)^3$ by excluding those that violate the positivity dictated by the $\{+,+,-\}$ sector. As for the $\{+,+,+\}$ sector, this is divided in two stages:
\begin{itemize}
\item Certain combinations of \pl coordinates set to zero will force any one of the $\Delta^{(i,j)}$s to explicitly violate the positivity conditions of the sector. We must remove all boundaries in $G_+ (0,4;1)^3$ where only negative terms survive for either $\Delta^{(1,2)}$ or $\Delta^{(1,3)}$, or where only positive terms survive for $\Delta^{(2,3)}$. For example, the boundary where $\Delta^{(2)}_{12} = \Delta^{(2)}_{14} = \Delta^{(2)}_{13} = \Delta^{(2)}_{24} = 0$ will force the $4\times 4$ minors to be
\begin{eqnarray} \label{eq:explsectorviolation}
\Delta^{(1,2)} &=& \Delta^{(1)}_{12} \Delta^{(2)}_{34} + 0 + \Delta^{(1)}_{14} \Delta^{(2)}_{23} + 0 - 0 - 0 \nonumber \\
\Delta^{(1,3)} &=& \Delta^{(1)}_{12} \Delta^{(3)}_{34} + \Delta^{(1)}_{34} \Delta^{(3)}_{12} + \Delta^{(1)}_{14} \Delta^{(3)}_{23} + \Delta^{(1)}_{23} \Delta^{(3)}_{14} - \Delta^{(1)}_{13} \Delta^{(3)}_{24} - \Delta^{(1)}_{24} \Delta^{(3)}_{13} \quad \quad  \nonumber \\
\Delta^{(2,3)} &=& 0 + \Delta^{(2)}_{34} \Delta^{(3)}_{12} + 0 + \Delta^{(2)}_{23} \Delta^{(3)}_{14} - 0 - 0 \; \; .
\end{eqnarray}
Here it is impossible for $\Delta^{(2,3)}<0$ on the support of positive \pl coordinates.
\item There are combinations of \pl coordinates which, when shut off, leave all three $\Delta^{(i,j)}$s containing both positive and negative terms, but where there are linear relations among $\Delta^{(1,2)}$, $\Delta^{(1,3)}$ and $\Delta^{(2,3)}$. An example of these $N \Delta^{(i,j)}$ scenarios with $N=3$ was given in \eref{eq:typeAequation}, which in the $\{+,+,+\}$ sector gave rise to a $\Gamma_1$ structure of Type A. In the $\{+,+,- \}$ sector, however, it may be that even the $\Gamma_0$ starting point is invalid, i.e.\ that \eref{eq:typeAequation} may not be satisfied on the non-zero domain of all three large minors. Indeed, in the $\{+,+,-\}$ sector \eref{eq:typeAequation} becomes
\begin{equation} \label{eq:violatedtypeA}
\lvert \Delta^{(1,2)} \rvert = - k_1 \lvert \Delta^{(2,3)} \rvert - k_2 \lvert \Delta^{(1,3)} \rvert \; \; , \quad k_1, k_2 >0
\end{equation}
which cannot possibly be satisfied when all three $\lvert \Delta^{(i,j)} \rvert \neq 0$. Hence, these boundaries must also be removed.
\end{itemize}
We note that in \eref{eq:violatedtypeA} it is still possible to shut off all three minors simultaneously, losing two d.o.f. While this is strictly speaking an effect belonging to the $\Gamma_1$ structure, it is important to discuss it in the $\Gamma_0$ structure, since throwing away sites in $\Gamma_0$ would forfeit our chance of obtaining this lower-dimensional point in $\Gamma_1$. Indeed, what we will do is discard this site from $\Gamma_0$, but remember to include its $\Gamma_1$ contribution to the total number of boundaries, as illustrated in \fref{fig:typeX}.

For the purpose of completeness, we remark that there are more restrictive scenarios yet, associated to relations of the schematic form
\begin{equation}
\Delta^{(1,2)} = k_1 \Delta^{(2,3)} - k_2 \Delta^{(1,3)} - k_3 \; \; , \quad k_1, k_2, k_3 >0 \; \; .
\end{equation}
In the $\{ +,+,+\}$ sector these give rise to a $\Gamma_1$ structure of Type B. In the $\{ +,+,-\}$ sector, however, this relation is more transparently written as
\begin{equation}
\lvert \Delta^{(1,2)} \rvert = - k_1 \lvert \Delta^{(2,3)} \rvert - k_2 \lvert \Delta^{(1,3)} \rvert - k_3
\end{equation}
where it is easy to see that not only is the relation impossible to satisfy when all three $\lvert \Delta^{(1,2)} \rvert \neq 0$, but it is in fact impossible to satisfy for any combination of $\Delta^{(i,j)} \to 0$. These boundaries may be safely removed from the $\Gamma_0$ structure without adversely affecting the $\Gamma_1$ structure; they behave precisely as those that violate the positivity requirements of the sector explicitly, as in \eref{eq:explsectorviolation}.

\paragraph{$\boldsymbol{\Gamma_1}$ Boundaries.}
We shall now characterize the possible $\Gamma_1$ structures emanating from the $\Gamma_0$ sites in this sector. We shall call the structure exemplified by \eref{eq:violatedtypeA} Type X, illustrated in \fref{fig:typeX}. As already discussed, this structure only contributes to $\Gamma_1$ without contributing to $\Gamma_0$. In \sref{sec:assemblingtypeA} we shall provide a geometric understanding of Type X boundaries, where we shall also see that they play a trivial role when assembling sectors, and are hence uninteresting in this context.

Performing a similar analysis to the one for the $\{ +,+,+\}$ sector yields the total $\Gamma_0 + \Gamma_1$ structure for the sector $\{+,+,-\}$, which is identical to that of the sectors $\{ +,-,+\}$ and $\{-,+,+ \}$. This structure is presented in \tref{tab:++-Gamma0Gamma1}.

\begin{table}
\begin{center}
\bigskip
{\footnotesize
\begin{tabular}{|@{\hskip 3pt}c@{\hskip 3pt}|c|c|c|c|c|c|c|c|c|c|c|c|}
\hline
$\boldsymbol{\{ + , + , - \}}$ & \multicolumn{9}{c|}{\textbf{3} $\boldsymbol{\Delta^{(i,j)}}$} & \multirow{2}{*}{\textbf{2} $\boldsymbol{\Delta^{(i,j)}}$} & \multirow{2}{*}{\textbf{1} $\boldsymbol{\Delta^{(i,j)}}$} & \multirow{2}{*}{\textbf{0} $\boldsymbol{\Delta^{(i,j)}}$} \\
 \cline{1-10}
\multicolumn{1}{|c|}{\textbf{Dim}} & \textbf{A} & \textbf{B} & \textbf{C} & \textbf{D} & \textbf{E} & \textbf{F} & \textbf{G} & \textbf{H} & \textbf{X} &  & &  \\
\hline
\multicolumn{1}{|c|}{\textbf{12}} & 0 & 0 & 0 & 0 & 0 & 0 & 0 & 1 & 0 & 0 & 0 & 0 \\
\hline 
\multicolumn{1}{|c|}{\textbf{11}} & 0 & 0 & 0 & 0 & 0 & 0 & 0 & 12 & 0 & 0 & 0 & 0 \\
\hline 
\multicolumn{1}{|c|}{\textbf{10}} & 0 & 0 & 0 & 0 & 0 & 0 & 0 & 78 & 0 & 0 & 0 & 0 \\
\hline 
\multicolumn{1}{|c|}{\textbf{9}} & 0 & 0 & 0 & 0 & 0 & 4 & 0 & 324 & 0 & 0 & 4 & 0 \\
\hline 
\multicolumn{1}{|c|}{\textbf{8}} & 0 & 0 & 48 & 8 & 0 & 12 & 0 & 726 & 0 & 98 & 36 & 0 \\
\hline 
\multicolumn{1}{|c|}{\textbf{7}} & 32 & 32 & 144 & 112 & 64 & 12 & 8 & 604 & 16 & 584 & 200 & 0 \\
\hline 
\multicolumn{1}{|c|}{\textbf{6}} & 96 & 64 & 144 & 80 & 32 & 2 & 0 & 144 & 48 & 1\,128 & 824 & 36 \\
\hline 
\multicolumn{1}{|c|}{\textbf{5}} & 96 & 0 & 24 & 0 & 0 & 0 & 0 & 0 & 48 & 832 & 1656 & 408 \\
\hline 
\multicolumn{1}{|c|}{\textbf{4}} & 16 & 0 & 0 & 0 & 0 & 0 & 0 & 0 & 8 & 240 & 1520 & 1\,364 \\
\hline 
\multicolumn{1}{|c|}{\textbf{3}} & 0 & 0 & 0 & 0 & 0 & 0 & 0 & 0 & 0 & 24 & 624 & 2\,160 \\
\hline 
\multicolumn{1}{|c|}{\textbf{2}} & 0 & 0 & 0 & 0 & 0 & 0 & 0 & 0 & 0 & 0 & 96 & 1\,856 \\
\hline 
\multicolumn{1}{|c|}{\textbf{1}} & 0 & 0 & 0 & 0 & 0 & 0 & 0 & 0 & 0 & 0 & 0 & 856 \\
\hline 
\multicolumn{1}{|c|}{\textbf{0}} & 0 & 0 & 0 & 0 & 0 & 0 & 0 & 0 & 0 & 0 & 0 & 170 \\
\hline 
\end{tabular}
}
\caption{\label{tab:++-Gamma0Gamma1}}
\end{center}
\end{table}

We remark on the similarity between \tref{tab:++-Gamma0Gamma1} and \tref{tab:+++Gamma0Gamma1}; it appears that specifying a different positivity does not alter the variety of $3\Delta^{(i,j)}$ in any meaningful way, aside from introducing the addition of Type X structures. In \sref{sec:+--and---} we shall find that this is also true for the remaining positivity sectors. The reason behind this similarity is simple: the linear relations among the $4 \times 4$ minors are independent of the positivity sector under consideration, since they are trivially true statements. Moreover, the same relations must appear in all sectors, since they involve those $\Delta^{(i,j)}$ with both positive and negative terms, which no sector will automatically exclude. The role of the positivity sector in these relations is to specify the domain under which these relations must operate, and in most cases there is enough leeway in the relation to be present in all sectors. In particular, since the sites in $\Gamma_0$ classified as $2\Delta^{(i,j)}$ all had the structure shown in \fref{fig:1Delta2Delta}, this must also be true in the remaining positivity sectors.

\paragraph{The Total Number of Boundaries.}
Let us now compute the total number of boundaries for these sectors. When computing the boundaries in $\Gamma_0$, we simply sum over the numbers in each row of \tref{tab:++-Gamma0Gamma1}, remembering to not include those in the column for Type X. For example, at dimension $d=4$ we have
\begin{equation}
\mathcal{N}^{(4)} = 16 + 240 + 1\,520 + 1\,364 = 3\,140 \; \; .
\end{equation}
The $\Gamma_0$ boundaries of all dimensions are presented in \tref{tab:++-strat}.

When computing the additional boundaries from $\Gamma_1$, we must remember that each instance of Type X contributes by 1 to the codimension-2 boundaries. Let us illustrate the counting by obtaining the number of 2-dimensional boundaries. Additionally to $\mathcal{N}^{(2)}$ we also get contributions from 3-dimensional $\Gamma_0$ boundaries: each $1 \Delta^{(i,j)}$ contributes by 1 to $d=2$, and each $2 \Delta^{(i,j)}$ contributes by 2. In total, we receive an additional contribution from $d=3$ of $(2 \cdot 24 + 624) = 672$. We also have contributions from 4-dimensional boundaries: each $2 \Delta^{(i,j)}$ will contribute by 1 to dimension $d=2$ and each Type A and Type X will also contribute by 1. Hence, we have an additional contribution of $16 + 8 +240 = 264$. We have no contributions from $d \geq 5$. Hence,
\begin{equation}
\mathfrak{N}^{(2)} = \mathcal{N}^{(2)} + 672 + 264 = 2\,888 \; \; .
\end{equation}
The final results are presented in \tref{tab:++-strat}. 

\begin{table}
\begin{center}
\bigskip 
{\small
\begin{tabular}{|c|c|c|c|}
\cline{2-4}
\multicolumn{1}{c|}{} & \multicolumn{2}{c|}{$\boldsymbol{\{ +, +, - \}}$} \\
\hline
\textbf{Dim} & $\boldsymbol{\mathcal{N}}$ & $\boldsymbol{\mathfrak{N}}$ \\
\hline
\textbf{12} &  1 & 1 \\
\hline
\textbf{11} & 12 & 15 \\
\hline
\textbf{10} & 78 & 117 \\
\hline
\textbf{9} & 332 & 603 \\
\hline
\textbf{8} & 928 & 2\,162 \\
\hline
\textbf{7} & 1\,792 & 5\,472 \\
\hline
\textbf{6} & 2\,550 & 9\,686 \\
\hline
\textbf{5} & 3\,016 & 11\,736 \\
\hline
\textbf{4} & 3\,140 & 9\,800 \\
\hline
\textbf{3} & 2\,808 & 6\,032 \\
\hline
\textbf{2} & 1\,952 & 2\,888 \\
\hline
\textbf{1} & 856 & 976 \\
\hline
\textbf{0} & 170 & 170 \\
\hline
\end{tabular}
}
\bigskip
\caption{Number of boundaries $\mathfrak{N}$ of various dimensions of the $\{ +,+,-\}$ sector of $G_+ (0,4;1)^3$. $\mathcal{N}$ is the number of boundaries in the $\Gamma_0$ structure.\label{tab:++-strat}}
\end{center}
\end{table}

In total we have $49\,658$ boundaries, where the largest contribution comes again from $\mathfrak{N}^{(5)} = 11\,736$. We compute the Euler number, and find
\begin{equation}
\mathcal{E} = 170 - 976 + \ldots -15 + 1 = -10 \; \; ,
\end{equation}
which we again remark is extremely small but still an order of magnitude greater than that of $G_+ (0,4;1)^3$.

\subsection{The $\{+,-,-\}$, $\{-,+,-\}$, $ \{-,-,+\}$ and $ \{-,-,-\}$ Sectors} 
\label{sec:+--and---}

We shall now present the stratification of the remaining positivity sectors in $G_+ (0,4;1)^3$. The analysis follows the same steps as those presented in \sref{sec:+--and---}; this section will therefore be rather brief. 

The positivity sectors $\{+,-,-\}$, $\{-,+,-\}$ and $ \{-,-,+\}$ all have identical stratifications, modulo swaps of labels, and we shall consequently only present one such table. A full analysis on the possible boundaries that are present in these sectors yields \tref{tab:+--Gamma0Gamma1}. As in the previous section, we have Type X boundaries contributing here.

Finally, the fully negative positivity sector $ \{-,-,-\}$ is given in \tref{tab:---Gamma0Gamma1}. We note that here there are no Type X structures. This is simply understood from the fact that when all three $\Delta^{(i,j)}$s are negative, the linear relations among the minors which give rise to Type X structures may be multiplied by an overall sign, yielding the same relation as for the $\{ +,+,+\}$ sector, cf.\ \eref{eq:violatedtypeA}.

\begin{table}
\begin{center}
\bigskip
{\footnotesize
\begin{tabular}{|@{\hskip 3pt}c@{\hskip 3pt}|c|c|c|c|c|c|c|c|c|c|c|c|}
\hline
$\boldsymbol{\{ + , - , - \}}$ & \multicolumn{9}{c|}{\textbf{3} $\boldsymbol{\Delta^{(i,j)}}$} & \multirow{2}{*}{\textbf{2} $\boldsymbol{\Delta^{(i,j)}}$} & \multirow{2}{*}{\textbf{1} $\boldsymbol{\Delta^{(i,j)}}$} & \multirow{2}{*}{\textbf{0} $\boldsymbol{\Delta^{(i,j)}}$} \\
 \cline{1-10}
\multicolumn{1}{|c|}{\textbf{Dim}} & \textbf{A} & \textbf{B} & \textbf{C} & \textbf{D} & \textbf{E} & \textbf{F} & \textbf{G} & \textbf{H} & \textbf{X} &  & &  \\
\hline
\multicolumn{1}{|c|}{\textbf{12}} & 0 & 0 & 0 & 0 & 0 & 0 & 0 & 1 & 0 & 0 & 0 & 0 \\
\hline 
\multicolumn{1}{|c|}{\textbf{11}} & 0 & 0 & 0 & 0 & 0 & 0 & 0 & 12 & 0 & 0 & 0 & 0 \\
\hline 
\multicolumn{1}{|c|}{\textbf{10}} & 0 & 0 & 0 & 0 & 0 & 0 & 0 & 78 & 0 & 0 & 0 & 0 \\
\hline 
\multicolumn{1}{|c|}{\textbf{9}} & 0 & 0 & 0 & 0 & 0 & 4 & 0 & 324 & 0 & 0 & 0 & 0 \\
\hline 
\multicolumn{1}{|c|}{\textbf{8}} & 0 & 8 & 48 & 0 & 4 & 12 & 0 & 726 & 0 & 100 & 2 & 0 \\
\hline 
\multicolumn{1}{|c|}{\textbf{7}} & 32 & 64 & 144 & 80 & 80 & 12 & 4 & 608 & 16 & 592 & 48 & 0 \\
\hline 
\multicolumn{1}{|c|}{\textbf{6}} & 96 & 80 & 144 & 64 & 40 & 2 & 0 & 144 & 48 & 1\,176 & 454 & 16 \\
\hline 
\multicolumn{1}{|c|}{\textbf{5}} & 96 & 0 & 24 & 0 & 0 & 0 & 0 & 0 & 48 & 872 & 1\,264 & 224 \\
\hline 
\multicolumn{1}{|c|}{\textbf{4}} & 16 & 0 & 0 & 0 & 0 & 0 & 0 & 0 & 8 & 240 & 1\,336 & 930 \\
\hline 
\multicolumn{1}{|c|}{\textbf{3}} & 0 & 0 & 0 & 0 & 0 & 0 & 0 & 0 & 0 & 24 & 592 & 1\,672 \\
\hline 
\multicolumn{1}{|c|}{\textbf{2}} & 0 & 0 & 0 & 0 & 0 & 0 & 0 & 0 & 0 & 0 & 96 & 1\,564 \\
\hline 
\multicolumn{1}{|c|}{\textbf{1}} & 0 & 0 & 0 & 0 & 0 & 0 & 0 & 0 & 0 & 0 & 0 & 768 \\
\hline 
\multicolumn{1}{|c|}{\textbf{0}} & 0 & 0 & 0 & 0 & 0 & 0 & 0 & 0 & 0 & 0 & 0 & 160 \\
\hline 
\end{tabular}
}
\caption{\label{tab:+--Gamma0Gamma1}}
\end{center}
\end{table}

\begin{table}
\begin{center}
\bigskip
{\footnotesize
\begin{tabular}{|@{\hskip 3pt}c@{\hskip 3pt}|c|c|c|c|c|c|c|c|c|c|c|}
\hline
$\boldsymbol{\{ - , - , - \}}$ & \multicolumn{8}{c|}{\textbf{3} $\boldsymbol{\Delta^{(i,j)}}$} & \multirow{2}{*}{\textbf{2} $\boldsymbol{\Delta^{(i,j)}}$} & \multirow{2}{*}{\textbf{1} $\boldsymbol{\Delta^{(i,j)}}$} & \multirow{2}{*}{\textbf{0} $\boldsymbol{\Delta^{(i,j)}}$} \\
 \cline{1-9}
\multicolumn{1}{|c|}{\textbf{Dim}} & \textbf{A} & \textbf{B} & \textbf{C} & \textbf{D} & \textbf{E} & \textbf{F} & \textbf{G} & \textbf{H} &  & &  \\
\hline
\multicolumn{1}{|c|}{\textbf{12}} & 0 & 0 & 0 & 0 & 0 & 0 & 0 & 1 & 0 & 0 & 0 \\
\hline 
\multicolumn{1}{|c|}{\textbf{11}} & 0 & 0 & 0 & 0 & 0 & 0 & 0 & 12 & 0 & 0 & 0 \\
\hline 
\multicolumn{1}{|c|}{\textbf{10}} & 0 & 0 & 0 & 0 & 0 & 0 & 0 & 78 & 0 & 0 & 0 \\
\hline 
\multicolumn{1}{|c|}{\textbf{9}} & 0 & 0 & 0 & 0 & 0 & 4 & 0 & 324 & 0 & 0 & 0 \\
\hline 
\multicolumn{1}{|c|}{\textbf{8}} & 0 & 0 & 48 & 12 & 0 & 12 & 0 & 726 & 102 & 6 & 0 \\
\hline 
\multicolumn{1}{|c|}{\textbf{7}} & 48 & 48 & 144 & 144 & 48 & 12 & 0 & 612 & 600 & 72 & 0 \\
\hline 
\multicolumn{1}{|c|}{\textbf{6}} & 144 & 96 & 144 & 120 & 0 & 2 & 0 & 144 & 1\,224 & 474 & 32 \\
\hline 
\multicolumn{1}{|c|}{\textbf{5}} & 144 & 0 & 24 & 0 & 0 & 0 & 0 & 0 & 912 & 1\,248 & 288 \\
\hline 
\multicolumn{1}{|c|}{\textbf{4}} & 24 & 0 & 0 & 0 & 0 & 0 & 0 & 0 & 240 & 1\,296 & 1\,008 \\
\hline 
\multicolumn{1}{|c|}{\textbf{3}} & 0 & 0 & 0 & 0 & 0 & 0 & 0 & 0 & 24 & 576 & 1\,680 \\
\hline 
\multicolumn{1}{|c|}{\textbf{2}} & 0 & 0 & 0 & 0 & 0 & 0 & 0 & 0 & 0 & 96 & 1\,524 \\
\hline 
\multicolumn{1}{|c|}{\textbf{1}} & 0 & 0 & 0 & 0 & 0 & 0 & 0 & 0 & 0 & 0 & 744 \\
\hline 
\multicolumn{1}{|c|}{\textbf{0}} & 0 & 0 & 0 & 0 & 0 & 0 & 0 & 0 & 0 & 0 & 156 \\
\hline 
\end{tabular}
}
\caption{\label{tab:---Gamma0Gamma1}}
\end{center}
\end{table}

We are now ready to present the total number of boundaries of each sector; these are shown in \tref{tab:+--and---strat}. The $\{+,-,-\}$, $\{-,+,-\}$ and $ \{-,-,+\}$ sectors each have a total of $46\,286$ boundaries, where the largest contribution comes from $\mathfrak{N}^{(5)} = 10\,962$. The Euler number is found to be
\begin{equation}
\mathcal{E} = -14 \; \; . 
\end{equation}

Finally, the $ \{-,-,-\}$ sector has a total of $47\,320$ boundaries, with the largest contributor being $\mathfrak{N}^{(5)} = 11\,418$. The Euler number for this sector is 
\begin{equation}
\mathcal{E} = -20 \; \; . 
\end{equation}

At the lower dimensionalities, we note that the more negative the sector under consideration is, the fewer the available boundaries are. This is due to the fact that each $\Delta^{(i,j)}$ is composed of four positive terms and two negative ones; hence, it is combinatorially easier to satisfy $\Delta^{(i,j)}>0$ than $\Delta^{(i,j)}<0$.

\begin{table}
\begin{center}
\bigskip 
{\small
\begin{tabular}{|c|c|c|c|}
\cline{2-4}
\multicolumn{1}{c|}{} & \multicolumn{2}{c|}{$\boldsymbol{\{ +, -, - \}}$} \\
\hline
\textbf{Dim} & $\boldsymbol{\mathcal{N}}$ & $\boldsymbol{\mathfrak{N}}$ \\
\hline
\textbf{12} &  1 & 1 \\
\hline
\textbf{11} & 12 & 15 \\
\hline
\textbf{10} & 78 & 117 \\
\hline
\textbf{9} & 328 & 599 \\
\hline
\textbf{8} & 900 & 2\,130 \\
\hline
\textbf{7} & 1\,664 & 5\,318 \\
\hline
\textbf{6} & 2\,216 & 9\,238 \\
\hline
\textbf{5} & 2\,480 & 10\,962 \\
\hline
\textbf{4} & 2\,522 & 8\,926 \\
\hline
\textbf{3} & 2\,288 & 5\,368 \\
\hline
\textbf{2} & 1\,660 & 2\,564 \\
\hline
\textbf{1} & 768 & 888 \\
\hline
\textbf{0} & 160 & 160 \\
\hline
\end{tabular}
\quad \quad \quad \quad
\begin{tabular}{|c|c|c|c|}
\cline{2-4}
\multicolumn{1}{c|}{} & \multicolumn{2}{c|}{$\boldsymbol{\{ -, -, - \}}$} \\
\hline
\textbf{Dim} & $\boldsymbol{\mathcal{N}}$ & $\boldsymbol{\mathfrak{N}}$ \\
\hline
\textbf{12} &  1 & 1 \\
\hline
\textbf{11} & 12 & 15 \\
\hline
\textbf{10} & 78 & 117 \\
\hline
\textbf{9} & 328 & 599 \\
\hline
\textbf{8} & 906 & 2\,136 \\
\hline
\textbf{7} & 1\,728 & 5\,398 \\
\hline
\textbf{6} & 2\,380 & 9\,544 \\
\hline
\textbf{5} & 2\,616 & 11\,418 \\
\hline
\textbf{4} & 2\,568 & 9\,188 \\
\hline
\textbf{3} & 2\,280 & 5\,376 \\
\hline
\textbf{2} & 1\,620 & 2\,508 \\
\hline
\textbf{1} & 744 & 864 \\
\hline
\textbf{0} & 156 & 156 \\
\hline
\end{tabular}
}
\bigskip
\caption{Number of boundaries $\mathfrak{N}$ of various dimensions of the $\{ +,-,-\}$ and $\{ -,-,-\}$ sectors of $G_+ (0,4;1)^3$. $\mathcal{N}$ is the number of boundaries in the $\Gamma_0$ structure.\label{tab:+--and---strat}}
\end{center}
\end{table}

\section{Assembling the Positivity Sectors} 
\label{sec:assembling}

We are now ready to explore how to glue positivity sectors together, in order to assemble the various spaces of interest. The techniques we develop in this section should be more generally applicable to this type of problem in the context of the amplituhedron and are therefore of relevance in a variety of situations, as discussed in \sref{sec:onshellinside} and \sref{sec:logsectors}. Here we shall primarily be interested in gluing the sectors of $G_+ (0,4;1)^3$ described in \sref{sec:3loopsectors}.

Generally, two things may happen when merging sectors:
\begin{itemize}
\item The $\Gamma_0$ structure may change, by the addition of $\Gamma_0$ sites that are not present in all sectors that are being merged. Hence, gluing the sectors will create a larger object able to probe both types of boundaries; this enlarges the total number of boundaries.
\item In a given $\Gamma_0$ site, i.e.\ in a site which has specified which \pl coordinates $\Delta^{(i)}_I$ are zero, the $\Gamma_1$ structure can change dramatically.
\end{itemize}
While the inclusion of additional $\Gamma_0$ sites is rather trivial to understand and implement, the changes to the $\Gamma_1$ structures are more challenging to deal with. The coming sections will take a pedestrian approach to resolving the issue of systematically dealing with the $\Gamma_1$ structures; we will begin by studying $\Gamma_0$ sites where all large minors $\Delta^{(i,j)}$ are independent, i.e.\ those described by \fref{fig:1Delta2Delta} and Type H in \fref{fig:3Ltrees}. We will then move on to consider $\Gamma_0$ sites where the large minors are not independent, and exemplify this with the study of Type A boundaries shown in \fref{fig:3Ltrees}. This will allow us in \sref{sec:assemblingalgorithm} to construct an algorithm for gluing sectors, which will be used in \sref{sec:formingallspaces} to study the geometry of $G_+ (0,4;1)^3$ as it is being assembled.

Before we begin, let us remark on the general structure of each of the large minors 
\begin{equation}
\label{eq:4x4repeat}
\Delta^{(i,j)} = \Delta^{(i)}_{12} \Delta^{(j)}_{34} + \Delta^{(i)}_{34} \Delta^{(j)}_{12} + \Delta^{(i)}_{14} \Delta^{(j)}_{23} + \Delta^{(i)}_{23} \Delta^{(j)}_{14} - \Delta^{(i)}_{13} \Delta^{(j)}_{24} - \Delta^{(i)}_{24} \Delta^{(j)}_{13}
\end{equation}
in the various positivity sectors. Depending on which $\Gamma_0$ site we are considering, different terms in \eref{eq:4x4repeat} may be present, and the following situations may arise for a given $\Delta^{(i,j)}$:
\begin{itemize}
\item $\Delta^{(i,j)}$ may be \textit{trivially positive}, from only having positive terms turned on. An example of this could be $\Delta^{(i,j)} = \Delta^{(i)}_{12} \Delta^{(j)}_{34}$.
\item It may be \textit{trivially negative}, from only having negative terms turned on, e.g.\ $\Delta^{(i,j)} = - \Delta^{(i)}_{13} \Delta^{(j)}_{24}$.
\item It may have both positive and negative terms turned on, and be constrained to the positive domain through a balance of these terms, e.g.\ $\Delta^{(i,j)} = \Delta^{(i)}_{12} \Delta^{(j)}_{34} - \Delta^{(i)}_{13} \Delta^{(j)}_{24} > 0$. We may say that $\Delta^{(i,j)}$ is \textit{non-trivially positive} here.
\item It may have both positive and negative terms turned on, but have the opposite balance of terms, and be constrained to be negative, e.g.\ $\Delta^{(i,j)} = \Delta^{(i)}_{12} \Delta^{(j)}_{34} - \Delta^{(i)}_{13} \Delta^{(j)}_{24} < 0$. $\Delta^{(i,j)}$ is then said to be \textit{non-trivially negative}.
\item It may be \textit{trivially zero}, from having none of the terms turned on, i.e.\ where we explicitly have $\Delta^{(i,j)} = 0$.
\item It may have both positive and negative terms turned on, but be zero through a precise balance of these terms, e.g.\ $\Delta^{(i,j)} = \Delta^{(i)}_{12} \Delta^{(j)}_{34} - \Delta^{(i)}_{13} \Delta^{(j)}_{24} = 0$. $\Delta^{(i,j)}$ is then said to be \textit{non-trivially zero}.
\end{itemize}

With the above in mind, we shall now move on to study how the stratification changes when assembling various positivity sectors.

\subsection{$0 \Delta^{(i,j)}$ Boundaries}
\label{sec:assembly0Delta}

All boundaries in $\Gamma_0$ with $0 \Delta^{(i,j)}$, as described by the right-most column of Tables \ref{tab:+++Gamma0Gamma1}, \ref{tab:++-Gamma0Gamma1}, \ref{tab:+--Gamma0Gamma1} and \ref{tab:---Gamma0Gamma1} only have $4 \times 4$ minors $\Delta^{(1,2)}$, $\Delta^{(1,3)}$ and $\Delta^{(2,3)}$ that are either trivially positive, trivially negative, or trivially zero. This is true for all sectors. Clearly, the trivially positive boundaries will be in different $\Gamma_0$ sites to the trivially negative sites. Hence, when assembling sectors, the $0 \Delta^{(i,j)}$ boundaries are particularly easy to deal with, and the only part of the procedure that requires care is to not double-count sites in $\Gamma_0$ that appear in multiple sectors.

\subsection{$1 \Delta^{(i,j)}$ Boundaries}
\label{sec:assembly1Delta}

For each $1 \Delta^{(i,j)}$ site in $\Gamma_0$, the $\Gamma_1$ structure may change when assembling positivity sectors. This happens when considering sectors in which a specific $\Gamma_0$ site has a $\Delta^{(i,j)}$ which is non-trivially positive in one sector and non-trivially negative in another. Both of these sectors will share the space where $\Delta^{(i,j)}$ is non-trivially zero. An example of this happens when we glue sectors $\{+,+,+\}$ and $\{-,+,+\}$ together. Both sectors have the $\Gamma_0$ site where $\Delta^{(1)}_{13} \neq 0$, $\Delta^{(1)}_{34} \neq 0$, $\Delta^{(2)}_{12} \neq 0$, $\Delta^{(2)}_{24} \neq 0$, $\Delta^{(3)}_{34} \neq 0$ and all other $\Delta^{(i)}_I = 0$. Here the $4 \times 4$ minors are
\begin{equation} \label{eq:example1Delta}
\Delta^{(1,2)} = \Delta^{(1)}_{34} \Delta^{(2)}_{12} - \Delta^{(1)}_{13} \Delta^{(2)}_{24} \quad \quad \quad \Delta^{(1,3)} = 0 \quad \quad \quad  \Delta^{(2,3)} = \Delta^{(2)}_{12} \Delta^{(3)}_{34} \quad ,
\end{equation}
and so we see that we are in a $1 \Delta^{(i,j)}$ boundary, where in the $\{+,+,+\}$ sector $\Delta^{(1,2)}$ is non-trivially positive and in the $\{-,+,+\}$ sector $\Delta^{(1,2)}$ is non-trivially negative. Both sectors share the subspace where $\Delta^{(1,2)}$ is non-trivially zero.

\begin{figure}[h]
\begin{center}
\includegraphics[width=\textwidth]{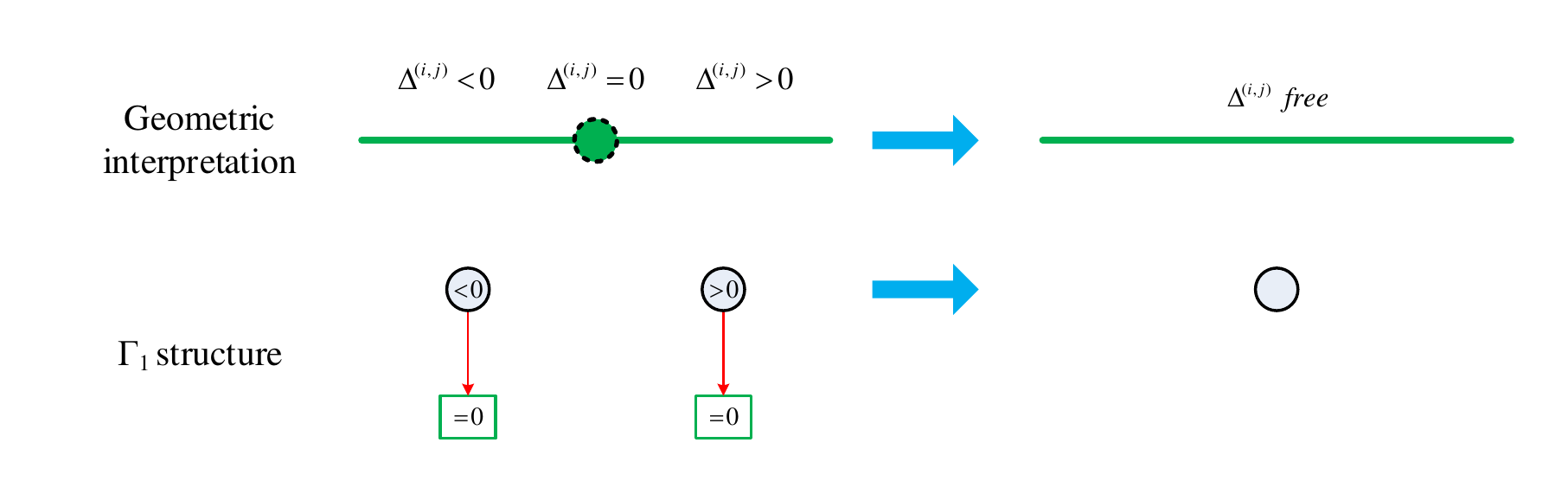}
\vspace{-0.8cm}
\caption{Assembly of a $\{+,\ast,\ast\}$ sector with a $\{-,\ast,\ast\}$ on a $1 \Delta^{(i,j)}$ site in $\Gamma_0$. The asterisks are free to be either plus or minus, because we consider a specific $\Gamma_0$ site which is present in both sectors, and which has only one non-trivial $\Delta^{(i,j)}$. The figure shows the geometric interpretation of this merging as well as the effect on the $\Gamma_1$ structure.}
\label{fig:merging1Delta}
\end{center}
\end{figure}

These three spaces should be merged into a single, continuous space, as illustrated by \fref{fig:merging1Delta}. The merging has a very simple effect: it wipes out the $\Gamma_1$ structure from this $\Gamma_0$ site, giving it the trivial structure of a $0 \Delta^{(i,j)}$ boundary.

\subsubsection{Full Stratification vs Mini Stratification}
\label{sec:assemblyRegions}

Let us briefly digress and discuss the effect of computing the full stratification of the amplituhedron instead of the mini stratification. Here we wish to argue that the generic process of assembling sectors of the amplituhedron may be insensitive to which of the two stratifications we are considering. This is due to the fact that the full stratification is a refinement of the mini stratification, obtained by giving certain boundaries in the mini stratification a multiplicity (which may contain lower-dimensional sub-boundaries), where all the new boundaries are specified by additional signs on combinations of \pl coordinates. However, all these new boundaries will have the same $\Delta^{(i)}_I$ and $\Delta^{(j,k)}$ labels as their mini-stratification counterparts.

\begin{figure}[h]
\begin{center}
\includegraphics[scale=0.5]{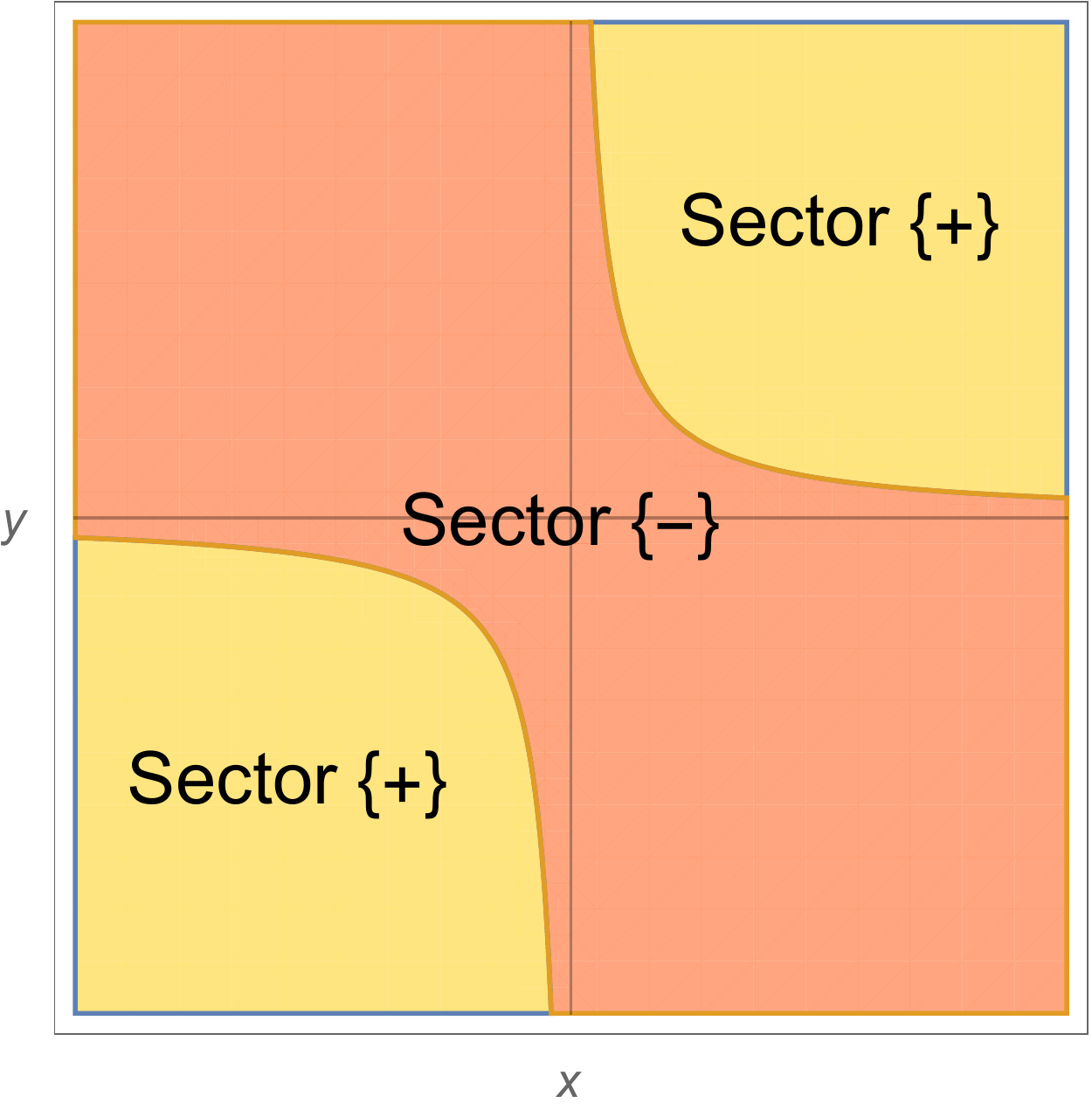}
\caption{Although there may be multiple regions in the space defined by an inequality $\Delta^{(1,2)}>0$, assembling this space with its complement $\Delta^{(1,2)}<0$ will produce a single space where $\Delta^{(1,2)}$ is free, as for the mini stratification.}
\label{fig:sectorregions}
\end{center}
\end{figure}

Since the assembly of positivity sectors is based on joining spaces defined by the signs of $4 \times 4$ minors, such an assembly is insensitive to the inner structure of these spaces. Expressed differently, the space $\Delta^{(1,2)}>0$ is always the complement space to $\Delta^{(1,2)}<0$, regardless of any multiplicity of solutions that constitute these spaces. For example, let us consider the space obtained by $\Delta_{23}^{(1)}=\Delta_{14}^{(1)}=0$, where $\Delta^{(1,2)}$ takes the schematic form in \eref{eq:regionsminor}, which we reproduce here for convenience,
\begin{equation}
\Delta^{(1,2)}= \frac{1}{k_1} \Big[  \left( \ldots \right)\left( \ldots \right)  -k_2 \Big] \, .
\end{equation}
As explained in \sref{sec:regionsintro}, $\Delta^{(1,2)}>0$ can be pictorially represented by \fref{fig:regions1and2}. Hence, $\Delta^{(1,2)}<0$ describes the complement of this space, as shown in \fref{fig:sectorregions}. In this way we see that when assembling the sectors $\{+ \}$ and $\{- \}$ we obtain a single space, where $\Delta^{(1,2)}$ is free, precisely as for the mini stratification.

\subsection{$2 \Delta^{(i,j)}$ Boundaries}
\label{sec:assembly2Delta}

\begin{figure}[h]
\begin{center}
\includegraphics[width=\textwidth]{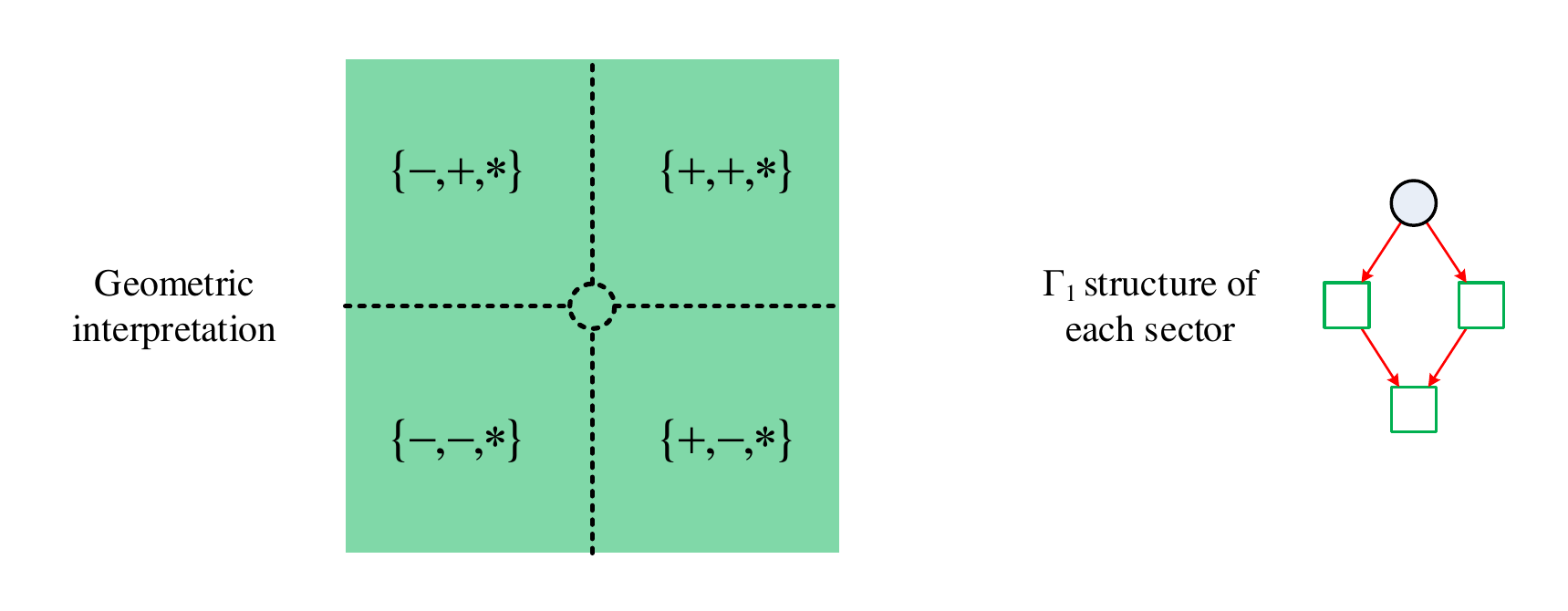}
\vspace{-0.8cm}
\caption{Since the large minors are independent in $2 \Delta^{(i,j)}$ sites in $\Gamma_0$, each sector has the geometric interpretation of filling a quadrant in a 2-dimensional Cartesian plane.}
\label{fig:2DeltaQuadrants}
\end{center}
\end{figure}

We shall now consider the slightly more involved cases of $\Gamma_0$ sites with $2 \Delta^{(i,j)}$, i.e.\ sites where two large minors are either non-trivially positive or non-trivially negative. Here there are four possible distinct scenarios: taking without loss of generality the two minors with both positive and negative terms to be $\Delta^{(1,2)}$ and $\Delta^{(1,3)}$, we either have the sectors $\{+,+,\ast\}$, $\{+,-,\ast\}$, $\{-,+,\ast\}$ or $\{-,-,\ast\}$, where the asterisks are free to be plus or minus. As for the case of $1 \Delta^{(i,j)}$ in \sref{sec:assembly1Delta}, these can be seen very geometrically: since $\Delta^{(1,2)}$ and $\Delta^{(1,3)}$ are completely independent, the four types of sectors can be seen as occupying the four quadrants in a 2-dimensional Cartesian plane, as illustrated in \fref{fig:2DeltaQuadrants}.

\begin{figure}[h]
\begin{center}
\includegraphics[scale=0.45]{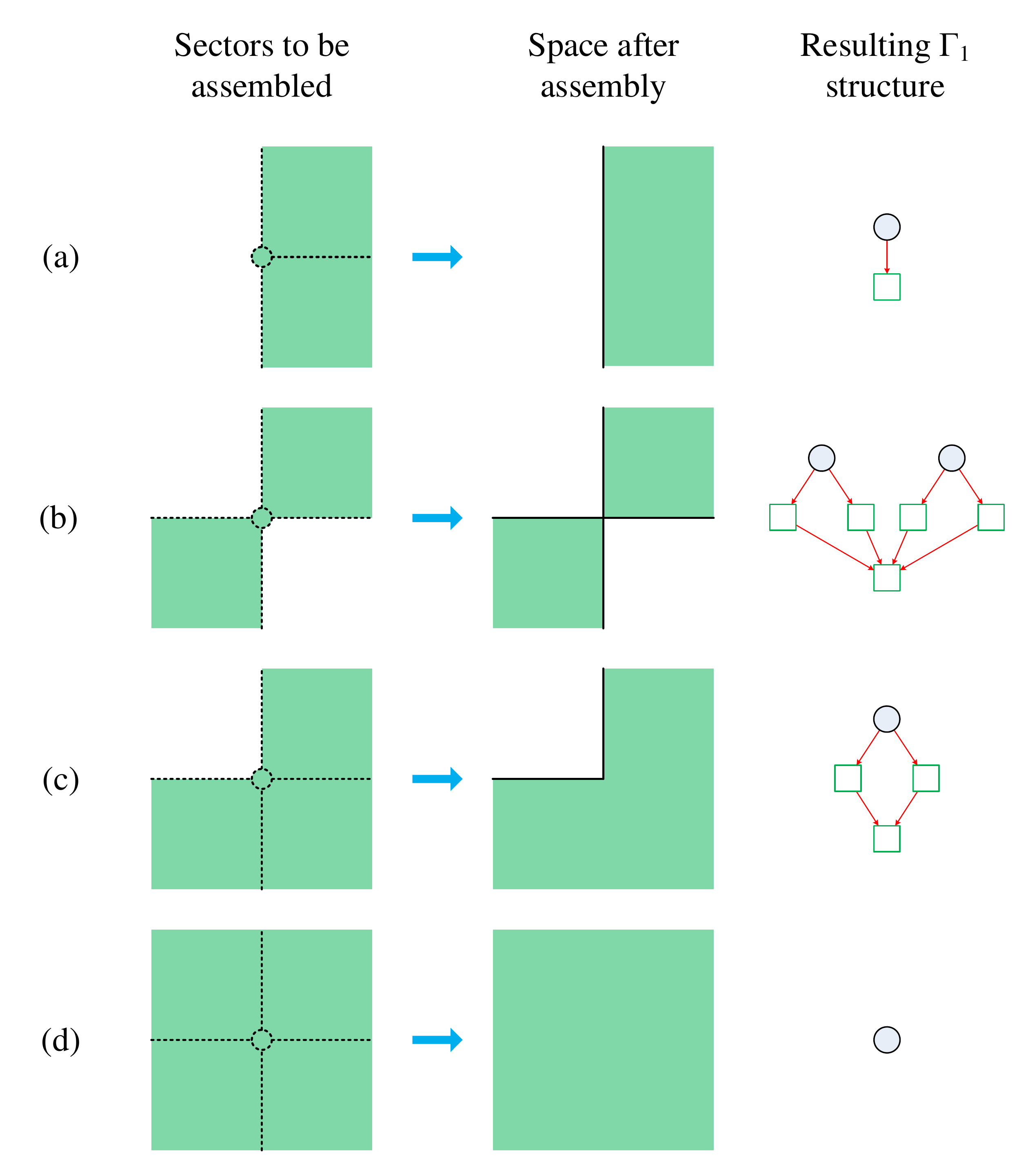}
\caption{Assembling various combinations of sectors in $2 \Delta^{(i,j)}$ sites in $\Gamma_0$. Each case leads to a different $\Gamma_1$ boundary structure after the merging.}
\label{fig:merging2Delta}
\end{center}
\end{figure}

In this way, it is easy to see that there are qualitatively four different ways to assemble different positivity sectors in a given $\Gamma_0$ site, each leading to a different $\Gamma_1$ structure. These are all shown in \fref{fig:merging2Delta}. As is clear from the figure, sometimes a codimension-1 boundary merges two of its parents, i.e.\ it merges two codimension-0 boundaries that both contain this codimension-1 object in their boundary structure. When this happens, the codimension-1 boundary disappears from the $\Gamma_1$ structure. This occurred already in \fref{fig:merging1Delta}; here however we see this effect happening at codimension 1, as in \fref{fig:merging2Delta}(a), (c) and (d), as well as codimension 2, as in \fref{fig:merging2Delta}(a) and (d).

As we shall see in \sref{sec:assemblingalgorithm}, much of the challenge in systematizing an algorithm for assembling positivity sectors lies in being able to determine when a codimension-$i$ boundary disappears from the $\Gamma_1$ structure, merging two of its parents. As an example of this difficulty, consider \fref{fig:merging2Delta}(a). Here we see that the point at the origin disappears from the $\Gamma_1$ structure, thus merging the two vertical semi-infinite lines into a single infinite line. However, in \fref{fig:merging2Delta}(b) we see that the same point appears on either end of the same two vertical lines, but here the origin does \textit{not} disappear, and its codimension-1 parents are not merged. The reason for this is that, in turn, the codimension-0 objects which are parent to the two codimension-1 parents are not merged by these codimension-1 parents. The algorithm we develop in \sref{sec:assemblingalgorithm} will carefully take this into account.

\subsection{$3 \Delta^{(i,j)}$ Boundaries}
\label{sec:assembly3Delta}

We shall now move on to the more diverse realm of sites in $\Gamma_0$ where all three $4 \times 4$ minors have both positive and negative terms. We shall begin with the simpler cases where the three minors may be shut off completely independently, i.e.\ boundaries of Type H, whose $\Gamma_1$ boundary structure is shown in \fref{fig:3Ltrees}. We will then move on to study cases where the minors are not all independent.

\subsubsection{Type H Boundaries}

\begin{figure}[h]
\begin{center}
\includegraphics[width=\textwidth]{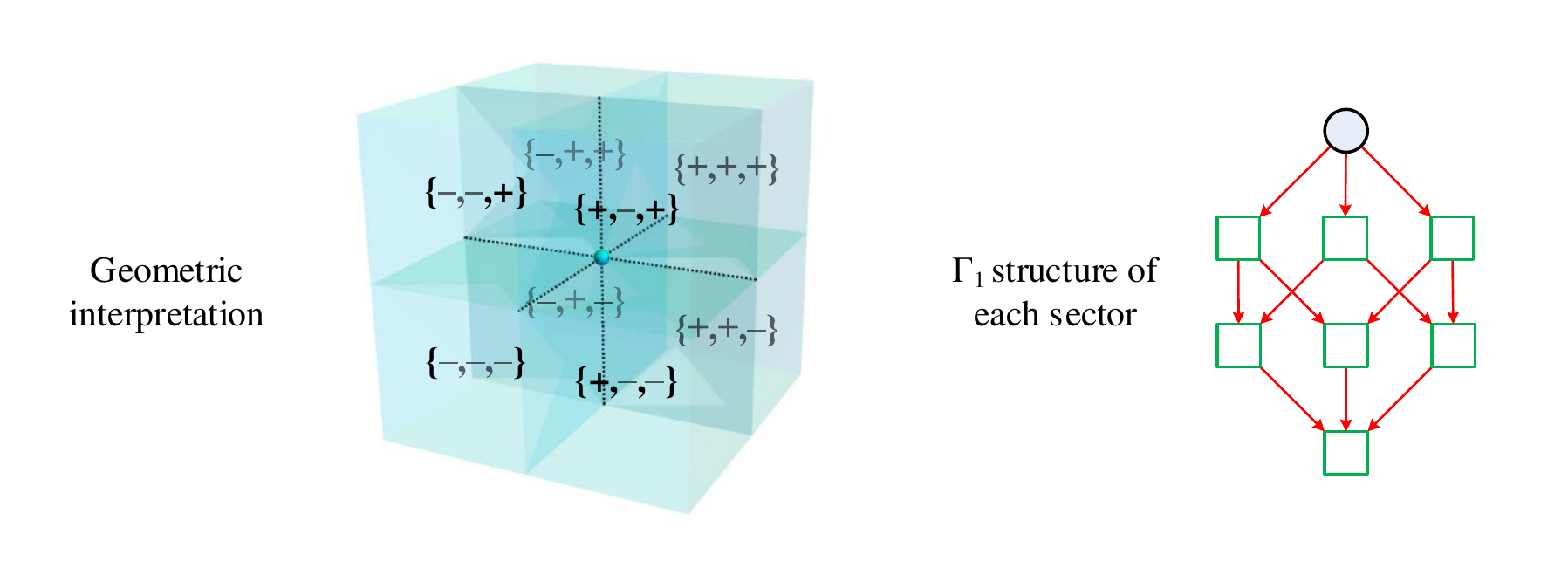}
\vspace{-0.8cm}
\caption{Since the large minors are independent in $3 \Delta^{(i,j)}$ sites of Type H in $\Gamma_0$, each sector has the geometric interpretation of filling a quadrant in a 3-dimensional Cartesian space.}
\label{fig:3DeltaQuadrants}
\end{center}
\end{figure}

Analogously to \sref{sec:assembly1Delta} and \sref{sec:assembly2Delta}, when the large minors are all independent they can be visualized as quadrants in a Cartesian coordinate space. Hence, we may see Type H boundaries as populating the eight quadrants of a 3-dimensional space, each quadrant containing a single positivity sector, as illustrated in \fref{fig:3DeltaQuadrants}.

Here there are many ways in which positivity sectors can be assembled. For completeness we provide a full list of cases with their resulting $\Gamma_1$ structure in Appendix \ref{app:typeH}. The challenges in systematizing a gluing algorithm described at the end of \sref{sec:assembly2Delta} are all the more prominent in these three-dimensional examples, which exhibit a greater level of complexity. As we shall see, for the point at the origin to vanish we will need to not only consult how the codimension-2 parents and in turn their codimension-1 parents glue together, but also how the codimension-0 objects are glued together.

\subsubsection{Type A Boundaries}
\label{sec:assemblingtypeA}

\begin{figure}[h]
\begin{center}
\includegraphics[width=\textwidth]{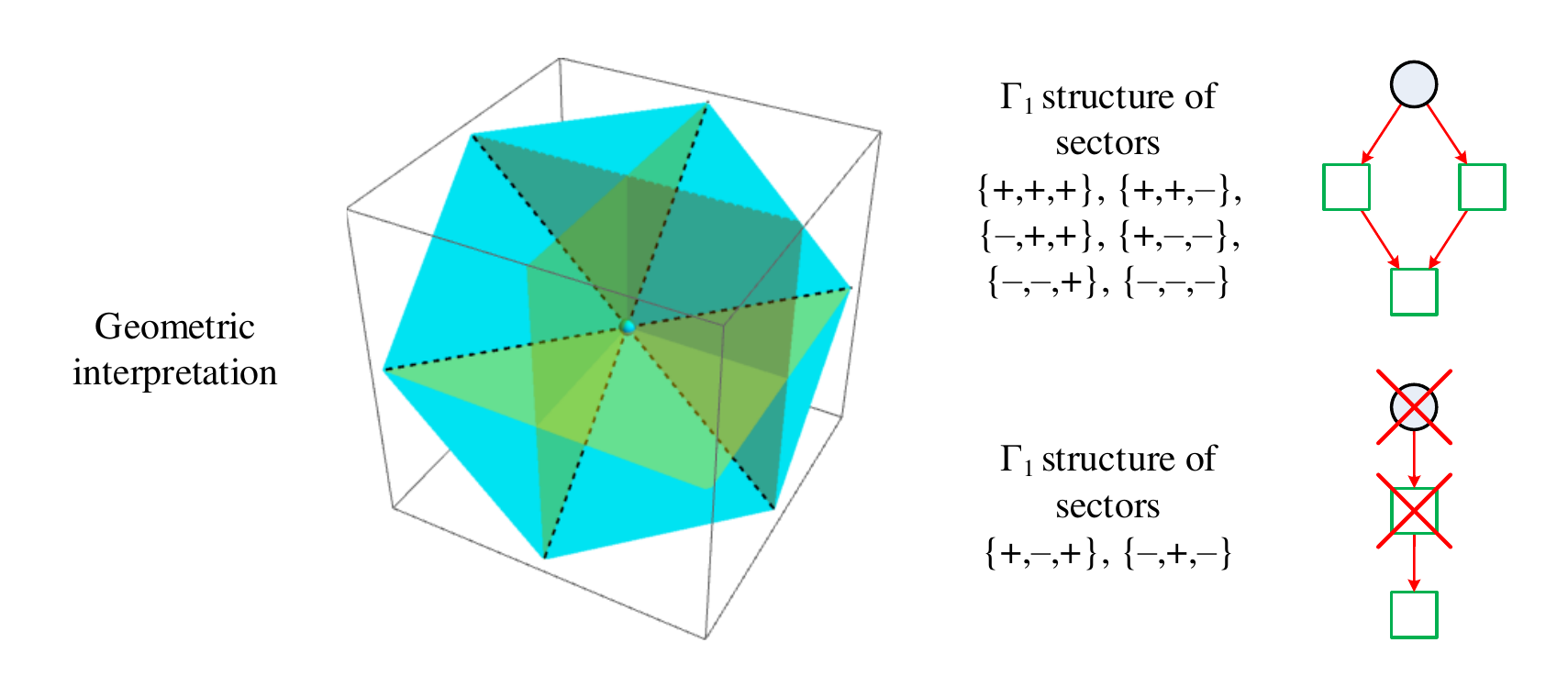}
\vspace{-0.8cm}
\caption{Since the large minors are not independent in $3 \Delta^{(i,j)}$ sites of Type A in $\Gamma_0$, each sector slices out a piece of a 2-dimensional plane, drawn in blue. The Cartesian axes are drawn in yellow. Certain quadrants contain a part of the plane, and hence have the $\Gamma_1$ structure of Type A from \fref{fig:3Ltrees}; those quadrants which do not contain the plane still include the origin, and for this reason have the structure of Type X from \fref{fig:typeX}.}
\label{fig:3DeltaAllTriangles}
\end{center}
\end{figure}

We are now ready to move on to cases where the $4 \times 4$ minors are not all independent. As is clear from the diversity of cases in \fref{fig:3Ltrees}, in general there are relations among the minors which cause them to only access a subspace of the full 3-dimensional Cartesian space. We shall exemplify their treatment by studying in detail Type A boundaries, which are subject to the relation 
\begin{equation} \label{eq:typeAplane}
\Delta^{(i,j)} = k_1 \Delta^{(j,k)} - k_2 \Delta^{(i,k)} \; \; , \quad k_1, k_2 >0 \; \; .
\end{equation}
The geometric interpretation of these types of boundaries is simple: while Type H sites span the full 8 quadrants of the Cartesian space, due to the complete independence of the three $4 \times 4$ minors $\Delta^{(1,2)}$, $\Delta^{(1,3)}$ and $\Delta^{(2,3)}$, Type A boundaries only span a two-dimensional subspace, determined by the equation \eref{eq:typeAplane}. This is illustrated in \fref{fig:3DeltaAllTriangles}. From the figure it is also clear what the geometric significance of Type X structures is: when a sector represents a quadrant which only touches the plane at the origin, its only contribution to $\Gamma_1$ will be a codimension-2 boundary, i.e.\ a single point.

As was the case for Type H boundaries, there are many ways in which the various sectors can be assembled. In Appendix \ref{app:typeA} we provide a full list of the the qualitatively different ways in which sectors may be assembled and their resulting $\Gamma_1$ structures. As for Type H boundaries, there are several subtleties that need to be addressed when constructing an algorithm for systematizing the assembly of sectors, in order to obtain the desired $\Gamma_1$ structure. We shall now move on to explain the precise algorithm that will yield the assembly described in this section, which will be put to use in \sref{sec:formingallspaces} to construct the full space of $G_+(0,4;1)^3$, as well as certain subspaces associated to terms in the three-loop log of the amplitude.

\section{An Algorithm for the Assembly of Sectors} 
\label{sec:assemblingalgorithm}

We are now ready to present an algorithm for merging positivity sectors of the amplituhedron. We stress that this algorithm is not prescriptive; it is proposed as one possible way of approaching the question, which will serve us as a tool when we study the spaces of interest in \sref{sec:formingallspaces}.\footnote{In fact, it is likely that this algorithm can be simplified using powerful computational tools such as the program Polymake \cite{polymake}.} This will allow us to explicitly study how the positivity sectors, which all have Euler number differing from 1, manage to glue together into $G_+(0,4;1)^L$, which has Euler number 1 for arbitrary loops $L$. 

\subsection{Merging $\Gamma_0$ Structures} 
\label{sec:Gamma0Gamma1labels}

The algorithm we shall now present describes the case where $L=3$, but its higher-$L$ generalization is straightforward. As we mentioned earlier, much of the difficulty in constructing the algorithm lies in correctly dealing with the $\Gamma_1$ structures. To this end, we begin our task by first creating useful labels that simultaneously characterize the $\Gamma_0$ and the $\Gamma_1$ structures. We can then deal with the $\Gamma_0$ part of the assembly of sectors, which will in turn allow us to focus on the task of merging the $\Gamma_1$ labels, for each site in $\Gamma_0$, as described in \sref{sec:assembling}. Hence, we begin with the following steps:
\begin{itemize}
\item In each positivity sector, the $\Gamma_0$ boundaries are already labeled by the set of non-vanishing \pl coordinates. We can then simply characterize the boundaries in the $\Gamma_1$ structure by additionally writing out explicitly the sign of each of the three $4 \times 4$ minors $\Delta^{(1,2)}$, $\Delta^{(1,3)}$ and $\Delta^{(2,3)}$. For example, a $1\Delta^{(i,j)}$ structure in a $\Gamma_0$ site, as shown in \fref{fig:1Delta2Delta}, could look like 
\begin{equation} \nonumber
\big\{ \Gamma_0 \text{ label}, \{\Delta^{(1,2)}=0, \Delta^{(1,3)}=0, \Delta^{(2,3)}>0 \}, \{\Delta^{(1,2)}=0, \Delta^{(1,3)}=0, \Delta^{(2,3)}=0 \} \big\} \; .
\end{equation}
As an additional example, a Type X boundary as shown in \fref{fig:typeX} would look like 
\begin{equation} \nonumber
\big\{ \Gamma_0 \text{ label}, \{\Delta^{(1,2)}=0, \Delta^{(1,3)}=0, \Delta^{(2,3)}=0 \} \big\} \; .
\end{equation}
\item We may now join together the $\Gamma_0$ structures of all the sectors that are to be assembled, ignoring duplicate $\Gamma_0$ labels.
\item For each $\Gamma_0$ label, we join together all the $\Gamma_1$ labels of all the sectors containing this $\Gamma_0$ site, ignoring duplicates.
\end{itemize}
We have now produced the $\Gamma_0$ structure of the final assembled result. At this stage, we have already successfully dealt with all $0\Delta^{(i,j)}$ boundaries as described in \sref{sec:assembly0Delta}. To give an example of what the above steps may produce, let us consider the simple case of a $\Gamma_0$ site with a $1\Delta^{(i,j)}$ $\Gamma_1$ structure, where we glue the $\{+,+,+\}$ and $\{-,+,+\}$ sectors together as in \eref{eq:example1Delta}. In these cases, the label that characterizes such a $\Gamma_0$ site in the resulting space would be
\begin{eqnarray}
\big\{ \Gamma_0 \text{ label}, \{\Delta^{(1,2)}>0, \Delta^{(1,3)}=0, \Delta^{(2,3)}>0 \}, \{\Delta^{(1,2)}<0, \Delta^{(1,3)}=0, \Delta^{(2,3)}>0 \}, \nonumber \\
 \{\Delta^{(1,2)}=0, \Delta^{(1,3)}=0, \Delta^{(2,3)}>0 \} \big\} \quad .\quad \quad \quad \quad \quad \quad \quad 
\end{eqnarray}

\subsection{Merging $\Gamma_1$ Structures} 
\label{sec:generalalgorithm}

The merging of $\Gamma_1$ structures has two sources of difficulty. This first is essentially a bookkeeping task: since the merging of boundaries at a given codimension depends on what has been merged at higher dimensions, we must deal with our boundaries in a particular order. The second, and harder, challenge is the precise algorithm for determining when a merge occurs. 

We shall begin by describing a solution to the first task, in generality, to be performed at each site in $\Gamma_0$, and demonstrate it at work in a concrete example. We thus propose the following:
\begin{itemize}
\item Split all boundaries in $\Gamma_1$ according to their codimension in the $\Gamma_0$ site.
\item Consider first the list of codimension-1 boundaries. For each such boundary, determine whether it merges two codimension-0 boundaries into a single boundary. We shall call these codimension-0 boundaries \textit{parents} of the codimension-1 boundary under consideration. We shall discuss in \sref{sec:mergingalgorithm} how to establish whether a merge happens.

An example of such a merge is seen in \fref{fig:merging2Delta}(a), where the horizontal 1-dimensional line merges the two areas. In our nomenclature, the two areas are both said to be parents of the horizontal line.
\item If the merge does occur, we perform the following operation on the lists of codimension-$i$ boundaries: merge the labels for the two codimension-0 parents into a single object which contains both labels, and which also contains the codimension-1 label representing the boundary that merged them. This object is declared to be codimension 0. For convenience we shall \textit{not} yet discard the codimension-1 boundary from the list of codimension-1 boundaries.
\item We now proceed to codimension 2, where we consider each boundary in turn, and see whether it merges any pair of boundaries at codimension 1. This is the reason why we did not remove any codimension-1 boundaries in the previous step, regardless of whether they were merged into a codimension-0 object: their removal would have complicated the search through codimension-1 parents which are merged by codimension-2 boundaries. We shall soon deal with this apparent double-counting of boundaries. This step is exemplified again in \fref{fig:merging2Delta}(a), where the point at the origin merges the two vertical lines into a single continuous vertical line.
\item If a codimension-2 boundary merges two codimension-1 parents, we do as we did earlier: we merge the labels for the two codimension-1 parents into a single object, and in this object we also place the codimension-2 boundary in question, without removing it from the list of codimension-2 boundaries.
\item We repeat the same procedure for codimension 3 which, if present, is the origin. 
\item Finally, we rectify the double-counting of boundaries---at this stage a codimension-$i$ boundary that merged two codimension-$(i-1)$ parents will still appear in the list of codimension-$i$ objects. This step consists in ensuring that the list of codimension-$i$ objects contains all the boundaries and sub-boundaries that participate in forming each object, and that the sub-boundaries are no longer present in the separate lists of objects at other codimensions.
\end{itemize}
We shall now study an example in detail to illustrate the practical implementation of the above strategy.

\begin{figure}[h]
\begin{center}
\includegraphics[width=\textwidth]{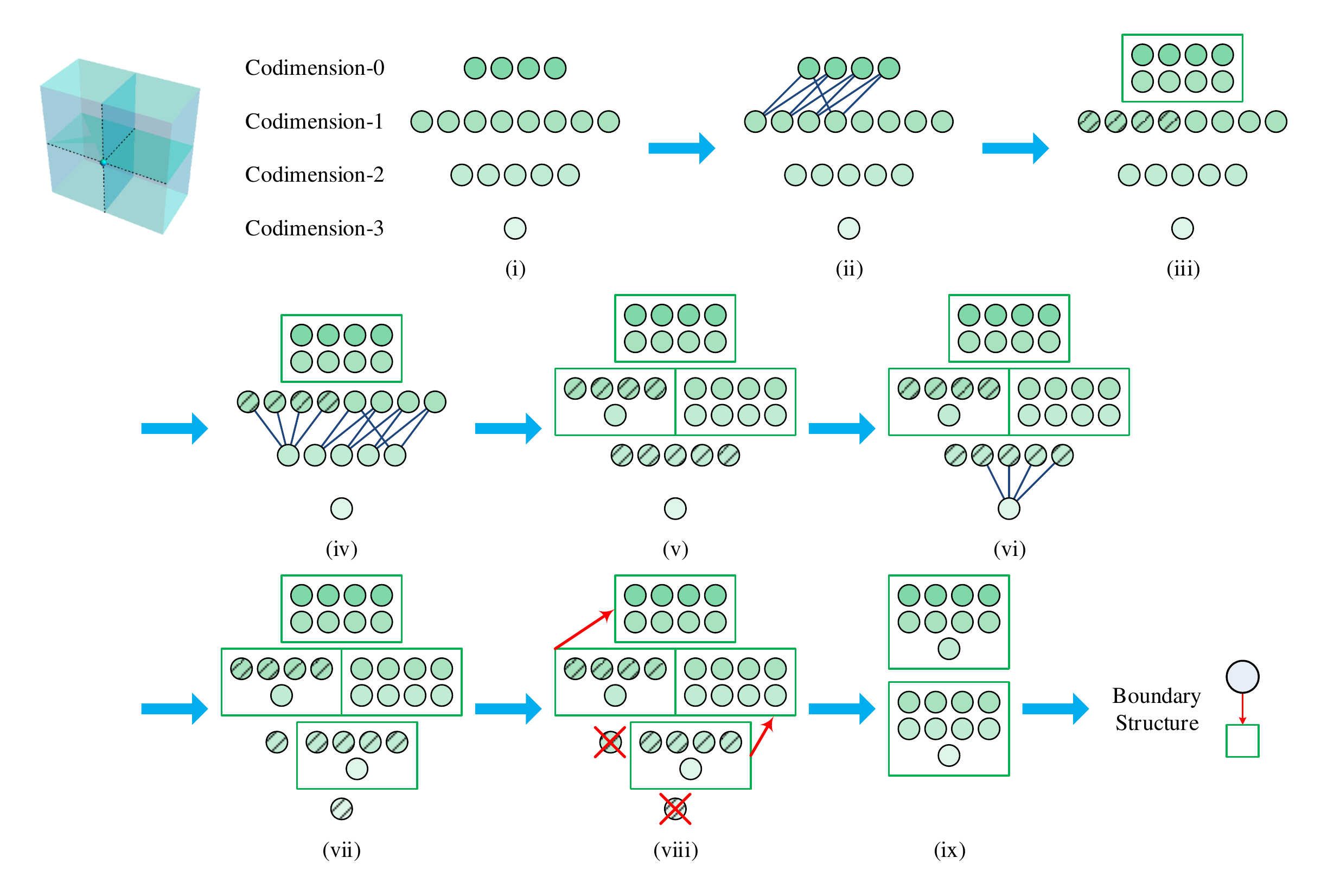}
\vspace{-0.8cm}
\caption{Schematic illustration of the assembly of positivity sectors for the example shown in \fref{fig:merging3Delta(2)}(g).}
\label{fig:organizationalgorithm}
\end{center}
\end{figure}

\paragraph{Example.} 
Let us assemble the positivity sectors $\{+,+,+\}$, $\{+,-,+\}$, $\{+,-,-\}$ and $\{+,+,-\}$, and study a Type H $3 \Delta^{(i,j)}$ site in $\Gamma_0$, as shown in \fref{fig:merging3Delta(2)}(g). After unifying the positivity sectors, we have the $\Gamma_1$ labels:
\vspace{-0.3cm}

{\footnotesize
\begin{equation} \label{eq:ExampleAlllabels}
\begin{array}{|c|c|}
\hline
\text{Codimension} & \Gamma_1 \text{ Labels} \\
\hline
\multirow{2}{*}{0} & \{\Delta^{(1,2)}>0, \Delta^{(1,3)}>0, \Delta^{(2,3)}>0 \}, \{\Delta^{(1,2)}>0, \Delta^{(1,3)}<0, \Delta^{(2,3)}>0 \} \\
  & \{\Delta^{(1,2)}>0, \Delta^{(1,3)}<0, \Delta^{(2,3)}<0 \}, \{\Delta^{(1,2)}>0, \Delta^{(1,3)}>0, \Delta^{(2,3)}<0 \}\\
\hline
\multirow{4}{*}{1} & \{\Delta^{(1,2)}>0, \Delta^{(1,3)}=0, \Delta^{(2,3)}>0 \}, \{\Delta^{(1,2)}>0, \Delta^{(1,3)}<0, \Delta^{(2,3)}=0 \} \\
  & \{\Delta^{(1,2)}>0, \Delta^{(1,3)}=0, \Delta^{(2,3)}<0 \}, \{\Delta^{(1,2)}>0, \Delta^{(1,3)}>0, \Delta^{(2,3)}=0 \} \\
  & \{\Delta^{(1,2)}=0, \Delta^{(1,3)}>0, \Delta^{(2,3)}>0 \}, \{\Delta^{(1,2)}=0, \Delta^{(1,3)}<0, \Delta^{(2,3)}>0 \} \\
  & \{\Delta^{(1,2)}=0, \Delta^{(1,3)}<0, \Delta^{(2,3)}<0 \}, \{\Delta^{(1,2)}=0, \Delta^{(1,3)}>0, \Delta^{(2,3)}<0 \} \\
\hline
\multirow{3}{*}{2} & \{\Delta^{(1,2)}>0, \Delta^{(1,3)}=0, \Delta^{(2,3)}=0 \}, \{\Delta^{(1,2)}=0, \Delta^{(1,3)}=0, \Delta^{(2,3)}>0 \} \\
  & \{\Delta^{(1,2)}=0, \Delta^{(1,3)}<0, \Delta^{(2,3)}=0 \}, \{\Delta^{(1,2)}=0, \Delta^{(1,3)}=0, \Delta^{(2,3)}<0 \} \\
  & \{\Delta^{(1,2)}=0, \Delta^{(1,3)}>0, \Delta^{(2,3)}=0 \} \\
\hline
3 & \{\Delta^{(1,2)}=0, \Delta^{(1,3)}=0, \Delta^{(2,3)}=0 \} \\
\hline
\end{array}
\end{equation}
}
where we have split the $\Gamma_1$ boundaries according to their codimension. The boundaries are schematically shown in \fref{fig:organizationalgorithm}(i), where the entire procedure for this example is illustrated.

The first step is to consider those codimension-1 boundaries that merge codimension-0 parents. In this example we have four relevant codimension-1 boundaries, representing the four internal faces; this is shown in step (ii) in the figure. The merges are connected in such a way that the labels for the four parents end up in a single object, which also includes these four codimension-1 boundaries, as shown in step (iii) in the figure. For later convenience, we also choose to mark these codimension-1 boundaries with black diagonal stripes.

We now move on to codimension-2 boundaries and perform the same operation. Here we have one boundary, labeled by $\{\Delta^{(1,2)}>0, \Delta^{(1,3)}=0, \Delta^{(2,3)}=0 \}$, which merges the four internal faces. We also have four boundaries which pairwise merge faces living on the $\Delta^{(1,2)}=0$ plane. This is shown in step (iv) in the figure. Step (v) then executes the merging of labels as described above, where we again mark with diagonal stripes those codimension-2 boundaries that participated in the merge.

We are finally ready to consider the codimension-3 boundary at the origin. This boundary has four parents, which are those lines living on the $\Delta^{(1,2)}=0$ plane, as seen in step (vi). Step (vii) again merges the labels as required, and marks the codimension-3 boundary with horizontal stripes to mark that it merged parent-boundaries.

The final two steps involve cleaning up the double counting of labels: when a boundary has stripes on it in the figure, the object in which it appears is subject to a vertical merging of labels. This implies for example the removal of the codimension-3 boundary, since it already appears in an object at codimension 2. Furthermore, at codimension 2 we have two objects, which both must be merged with objects at higher dimensions. This procedure results in what is shown in \fref{fig:organizationalgorithm}(ix), which we see only has two objects: one at codimension 0, containing all boundaries with $\Delta^{(1,2)}>0$, and one at codimension 1, containing all boundaries with $\Delta^{(1,2)}=0$, in accordance to \fref{fig:merging3Delta(2)}(g).

\subsubsection{Determining the Merging of Parent-Boundaries} 
\label{sec:mergingalgorithm}

The above algorithm describes a useful bookkeeping strategy that enables the successful assembly of different positivity sectors. The trickier part of the assembly lies of course in determining whether a codimension-$i$ boundary does indeed merge two codimension-$(i-1)$ parents. We shall now present an algorithm that determines precisely this. 

As mentioned previously, the difficulty is largely caused by the need to recursively consider the gluing of higher-dimensional boundaries in $\Gamma_1$ in order to determine whether a codimension-$i$ boundary merges two codimension-$(i-1)$ boundaries; the same codimension-$(i-1)$ boundaries may or may not be merged by the codimension-$i$ boundary, depending on the behavior of the codimension-$(i-2)$ boundaries and the codimension-$(i-3)$ boundaries. For an example of this it is sufficient to consider the example in \fref{fig:merging3Delta(1)}(e): here the codimension-3 point at the origin does not join the two vertical lines, as it instead does in \fref{fig:merging3Delta(1)}(a), simply because of how the codimension-1 faces glue up.

For the examples arising in the context of assembling positivity sectors in the amplituhedron, there are only a few conditions that must be checked in order to declare a merging of codimension-$(i-1)$ boundaries. Heuristically, a codimension-$i$ boundary $\mathcal{B}$ gets subsumed by the merging of two of its parents if there are no other kinks around $\mathcal{B}$, which happens if all codimension-$(i-1)$ boundaries ending on $\mathcal{B}$ behave in a specific way. To this end, our algorithm begins with the following steps:
\begin{itemize}
\item Consider all codimension-$(i-1)$ boundaries that contain $\mathcal{B}$ as a sub-boundary. We shall call this set $\mathcal{P}$. These boundaries reach $\mathcal{B}$ by shutting off some collection of $4 \times 4$ minors. For the example in \fref{fig:merging3Delta(1)}(e), the boundary $\mathcal{B} = \{\Delta^{(1,2)}=0, \Delta^{(1,3)}=0, \Delta^{(2,3)}=0 \}$ has six potential parents in $\mathcal{P}$, which are the six lines ending at the origin.
\item The first condition for the merging to occur is that there must exist a pair of boundaries in $\mathcal{P}$ whose labels share the same sign for all $\Delta^{(i,j)}$, except for the $\Delta^{(i,j)}$'s which are shut off to reach $\mathcal{B}$, which must instead have the opposite sign. In the example in \fref{fig:merging3Delta(1)}(e), the two vertical lines have labels $\{\Delta^{(1,2)}=0, \Delta^{(1,3)}=0, \Delta^{(2,3)}>0 \}$ and $\{\Delta^{(1,2)}=0, \Delta^{(1,3)}=0, \Delta^{(2,3)}<0 \}$, and reach $\mathcal{B}$ by shutting off $\Delta^{(1,2)}$. Hence, they do indeed form such a pair in $\mathcal{P}$, and should hence be considered for a merging. In this example, $\mathcal{P}$ has two additional pairs, formed by pairing up the remaining four lines in a similar way.
\end{itemize}

We must now investigate whether each of the pairs $p$ in $\mathcal{P}$ has the potential to be merged, i.e.\ whether the pairs satisfy additional conditions necessary for a correct merging. These conditions are the following:
\begin{itemize}
\item All codimension-$(i-2)$ boundaries $\mathcal{G}_p$ ending in boundaries in a given $p$ must glue up in a certain way, which we shall soon describe. This condition does not apply if the set $\mathcal{G}_p$ is empty.
\item The codimension-$(i-3)$ boundaries ending in boundaries in $\mathcal{G}_p$ must also glue up in a certain way. Again, if there are no such codimension-$(i-3)$ boundaries, this condition does not apply.
\end{itemize}
The precise condition on the boundaries in $\mathcal{G}_p$ is similar to that forming the pair $p$: each of the boundaries in $\mathcal{G}_p$ must have an ``opposite partner'' in $\mathcal{G}_p$, where opposite partners are defined as having the same signs for all minors, except those minors that are turned off when reaching $\mathcal{B}$ from a boundary in $p$. We do not impose this condition on those boundaries in $\mathcal{G}_p$ that have been subsumed in the merging of higher-dimensional boundaries.

Let us again illustrate this requirement with the example in \fref{fig:merging3Delta(1)}(e). Let us consider the pair $p$ composed of the two vertical lines. The line $\{\Delta^{(1,2)}=0, \Delta^{(1,3)}=0, \Delta^{(2,3)}>0 \} \in p$ contributes to $\mathcal{G}_p$ with the faces
\begin{eqnarray}
\big\{ \{\Delta^{(1,2)}>0, \Delta^{(1,3)}=0, \Delta^{(2,3)}>0 \} , \{\Delta^{(1,2)}=0, \Delta^{(1,3)}>0, \Delta^{(2,3)}>0 \} , \, \nonumber \\
\{\Delta^{(1,2)}<0, \Delta^{(1,3)}=0, \Delta^{(2,3)}>0 \} , \{\Delta^{(1,2)}=0, \Delta^{(1,3)}<0, \Delta^{(2,3)}>0 \} \big\}
\end{eqnarray}
and the line $\{\Delta^{(1,2)}=0, \Delta^{(1,3)}=0, \Delta^{(2,3)}<0 \} \in p$ contributes to $\mathcal{G}_p$ with the faces
\begin{equation}
\big\{ \{\Delta^{(1,2)}>0, \Delta^{(1,3)}=0, \Delta^{(2,3)}<0 \} , \{\Delta^{(1,2)}=0, \Delta^{(1,3)}>0, \Delta^{(2,3)}<0 \} \big\} \; .
\end{equation}
None of the boundaries in $\mathcal{G}_p$ participated in the merging of higher-dimensional boundaries, so they are all subject to the condition we have just laid out. In this example the minors in the label $\{\Delta^{(1,2)}>0, \Delta^{(1,3)}=0, \Delta^{(2,3)}>0 \} \in \mathcal{G}_p$ only differ from $\{\Delta^{(1,2)}>0, \Delta^{(1,3)}=0, \Delta^{(2,3)}<0 \} \in \mathcal{G}_p$ by the sign of $\Delta^{(2,3)}$, which is the minor which was shut off from the boundaries in $p$ to reach $\mathcal{B}$. Hence, we affirm that these boundaries in $\mathcal{G}_p$ have satisfied the condition above. However, the boundaries $\{\Delta^{(1,2)}<0, \Delta^{(1,3)}=0, \Delta^{(2,3)}>0 \} \in \mathcal{G}_p$ and $\{\Delta^{(1,2)}=0, \Delta^{(1,3)}<0, \Delta^{(2,3)}>0 \} \in \mathcal{G}_p$ do not have opposite partners $\{\Delta^{(1,2)}<0, \Delta^{(1,3)}=0, \Delta^{(2,3)}<0 \}$ and $\{\Delta^{(1,2)}=0, \Delta^{(1,3)}<0, \Delta^{(2,3)}<0 \}$. Hence, we declare that the condition as a whole has been violated, since \textit{all} boundaries in $\mathcal{G}_p$ must satisfy it. This stops us from proceeding with the merging: the point at the origin should be counted separately in $\Gamma_1$, and not be subsumed as part of a continuous higher-dimensional boundary, as can indeed be verified by \fref{fig:merging3Delta(1)}(e).

Assuming that the conditions on  the boundaries in $\mathcal{G}_p$ are all satisfied, as is the case in \fref{fig:merging3Delta(1)}(a), we proceed with the condition on the codimension-$(i-3)$ boundaries ending in each pair of ``opposite partners'' in $\mathcal{G}_p$. The condition simply requires that there must exist opposite partners even amongst these codimension-$(i-3)$ boundaries. In contrast to the requirement on $\mathcal{G}_p$, at codimension $(i-3)$ it is sufficient that there exists one such pair. For example, in \fref{fig:merging3Delta(1)}(a) we have in $\mathcal{G}_p$ the boundaries $\{\Delta^{(1,2)}>0, \Delta^{(1,3)}=0, \Delta^{(2,3)}>0 \}$ and $\{\Delta^{(1,2)}>0, \Delta^{(1,3)}=0, \Delta^{(2,3)}<0 \}$ which satisfy the condition on codimension-$(i-2)$ boundaries. The codimension-$(i-3)$ boundaries that end in these boundaries in $\mathcal{G}_p$ are $\{\Delta^{(1,2)}>0, \Delta^{(1,3)}>0, \Delta^{(2,3)}>0 \}$ and $\{\Delta^{(1,2)}>0, \Delta^{(1,3)}>0, \Delta^{(2,3)}<0 \}$, which indeed only differ in the sign of $\Delta^{(2,3)}$ and are hence seen to satisfy the condition on the codimension-$(i-3)$ boundaries.\footnote{We note that not imposing this condition on codimension-$(i-3)$ boundaries would yield an incorrect $\Gamma_1$ structure in the example in \fref{fig:merging3Delta(3)}(l).}

Before we are ready to declare that $\mathcal{B}$ merges the two boundaries in $p \in \mathcal{P}$, we must impose one final condition. This last condition takes into account the fact that even though the boundaries in $p$ appear to glue up well by studying the signs of the boundaries in $\mathcal{G}_p$ as well as those at codimension-$(i-3)$, they might still not glue up as expected. An example of this is provided in \fref{fig:merging3DeltaTriangles(1)}(c), where the point at the origin $\mathcal{B}$ has two lines $p$ which are on opposite sides of this point, the condition on $\mathcal{G}_p$ is not violated since the blue areas are ``opposite partners'', and there are no codimension-$(i-3)$ boundaries on which to apply any conditions. However, we clearly see that the point at the origin should be present in the final $\Gamma_1$ structure and not subsumed by the two lines in $p$. 

The final condition that ensures a correct merging is to impose an additional check on opposite partners in $\mathcal{G}_p$: they must have been glued together into a single object by the mergings that occurred at higher dimensions. In \fref{fig:merging3DeltaTriangles(1)}(e), for example, the lines have merged the areas into a single object. Thus, we see that in this example the point at the origin should participate in the merging of a pair of lines.

We are finally ready to determine whether $\mathcal{B}$ merges its parents. To do so, we should consider all pairs $p \in \mathcal{P}$ which \textit{do not} satisfy the conditions above. From those pairs we remove those codimension-$(i-1)$ boundaries that participated in a higher-dimensional merge. If this leaves us with no boundaries remaining, we may proceed with the merge, as detailed in \sref{sec:generalalgorithm}.

\paragraph{Summary of the Merging Algorithm.} 
Due to the technical nature of the algorithm we have just expounded, we shall here summarize its steps, to be performed starting at the highest dimension and proceeding to the lowest:
\begin{itemize}
\item Consider in turn each boundary $\mathcal{B}$ at a given codimension $i$.
\item Form the set $\mathcal{P}$ of codimension-$(i-1)$ boundaries that have $\mathcal{B}$ as a sub-boundary.
\item Find pairs of boundaries $p \in \mathcal{P}$, determined by having identical signs for all minors, except those minors required to access $\mathcal{B}$, which must have opposite signs.
\item For each pair $p$ form the list $\mathcal{G}_p$ of all codimension-$(i-2)$ boundaries ending in $p$, where we do not include in $\mathcal{G}_p$ those boundaries that merged two codimension-$(i-3)$ boundaries.
\item For each pair $p$, check the first condition: all boundaries in $\mathcal{G}_p$, if any, must be paired up with opposite partners from the same set $\mathcal{G}_p$. In other words, for each boundary in $\mathcal{G}_p$ there must be another boundary in $\mathcal{G}_p$ with the same signs for all minors, except those which are shut off by boundaries in $p$ to reach $\mathcal{B}$.
\item Now check the second condition, which is a condition on each of the pairs in $\mathcal{G}_p$: if there are any codimension-$(i-3)$ boundaries ending on boundaries in the pair in $\mathcal{G}_p$, we must find at least one pair of opposite partners, defined in the same way as for the pairs in $\mathcal{G}_p$.
\item Check the third condition, which is also a condition on boundaries in $\mathcal{G}_p$: all pairs in $\mathcal{G}_p$ as formed above must have been merged into a single object by codimension-$(i-1)$ boundaries.
\item Finally, a merge happens if there is at least one pair $p$ which satisfies all conditions above, and the other pairs either satisfy the conditions too, or have in turn merged higher-dimensional boundaries.
\end{itemize}

\section{Forming Spaces by Assembling Sectors} 
\label{sec:formingallspaces}

We are now ready to assemble the positivity sectors of the amplituhedron. In this section we present the assembly of $G_+(0,4;1)^3$, and provide some intuitive geometric understanding of how the Euler numbers can simplify when assembling positivity sectors. In order to gain a better understanding of the geometry of $G_+(0,4;1)^3$ we shall display the geometry of an intermediate stage of the assembly process. This stage is in fact associated with a term contributing to the three-loop log of the amplitude, given by
\begin{equation} \label{eq:3looplogterms}
\frac{A_1^3}{3} - A_1 A_2 + A_3 \; .
\end{equation}
As mentioned in \sref{sec:logsectors}, these terms do not appear to have a simple geometric interpretation when summed together; however, neglecting prefactors they individually correspond to interesting spaces. The term $A_1^3$ simply corresponds to $G_+ (0,4;1)^3$, which contains all 8 positivity sectors. $A_1 A_2$ on the other hand corresponds to the product of a two-loop amplitude, which is the $\{ + \}$ sector inside $G_+ (0,4;1)^2$, with a one-loop amplitude. Embedded within the larger space of $G_+ (0,4;1)^3$, we see that $A_1 A_2$ only specifies one of the three $\Delta^{(i,j)}$, which we may take without loss of generality to be the first, i.e.\ $\Delta^{(1,2)}$. Hence, $A_1 A_2$ contains the 4 sectors $\{+ , + , + \}$, $\{+ , + , - \}$, $\{+ , - , + \}$ and $\{+ , - , - \}$. Finally, $A_3$ corresponds to $G_+ (0,4;3)$, which is given by the $\{ +, +, +\}$ sector.

\subsection{Forming $G_+(0,4;1)^3$} 
\label{sec:formingcubeof1loop}

We shall now proceed to construct the spaces corresponding to terms in \eref{eq:3looplogterms}. Since the term $A_3$ simply corresponds to the single sector $\{ +, +, +\}$, the boundaries are described in \tref{tab:+++strat}, which we reproduce for convenience in the left-hand column of \tref{tab:oneLcubestratFinal}. 

The structure of $A_1 A_2$ and $A_1^3$, on the other hand, does require us to use the algorithm described in \sref{sec:assemblingalgorithm}. This section will implement the algorithm in order to present the boundary structures of the resulting geometries.

\paragraph{Forming the Geometry of $A_1 A_2$.}
As already described in \sref{sec:possectors}, to form the cube of the one-loop amplitude we need to combine together all 8 positivity sectors. An intermediate step towards this goal, which still has a physical counterpart in terms of contributions towards the three-loop log of the amplitude, is to assemble the sectors $\{+ , + , + \}$, $\{+ , + , - \}$, $\{+ , - , + \}$ and $\{+ , - , - \}$, which together make up the term $A_1 A_2$ in \eref{eq:3looplogterms}. Here we expect all $1 \Delta^{(i,j)}$ sites in $\Gamma_0$ with $\Delta^{(i,j)} = \Delta^{(1,3)}$ or $\Delta^{(i,j)} = \Delta^{(2,3)}$ to combine as shown in \fref{fig:merging1Delta}. Also, all $2 \Delta^{(i,j)}$ sites should combine as shown in \fref{fig:merging2Delta}(d) if the two $4 \times 4$ minors are $\Delta^{(1,3)}$ and $\Delta^{(2,3)}$, and as shown in \fref{fig:merging2Delta}(a) if one of the $4 \times 4$ minors is $\Delta^{(1,2)}$. Finally, all $3 \Delta^{(i,j)}$ sites in $\Gamma_0$ of Type H will combine as shown in \fref{fig:merging3Delta(2)}(g), and of Type A as shown in \fref{fig:merging3DeltaTriangles(1)}(e).

The correct application of the algorithm in \sref{sec:assemblingalgorithm} shows that this is indeed the case. The reader may explicitly verify this using the supporting Mathematica files, whose use is described in Appendix \ref{app:supportfiles}. We are thus left with the boundary structure shown in the middle column of \tref{tab:oneLcubestratFinal}.

\begin{table}
\begin{center}
\bigskip 
{\small
\begin{tabular}{|c|c|c|}
\cline{2-3}
\multicolumn{1}{c|}{} & $\boldsymbol{A_3}$ \\
\hline
\textbf{Dim} & $\boldsymbol{\mathfrak{N}}$ \\
\hline
\textbf{12} & 1 \\
\hline
\textbf{11}& 15 \\
\hline
\textbf{10} & 117 \\
\hline
\textbf{9} & 611 \\
\hline
\textbf{8} & 2\,244 \\
\hline
\textbf{7} & 5\,908 \\
\hline
\textbf{6} & 10\,996 \\
\hline
\textbf{5} & 13\,956 \\
\hline
\textbf{4} & 12\,044 \\
\hline
\textbf{3} & 7\,488 \\
\hline
\textbf{2} & 3\,504 \\
\hline
\textbf{1} & 1\,128 \\
\hline
\textbf{0} & 186 \\
\hline
\multicolumn{3}{c}{} \\
\multicolumn{3}{c}{$\boldsymbol{\mathcal{E}_{A_3} = -14}$}
\end{tabular}
\quad \quad \begin{tabular}{|c|c|c|}
\cline{2-3}
\multicolumn{1}{c|}{} & $\boldsymbol{A_1 A_2}$ \\
\hline
\textbf{Dim} & $\boldsymbol{\mathfrak{N}}$ \\
\hline
\textbf{12} &  1 \\
\hline
\textbf{11} & 13 \\
\hline
\textbf{10} & 90 \\
\hline
\textbf{9} & 418 \\
\hline
\textbf{8} & 1\,388 \\
\hline
\textbf{7} & 3\,408 \\
\hline
\textbf{6} & 6\,268 \\
\hline
\textbf{5} & 8\,620 \\
\hline
\textbf{4} & 8\,830 \\
\hline
\textbf{3} & 6\,644 \\
\hline
\textbf{2} & 3\,556 \\
\hline
\textbf{1} & 1\,224 \\
\hline
\textbf{0} & 204 \\
\hline
\multicolumn{3}{c}{} \\
\multicolumn{3}{c}{$\boldsymbol{\mathcal{E}_{A_1 A_2} = 10}$}
\end{tabular}
\quad \quad \begin{tabular}{|c|c|c|}
\cline{2-3}
\multicolumn{1}{c|}{} & $\boldsymbol{A_1^3 = G_+(0,4;1)^3}$ \\
\hline
\textbf{Dim} & $\boldsymbol{\mathfrak{N}}$ \\
\hline
\textbf{12} &  1 \\
\hline
\textbf{11} & 12 \\
\hline
\textbf{10} & 78 \\
\hline
\textbf{9} & 340 \\
\hline
\textbf{8} & 1\,086 \\
\hline
\textbf{7} & 2\,640 \\
\hline
\textbf{6} & 4\,960 \\
\hline
\textbf{5} & 7\,200 \\
\hline
\textbf{4} & 7\,956 \\
\hline
\textbf{3} & 6\,480 \\
\hline
\textbf{2} & 3\,672 \\
\hline
\textbf{1} & 1\,296 \\
\hline
\textbf{0} & 216 \\
\hline
\multicolumn{3}{c}{} \\
\multicolumn{3}{c}{$\boldsymbol{\mathcal{E}_{A_1^3} = 1}$}
\end{tabular}
}
\bigskip
\caption{Number of boundaries of the cube of the 1-loop stratification of the amplituhedron, and of the geometries corresponding to intermediate stages in the assembly of $G_+(0,4;1)^3$. Each of these intermediate stages is associated to a term in the three-loop log of the amplitude, as given in \eref{eq:3looplogterms}, excluding prefactors.\label{tab:oneLcubestratFinal}}
\end{center}
\end{table}

\paragraph{Forming the Geometry of $A_1^3$.} 
We shall now assemble all 8 positivity sectors. In so doing, we expect all $1 \Delta^{(i,j)}$ sites in $\Gamma_0$ to combine as shown in \fref{fig:merging1Delta}, all $2 \Delta^{(i,j)}$ sites in $\Gamma_0$ to combine as shown in \fref{fig:merging2Delta}(d), and all $3 \Delta^{(i,j)}$ sites in $\Gamma_0$ to combine as exemplified for Type H boundaries in \fref{fig:merging3Delta(4)}(t) and for Type A boundaries in \fref{fig:merging3DeltaTriangles(2)}(k). Thus, all $\Gamma_1$ structures at all $\Gamma_0$ sites should trivialize, and we expect to only be left with the $\Gamma_0$ boundary structure. Using the supporting Mathematica files, we see that this is indeed the correct conclusion. In this way we obtain the boundary structure shown in the right-hand column in \tref{tab:oneLcubestratFinal}, which is seen to be identical to that of \tref{tab:oneLcubestrat}.

\subsubsection{Simplification of the Euler number} 
\label{sec:Eulersimplification}
Let us comment on the Euler number of $G_+(0,4;1)^3$. The positivity sectors required to assemble this space all have Euler number different from one, but the final assembled result has Euler number $\mathcal{E}=1$. This may be intuitively understood as follows: $G_+(0,4;1)^3$ is a larger space which fits all 8 positivity sectors. This space has a geometrically simple boundary structure, whose simplicity follows from the simplicity of the positroid stratification. This boundary structure may be seen as the external boundary of the space that encloses all 8 positivity sectors. From this point of view, each of the 8 positivity sectors behaves in a geometrically simple way along these external boundaries, and all of the complexity of each sector stems from their $\Gamma_1$ structure, which seen from the point of view of $G_+(0,4;1)^3$ is an internal structure. When assembling together the positivity sectors, these complicated internal boundaries piece together and disappear, leaving us with the simple external boundary of $G_+(0,4;1)^3$. 

To offer a true analogy, assembling sectors of the amplituhedron is like combining pieces of a children's puzzle, whose internal edges are ragged and complicated, but when assembled yield a simple rectangular shape, which is the external boundary of the complete puzzle. In the amplituhedron, the ragged internal edges are pictorially understood through the figures in \sref{sec:assembling} and in Appendices \ref{app:typeH} and \ref{app:typeA}. As for a puzzle, when all sectors are combined the ragged edges always cancel out, which is the phenomenon shown in the Figures \ref{fig:merging1Delta}, \ref{fig:merging2Delta}(d), \ref{fig:merging3Delta(4)}(t) and \ref{fig:merging3DeltaTriangles(2)}(k).

In fact, we indeed see that the intermediate stages in the assembly of $G_+(0,4;1)^3$ have Euler numbers that get closer to 1 the more sectors we include, $\mathcal{E}_{A_3}$ being further away from $\mathcal{E}_{A_1^3}$ than $\mathcal{E}_{A_1 A_2}$. We also note that very low-dimensional boundaries of $A_3$ and $A_1 A_2$ are for the most part also boundaries of $G_+(0,4;1)^3$, while high-dimensional boundaries are often internal in the space of $G_+(0,4;1)^3$. Thus, when assembling many sectors together, we generally expect many high-dimensional boundaries to disappear, in accordance with the discussion above. Conversely, we also expect the number of low-dimensional boundaries to grow, since we are able to access increasingly many boundaries belonging to $G_+(0,4;1)^3$. \tref{tab:oneLcubestratFinal} shows that this is indeed the case.

In order to facilitate the explicit construction of the assembly algorithm, and to further confirm the geometric interpretation of the assembly process offered in this section, we present in Appendix \ref{app:forming3loophard} the explicit geometric results obtained when gluing together a complicated set of sectors.

\section{Conclusion} 
\label{sec:conclusion}

This work focused on the study of gluing together generalized amplituhedron-like spaces, which we call positivity sectors and define through the specification of $2 \times 2$ and $4 \times 4$ minors of the amplituhedron. Positivity sectors are seen to be part of a larger space, $G_+(0,4;1)^L$, whose sign of the $4 \times 4$ minors is undetermined and hence free. The full boundary structure of the 8 different positivity sectors of the three-loop amplituhedron was presented in detail, along with the associated Euler numbers.

We carefully discussed the process of gluing together sectors of the amplituhedron, in particular in the non-trivial cases where there are multiple $4 \times 4$ minors, and provided a useful geometric picture of such gluing. This allowed us to present in detail an algorithm for gluing the spaces of interest. This algorithm was tested on large classes of examples, provided in the article and the appendices, as well as in the construction of the cube of the one-loop amplitude, whose boundary structure is simple and known.

We also used our algorithm to construct spaces appearing at intermediate stages of the assembly of the cube of the one-loop amplitude, corresponding to various terms contributing to the three-loop log of the amplitude.  We also provide supporting Mathematica files, with brief documentation, to facilitate future studies of the amplituhedron and its positivity sectors.

The utility of our results is twofold; on the one hand, the process of gluing spaces with definite signs is crucial to the construction of the amplitude through triangulations of the amplituhedron. The BCFW recursion relations form one such triangulation, but the amplituhedron allows for the possibility of different, more efficient triangulations of the space. On the other hand, we have constructed spaces of physical interest, i.e.\ the cube of the one-loop amplitude and terms in the three-loop log of the amplitude. In fact, with minor modifications the techniques presented in this paper should be more generally valid in the construction of the $L^{th}$ power of the one-loop amplituhedron, and any relevant subspace. We expect the results in this article to play a useful role in future work to firmly place the amplituhedron as a particle physics tool for the study of scattering amplitudes.

\section*{Acknowledgements}

I would like to thank S.\ Franco, A.\ Mariotti and J.\ Trnka for useful discussions in research preceding this work. I would like to also thank J.\ Trnka for providing very useful feedback on the article prior to submission. This work was in part supported by the U.K. Science and Technology Facilities Council (STFC) STEP award. I wish to thank the Institute for Particle Physics Phenomenology, Durham University, as well as the Dipartimento di Fisica, Universit\'{a} di Torino, where this research was conducted.

\newpage


\appendix

\section{Assembly of Type H Boundaries}
\label{app:typeH}

\begin{figure}[htb]
\begin{center}
\includegraphics[scale=0.55]{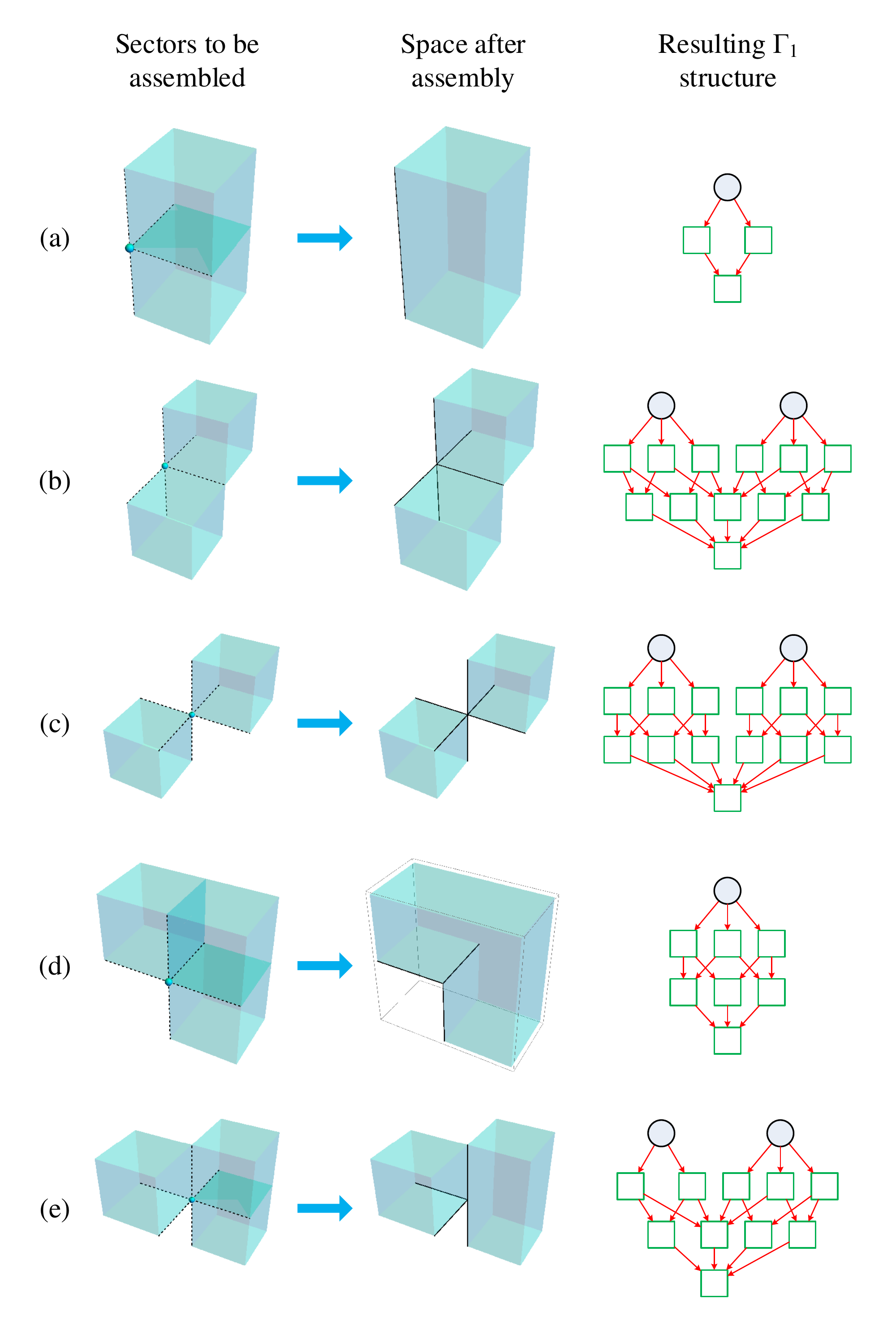}
\vspace{-0.8cm}
\caption{}
\label{fig:merging3Delta(1)}
\end{center}
\end{figure}

\begin{figure}[htb]
\begin{center}
\includegraphics[scale=0.65]{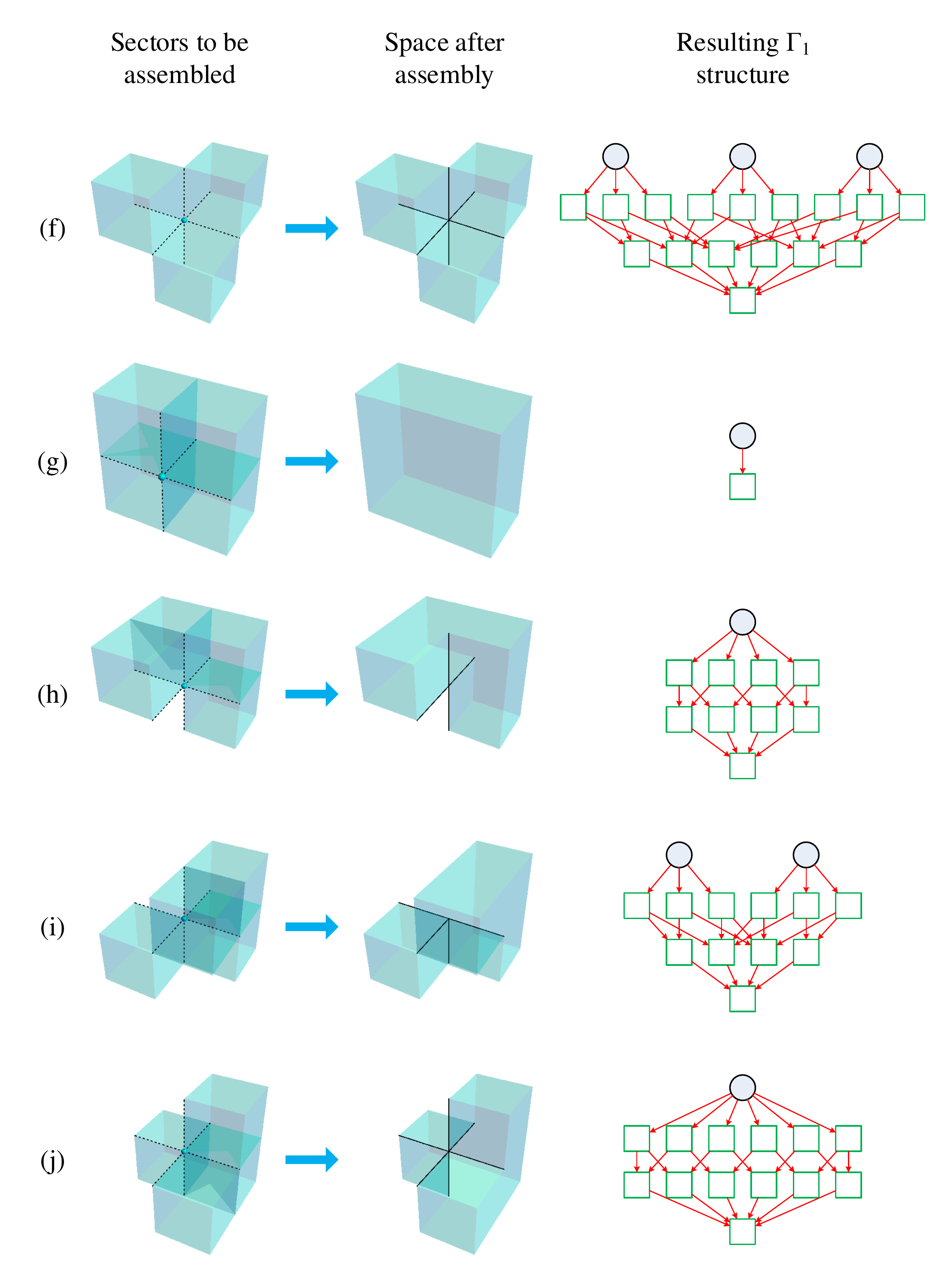}
\vspace{-0.8cm}
\caption{}
\label{fig:merging3Delta(2)}
\end{center}
\end{figure}

\begin{figure}[htb]
\begin{center}
\includegraphics[scale=0.65]{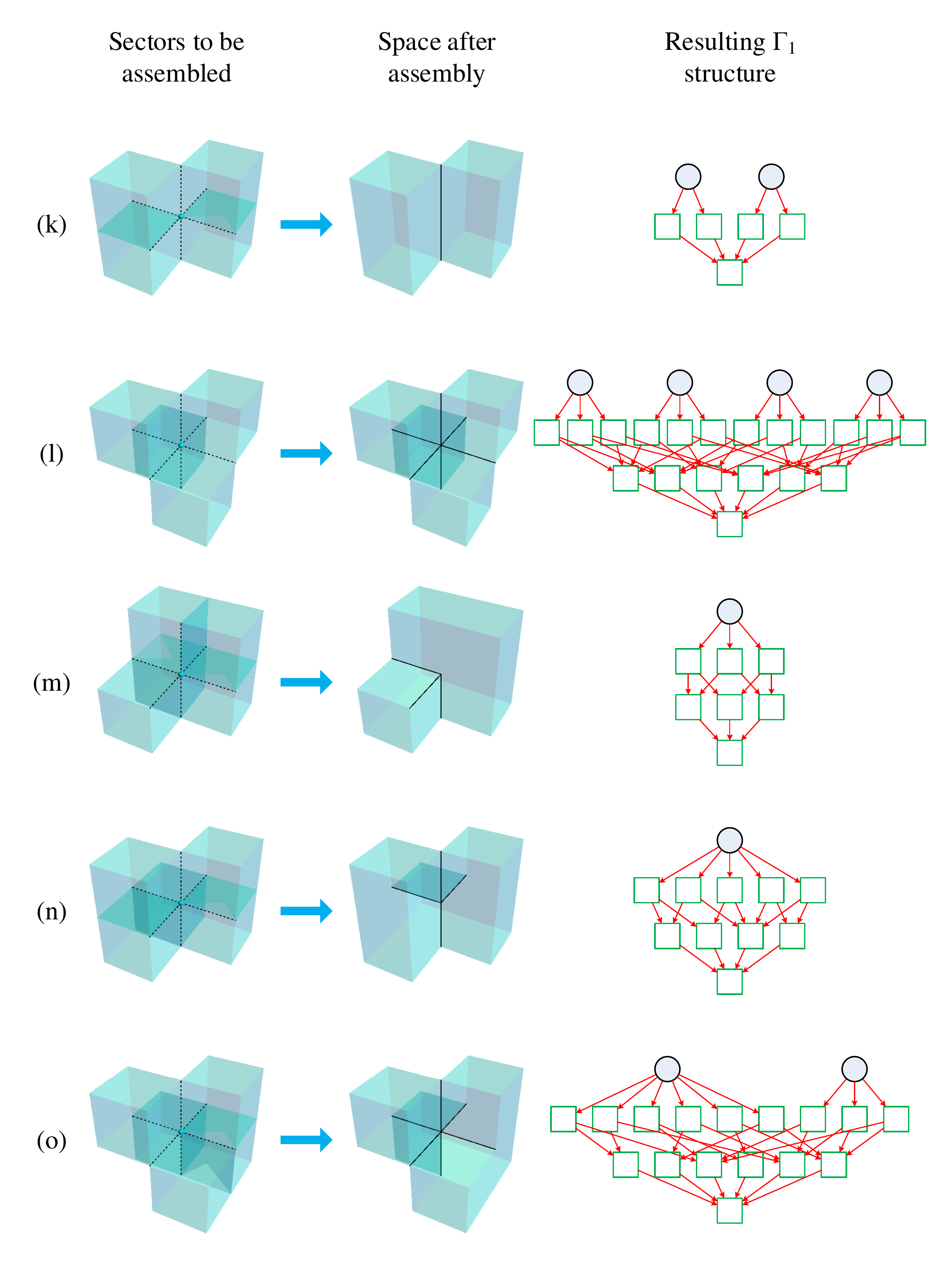}
\vspace{-0.8cm}
\caption{}
\label{fig:merging3Delta(3)}
\end{center}
\end{figure}

\begin{figure}[htb]
\begin{center}
\includegraphics[scale=0.65]{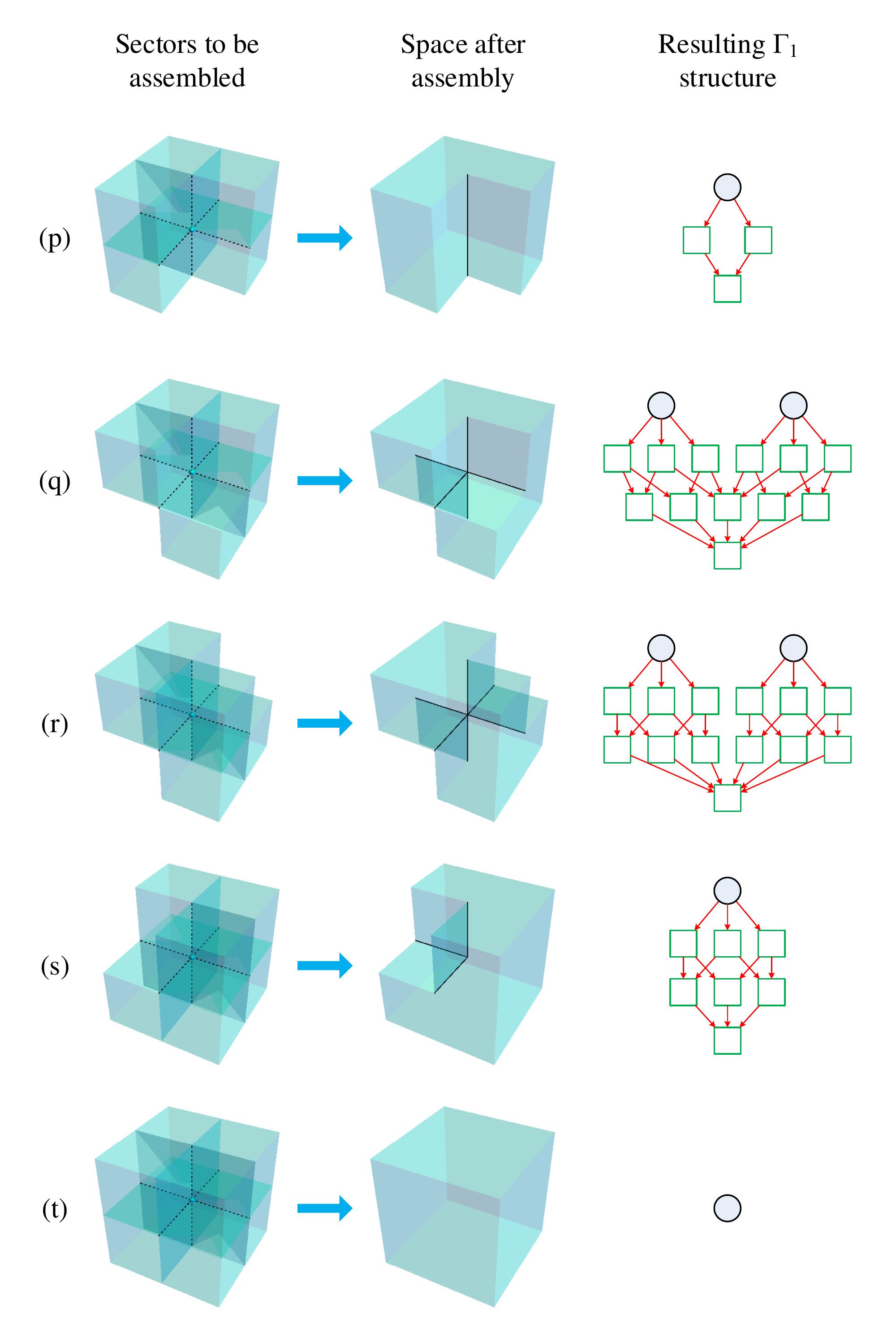}
\vspace{-0.8cm}
\caption{}
\label{fig:merging3Delta(4)}
\end{center}
\end{figure}

In this appendix we display all possible configurations in which positivity sectors can be assembled for Type H boundaries, as described in \sref{sec:assembly3Delta}. The different cases lead to a different $\Gamma_1$ boundary structure after the merging, as illustrated in Figures \ref{fig:merging3Delta(1)}, \ref{fig:merging3Delta(2)}, \ref{fig:merging3Delta(3)} and \ref{fig:merging3Delta(4)}.

\section{Assembly of Type A Boundaries}
\label{app:typeA}

\begin{figure}[htb]
\begin{center}
\includegraphics[scale=0.55]{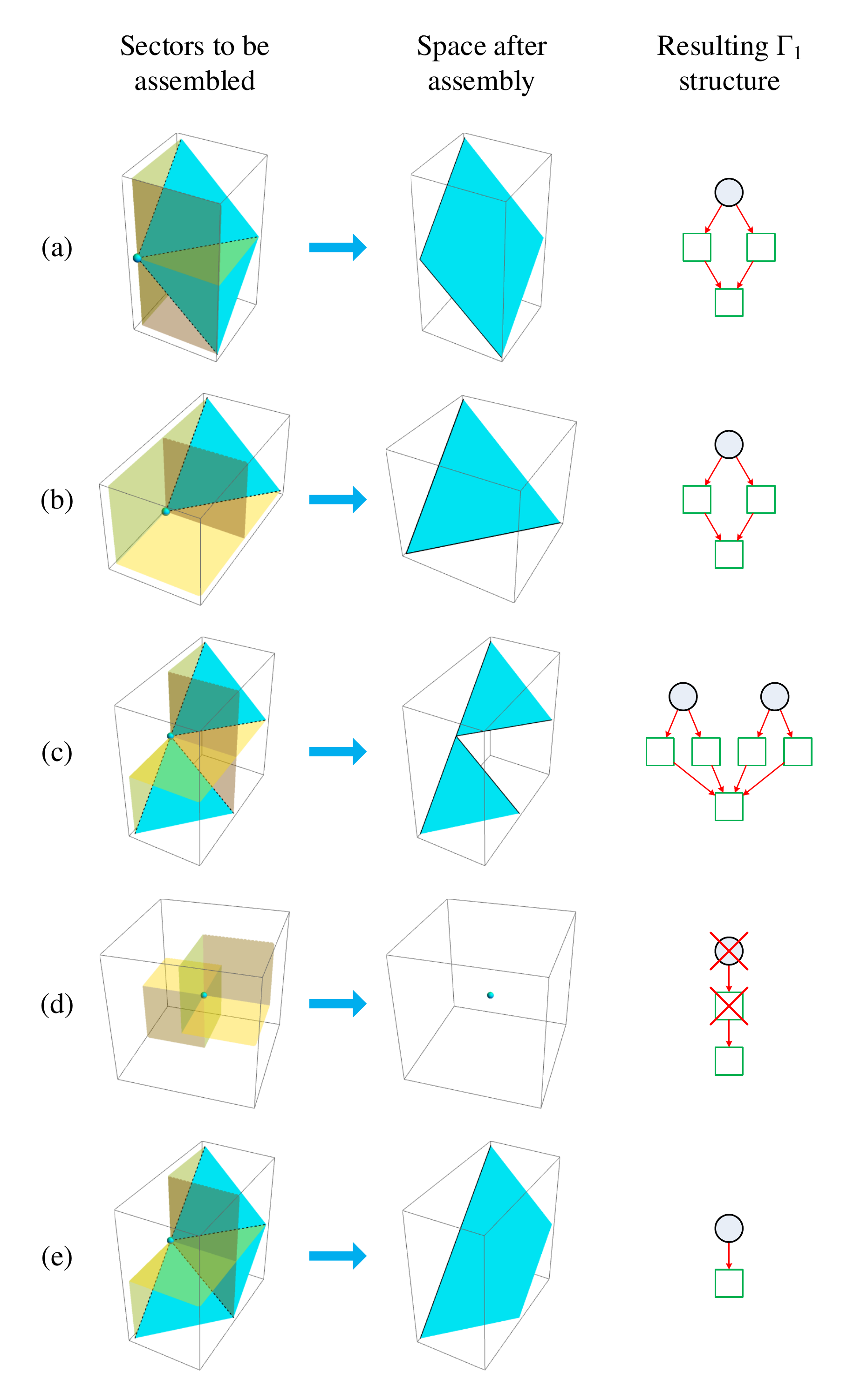}
\vspace{-0.8cm}
\caption{}
\label{fig:merging3DeltaTriangles(1)}
\end{center}
\end{figure}

\begin{figure}[htb]
\begin{center}
\includegraphics[scale=0.55]{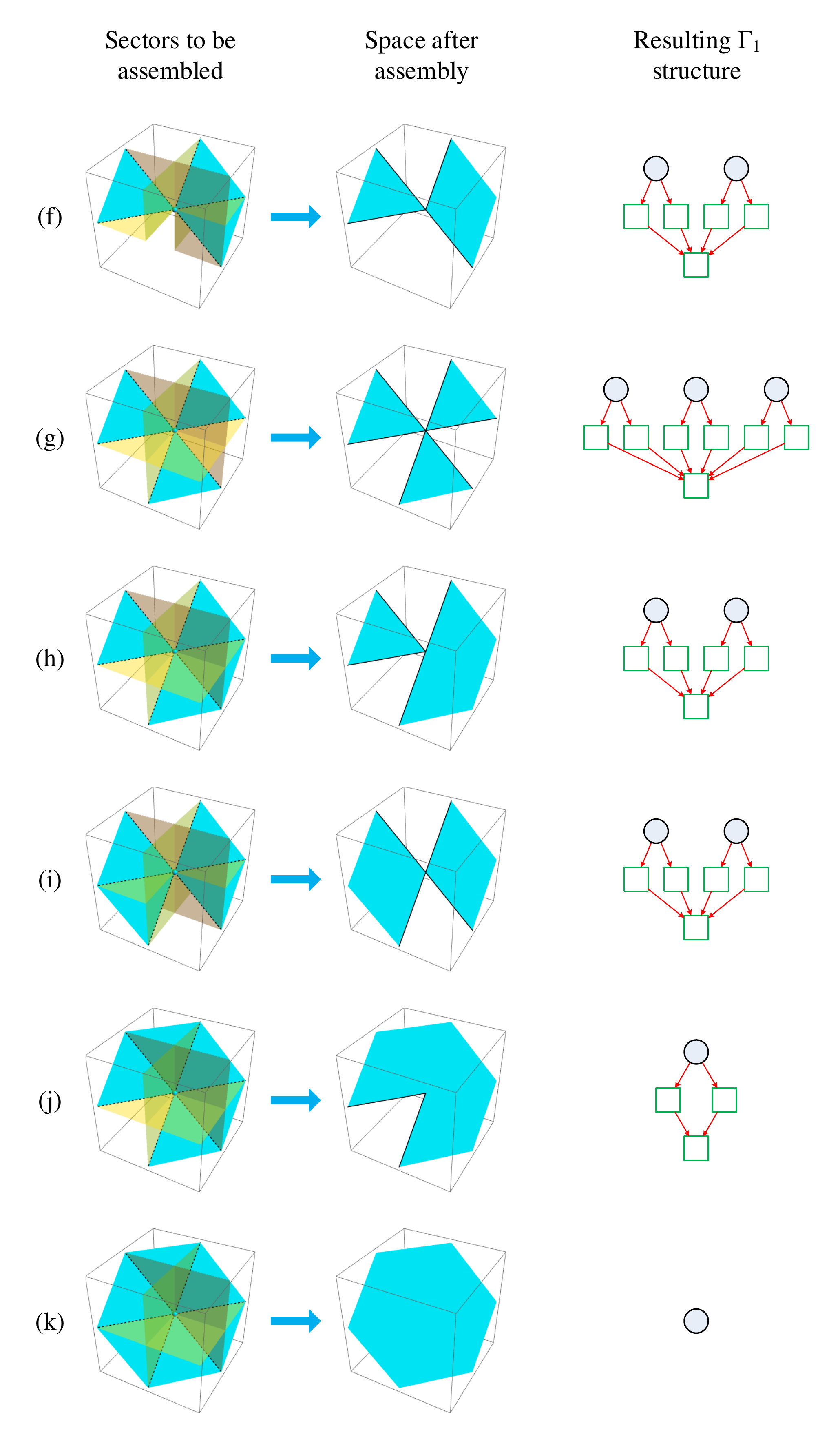}
\vspace{-0.8cm}
\caption{}
\label{fig:merging3DeltaTriangles(2)}
\end{center}
\end{figure}

This appendix displays all possible configurations in which we may assemble together positivity sectors for Type A boundaries. The information is fully contained in Figures \ref{fig:merging3DeltaTriangles(1)} and \ref{fig:merging3DeltaTriangles(2)}.

\section{Explicit Results from the Assembly Algorithm}
\label{app:forming3loophard}

This appendix contains very explicit results obtained from assembling a complicated assortment of sectors, which we chose to be $\{ + , + , + \}$, $\{ - , + , + \}$, $\{ - , + , - \}$, $\{ - , - , + \}$ and $\{ - , - , - \}$. Using the supporting files described in Appendix \ref{app:supportfiles} and applying the algorithm presented in \sref{sec:assemblingalgorithm}, we obtain the results presented in \tref{tab:3LhardGamma0Gamma1}.

\bigskip
\begin{table}[htb]
\begin{center}
{\footnotesize
\begin{tabular}{|c@{\hskip 3pt}|@{\hskip 0pt}c@{\hskip 0pt}|@{\hskip 0pt}c@{\hskip 0pt}|@{\hskip 0pt}c@{\hskip 0pt}|@{\hskip 0pt}c@{\hskip 0pt}|@{\hskip 0pt}c@{\hskip 0pt}|@{\hskip 0pt}c@{\hskip 0pt}|@{\hskip 0pt}c@{\hskip 0pt}|@{\hskip 0pt}c@{\hskip 0pt}|@{\hskip 0pt}c@{\hskip 0pt}|@{\hskip 0pt}c@{\hskip 0pt}|@{\hskip 0pt}c@{\hskip 0pt}|@{\hskip 0pt}c@{\hskip 0pt}|@{\hskip 0pt}c@{\hskip 0pt}|@{\hskip 0pt}c@{\hskip 0pt}|@{\hskip 0pt}c@{\hskip 0pt}|@{\hskip 0pt}c@{\hskip 0pt}|}
\cline{2-17}
\multicolumn{1}{@{\hskip 0pt}c@{\hskip 0pt}|}{} & \multicolumn{10}{c|}{\textbf{3} $\boldsymbol{\Delta^{(i,j)}}$} & \multicolumn{3}{c|}{\textbf{2} $\boldsymbol{\Delta^{(i,j)}}$} & \multicolumn{2}{c|}{\textbf{1} $\boldsymbol{\Delta^{(i,j)}}$} & \ \textbf{0} $\boldsymbol{\Delta^{(i,j)}}$ \ \\
\cline{2-17}
\multicolumn{1}{@{\hskip 0pt}c@{\hskip 0pt}|}{} & $\left( \begin{array}{@{\hskip 1pt}c@{\hskip 1pt}} 1\\ 2\\ 1\\ 0\\ \end{array} \right)$ & $\left( \begin{array}{@{\hskip 1pt}c@{\hskip 1pt}} 1\\ 3\\ 2\\ 0\\ \end{array} \right)$ & $\left( \begin{array}{@{\hskip 1pt}c@{\hskip 1pt}} 1\\ 3\\ 2\\ 1\\ \end{array} \right)$ & $\left( \begin{array}{@{\hskip 1pt}c@{\hskip 1pt}} 1\\ 3\\ 3\\ 0\\ \end{array} \right)$ & $\left( \begin{array}{@{\hskip 1pt}c@{\hskip 1pt}} 1\\ 3\\ 3\\ 1\\ \end{array} \right)$ & $\left( \begin{array}{@{\hskip 1pt}c@{\hskip 1pt}} 1\\ 3\\ 4\\ 0\\ \end{array} \right)$ & $\left( \begin{array}{@{\hskip 1pt}c@{\hskip 1pt}} 1\\ 4\\ 2\\ 0\\ \end{array} \right)$ & $\left( \begin{array}{@{\hskip 1pt}c@{\hskip 1pt}} 1\\ 4\\ 3\\ 0\\ \end{array} \right)$ & $\left( \begin{array}{@{\hskip 1pt}c@{\hskip 1pt}} 2\\ 3\\ 1\\ 0\\ \end{array} \right)$ & $\left( \begin{array}{@{\hskip 1pt}c@{\hskip 1pt}} 2\\ 4\\ 1\\ 0\\ \end{array} \right)$ & $\left( \begin{array}{@{\hskip 1pt}c@{\hskip 1pt}} 1\\ 0\\ 0\\ 0\\ \end{array} \right)$ & $\left( \begin{array}{@{\hskip 1pt}c@{\hskip 1pt}} 1\\ 1\\ 0\\ 0\\ \end{array} \right)$ & $\left( \begin{array}{@{\hskip 1pt}c@{\hskip 1pt}} 1\\ 2\\ 1\\ 0\\ \end{array} \right)$ & $\left( \begin{array}{@{\hskip 1pt}c@{\hskip 1pt}} 1\\ 0\\ 0\\ 0\\ \end{array} \right)$ & $\left( \begin{array}{@{\hskip 1pt}c@{\hskip 1pt}} 1\\ 1\\ 0\\ 0\\ \end{array} \right)$ & $\left( \begin{array}{@{\hskip 1pt}c@{\hskip 1pt}} 1\\ 0\\ 0\\ 0\\ \end{array} \right)$ \\
\hline
\multicolumn{1}{|@{\hskip 0pt}c@{\hskip 0pt}|}{\textbf{\ 12 \ }} & 0 & 0 & 0 & 0 & 1 & 0 & 0 & 0 & 0 & 0 & 0 & 0 & 0 & 0 & 0 & 0 \\
\hline 
\multicolumn{1}{|@{\hskip 0pt}c@{\hskip 0pt}|}{\textbf{11}} & 0 & 0 & 0 & 0 & 12 & 0 & 0 & 0 & 0 & 0 & 0 & 0 & 0 & 0 & 0 & 0 \\
\hline 
\multicolumn{1}{|@{\hskip 0pt}c@{\hskip 0pt}|}{\textbf{10}} & 0 & 0 & 0 & 0 & 78 & 0 & 0 & 0 & 0 & 0 & 0 & 0 & 0 & 0 & 0 & 0 \\
\hline 
\multicolumn{1}{|@{\hskip 0pt}c@{\hskip 0pt}|}{\textbf{9}} & 0 & 0 & 0 & 0 & 324 & 4 & 0 & 0 & 0 & 0 & 0 & 0 & 0 & 4 & 8 & 0 \\
\hline 
\multicolumn{1}{|@{\hskip 0pt}c@{\hskip 0pt}|}{\textbf{8}} & 8 & 0 & 0 & 16 & 726 & 12 & 0 & 32 & 4 & 0 & 34 & 52 & 88 & 40 & 74 & 0 \\
\hline 
\multicolumn{1}{|@{\hskip 0pt}c@{\hskip 0pt}|}{\textbf{7}} & 96 & 80 & 12 & 64 & 600 & 12 & 16 & 128 & 32 & 16 & 200 & 272 & 512 & 248 & 352 & 0 \\
\hline 
\multicolumn{1}{|@{\hskip 0pt}c@{\hskip 0pt}|}{\textbf{6}} & 176 & 64 & 0 & 48 & 144 & 2 & 32 & 96 & 40 & 48 & 408 & 496 & 920 & 1\,084 & 1\,122 & 200 \\
\hline 
\multicolumn{1}{|@{\hskip 0pt}c@{\hskip 0pt}|}{\textbf{5}} & 96 & 0 & 0 & 8 & 0 & 0 & 0 & 16 & 0 & 48 & 304 & 272 & 624 & 2\,232 & 1\,736 & 1\,464 \\
\hline 
\multicolumn{1}{|@{\hskip 0pt}c@{\hskip 0pt}|}{\textbf{4}} & 16 & 0 & 0 & 0 & 0 & 0 & 0 & 0 & 0 & 8 & 80 & 16 & 168 & 1\,992 & 1\,156 & 3\,888 \\
\hline 
\multicolumn{1}{|@{\hskip 0pt}c@{\hskip 0pt}|}{\textbf{3}} & 0 & 0 & 0 & 0 & 0 & 0 & 0 & 0 & 0 & 0 & 8 & 0 & 16 & 768 & 320 & 4\,936 \\
\hline 
\multicolumn{1}{|@{\hskip 0pt}c@{\hskip 0pt}|}{\textbf{2}} & 0 & 0 & 0 & 0 & 0 & 0 & 0 & 0 & 0 & 0 & 0 & 0 & 0 & 112 & 32 & 3\,392 \\
\hline 
\multicolumn{1}{|@{\hskip 0pt}c@{\hskip 0pt}|}{\textbf{1}} & 0 & 0 & 0 & 0 & 0 & 0 & 0 & 0 & 0 & 0 & 0 & 0 & 0 & 0 & 0 & 1\,280 \\
\hline 
\multicolumn{1}{|@{\hskip 0pt}c@{\hskip 0pt}|}{\textbf{0}} & 0 & 0 & 0 & 0 & 0 & 0 & 0 & 0 & 0 & 0 & 0 & 0 & 0 & 0 & 0 & 216 \\
\hline 
\end{tabular}
}
\caption{Number of boundaries with $N \Delta^{(i,j)}$ minors that may naively be set to zero. Here we do not split the various scenarios that can occur into Types, as we did in \sref{sec:3loopsectors}, but rather specify the scenarios in terms of their boundary structure. For example, the last column in the $3 \Delta^{(i,j)}$ segment of the table has a boundary structure as shown in \fref{fig:merging3Delta(3)}(k), i.e.\ it has 2 codimension-0 boundaries, 4 codimension-1 boundaries and 1 codimension-2 boundary.\label{tab:3LhardGamma0Gamma1} }
\end{center}
\end{table}

We may now add up the number of boundaries of various dimensions, to obtain the final boundary structure of the resulting space. We remind the reader that this boundary structure describes the mini stratification. Following the discussion in \sref{sec:formingcubeof1loop} on the geometric significance of assembling positivity sectors, it is interesting to compute both the $\Gamma_0$ stratification after assembly, representing the boundaries shared with $G_+(0,4;1)^3$, as well as the full boundary structure given by $\Gamma_0 + \Gamma_1$. Both are shown in \tref{tab:3LhardStructure}.

\begin{table}[htb]
\begin{center}
\bigskip
{\small
\begin{tabular}{|c|c|c|c|}
\cline{2-4}
\multicolumn{1}{c|}{} & \multicolumn{2}{c|}{\textbf{Geometry of}} \\
\multicolumn{1}{c|}{} & \multicolumn{2}{c|}{\textbf{resulting space}} \\
\hline
\textbf{Dim} & $\boldsymbol{\mathcal{N}}$ & $\boldsymbol{\mathfrak{N}}$ \\
\hline
\textbf{12} & 1 & 1 \\
\hline
\textbf{11} & 12 & 15 \\
\hline
\textbf{10} & 78 & 117 \\
\hline
\textbf{9} & 340 & 611 \\
\hline
\textbf{8} & 1\,086 & 2\,328 \\
\hline
\textbf{7} & 2\,640 & 6\,474 \\
\hline
\textbf{6} & 4\,880 & 12\,642 \\
\hline
\textbf{5} & 6\,800 & 16\,278 \\
\hline
\textbf{4} & 7\,324 & 13\,920 \\
\hline
\textbf{3} & 6\,048 & 8\,604 \\
\hline
\textbf{2} & 3\,536 & 4\,080 \\
\hline
\textbf{1} & 1\,280 & 1\,328 \\
\hline
\textbf{0} & 216 & 216 \\
\hline
\end{tabular}
}
\caption{Number of boundaries obtained when assembling the sectors $\{ + , + , + \}$, $\{ - , + , + \}$, $\{ - , + , - \}$, $\{ - , - , + \}$ and $\{ - , - , - \}$.\label{tab:3LhardStructure}}
\end{center}
\end{table}

Finally, we compute the Euler number for the resulting object:
\begin{equation}
\mathcal{E} = 216 - 1\,328 + \ldots - 15 + 1 = -6 \; \; ,
\end{equation}
whose smallness is a result of the geometric nature of the amplituhedron, but whose value differing from 1 is indicative of a relatively complex internal structure in the larger space $G_+(0,4;1)^3$. A quick comparison with \tref{tab:oneLcubestratFinal} shows that including more sectors does indeed result in geometric objects with Euler number closer to 1. 

\section{Supporting Files}
\label{app:supportfiles}

With this article we have also made available files containing all boundaries of all positivity sectors of the three-loop amplituhedron, i.e.\ all boundaries presented in \sref{sec:3loopsectors}. These files can be downloaded together with the arXiv source file. Each file contains the boundaries of a given positivity sector, at all dimensions, for any choice of $N \Delta^{(i,j)}$ sites in $\Gamma_0$.\footnote{We note that the Type X boundaries are not placed at codimension $2$ w.r.t.\ the naive counting of dimensions by looking at the \pl coordinates; expressed differently, the dimension counting completely neglects the $\Gamma_1$ structure, as is done in the tables in \sref{sec:3loopsectors}.}

The naming format of the files is the following. Each file begins with three letters, describing the positivity sector under consideration, followed by the word $\mathtt{Boundaries}$; the letter ``p'' signifies $(+1)$ and ``m'' signifies $(-1)$. For example, the sector $\{+,-,+ \}$ is given by file $\mathtt{pmpBoundaries}$.

The files provided are stored as compressed Mathematica variables; to unpackage them correctly, we have provided a Mathematica notebook named $\mathtt{LoadBoundaries.nb}$. Loading the boundaries of a chosen positivity sector into Mathematica is then very simple:
\begin{itemize}
\item The notebook $\mathtt{LoadBoundaries.nb}$ should be in the same folder as the files containing the boundaries of the positivity sector.
\item The first cell of $\mathtt{LoadBoundaries.nb}$ should be evaluated. This creates a simple function called $\mathtt{loadBoundaries}$ that can correctly load the boundaries of a positivity sector.
\item In the second cell, use the function $\mathtt{loadBoundaries}$ to load the boundaries of the desired positivity sector. $\mathtt{loadBoundaries}$ takes as input argument the file name of the sector under consideration. For example, if we wish to load the sector $\{+,-,+ \}$, we evaluate the expression
\begin{flalign*}
& \quad \quad \quad \mathtt{\displaystyle \color[RGB]{69,78,153} In[2]:=} \; \; \mathtt{loadBoundaries[''} \mathtt{\displaystyle \color[RGB]{102,102,102} pmpBoundaries} \mathtt{''];} & \nonumber
\end{flalign*}
This will create a variable named $\mathtt{boundaries}$ which contains all boundaries of the positivity sector.
\item To explicitly view the boundaries, two numbers must be chosen: the number $N$ in $N \Delta^{(i,j)}$, and the dimension of the sites in $\Gamma_0$ we are interested in. These numbers are given as inputs to the variable $\mathtt{boundaries}$. For example, if we are interested in 8-dimensional $3 \Delta^{(i,j)}$ boundaries, we evaluate
\begin{flalign*}
& \quad \quad \quad \mathtt{\displaystyle \color[RGB]{69,78,153} In[3]:=} \; \; \mathtt{boundaries[3,8]} & \nonumber
\end{flalign*}
which outputs the required boundaries.
\end{itemize}

The structure of each $\texttt{boundaries[} i \texttt{,} j \texttt{]}$ is a Mathematica list. Each element in this list will have the structure $\mathtt{\{}\Gamma_0 \text{ label}, \Gamma_1 \text{ boundary structure}\mathtt{\}}$. The $\Gamma_0$ label is a Mathematica list of \pl coordinates which are set to zero in this $\Gamma_0$ site, where the \pl coordinate $\Delta^{(i)}_{I_1 I_2}$ is written as $\texttt{pluck[} i \texttt{][} I_1 \texttt{,} I_2 \texttt{]}$. For example, the \pl coordinate $\Delta^{(2)}_{14}$ is written as $\texttt{pluck[2][1,4]}$.

The $\Gamma_1$ boundary structure is also presented as a list of elements. Each element is in turn a list of $4 \times 4$ minors  $\Delta^{(i,j)}$, written as c,  with their positivity domain, followed by a number which denotes the codimension of this element in $\Gamma_1$. For example, a codimension-1 site in $\Gamma_1$ could be
\begin{equation}
\texttt{\{delta[1,2]>0, delta[1,3]>0, delta[2,3]=0, 1\}} \quad .
\end{equation}

Thus, an example of an element in the list that comes from loading the file $\mathtt{pmpBoundaries}$ and evaluating $\mathtt{boundaries[3,7]}$ could be
{\small
\begin{align}
\texttt{ \{\{}&\texttt{pluck[1][1,4],pluck[2][1,2],pluck[2][1,4],pluck[2][2,4],pluck[3][1,3],} \nonumber \\ 
&\texttt{pluck[3][1,4],pluck[3][3,4]\}, \{\{delta[1,2]>0,delta[1,3]<0,delta[2,3]>0,0\},} \nonumber \\
&\texttt{\{delta[1,2]>0,delta[1,3]==0,delta[2,3]>0,1\},\{delta[1,2]>0,delta[1,3]<0,} \nonumber \\
&\texttt{delta[2,3]==0,1\},\{delta[1,2]==0,delta[1,3]==0,delta[2,3]==0,2\}\}\}}
\end{align}
}
which as we see is a Type A boundary.


\bibliographystyle{JHEP}
\bibliography{paper}

\providecommand{\href}[2]{#2}\begingroup\raggedright\begin{thebibliography}{10}

\bibitem{Bern:1994zx}
Z.~Bern, L.~J. Dixon, D.~C. Dunbar, and D.~A. Kosower, {\it {One loop n point
  gauge theory amplitudes, unitarity and collinear limits}},  {\em Nucl.Phys.}
  {\bf B425} (1994) 217--260,
  [\href{http://xxx.lanl.gov/abs/hep-ph/9403226}{{\tt hep-ph/9403226}}].

\bibitem{Bern:1994cg}
Z.~Bern, L.~J. Dixon, D.~C. Dunbar, and D.~A. Kosower, {\it {Fusing gauge
  theory tree amplitudes into loop amplitudes}},  {\em Nucl.Phys.} {\bf B435}
  (1995) 59--101, [\href{http://xxx.lanl.gov/abs/hep-ph/9409265}{{\tt
  hep-ph/9409265}}].

\bibitem{Cachazo:2004kj}
F.~Cachazo, P.~Svrcek, and E.~Witten, {\it {MHV vertices and tree amplitudes in
  gauge theory}},  {\em JHEP} {\bf 0409} (2004) 006,
  [\href{http://xxx.lanl.gov/abs/hep-th/0403047}{{\tt hep-th/0403047}}].

\bibitem{Britto:2004nc}
R.~Britto, F.~Cachazo, and B.~Feng, {\it {Generalized unitarity and one-loop
  amplitudes in N=4 super-Yang-Mills}},  {\em Nucl.Phys.} {\bf B725} (2005)
  275--305, [\href{http://xxx.lanl.gov/abs/hep-th/0412103}{{\tt
  hep-th/0412103}}].

\bibitem{Britto:2004ap}
R.~Britto, F.~Cachazo, and B.~Feng, {\it {New recursion relations for tree
  amplitudes of gluons}},  {\em Nucl.Phys.} {\bf B715} (2005) 499--522,
  [\href{http://xxx.lanl.gov/abs/hep-th/0412308}{{\tt hep-th/0412308}}].

\bibitem{Britto:2005fq}
R.~Britto, F.~Cachazo, B.~Feng, and E.~Witten, {\it {Direct proof of tree-level
  recursion relation in Yang-Mills theory}},  {\em Phys.Rev.Lett.} {\bf 94}
  (2005) 181602, [\href{http://xxx.lanl.gov/abs/hep-th/0501052}{{\tt
  hep-th/0501052}}].

\bibitem{Bern:2005iz}
Z.~Bern, L.~J. Dixon, and V.~A. Smirnov, {\it {Iteration of planar amplitudes
  in maximally supersymmetric Yang-Mills theory at three loops and beyond}},
  {\em Phys.Rev.} {\bf D72} (2005) 085001,
  [\href{http://xxx.lanl.gov/abs/hep-th/0505205}{{\tt hep-th/0505205}}].

\bibitem{Dixon:1996wi}
L.~J. Dixon, {\it {Calculating scattering amplitudes efficiently}},
  \href{http://xxx.lanl.gov/abs/hep-ph/9601359}{{\tt hep-ph/9601359}}.

\bibitem{Beisert:2010jr}
N.~Beisert, C.~Ahn, L.~F. Alday, Z.~Bajnok, J.~M. Drummond, et~al., {\it
  {Review of AdS/CFT Integrability: An Overview}},  {\em Lett.Math.Phys.} {\bf
  99} (2012) 3--32, [\href{http://xxx.lanl.gov/abs/1012.3982}{{\tt
  arXiv:1012.3982}}].

\bibitem{Drummond:2011ic}
J.~Drummond, {\it {Tree-level amplitudes and dual superconformal symmetry}},
  {\em J.Phys.} {\bf A44} (2011) 454010,
  [\href{http://xxx.lanl.gov/abs/1107.4544}{{\tt arXiv:1107.4544}}].

\bibitem{Elvang:2013cua}
H.~Elvang and Y.-t. Huang, {\it {Scattering Amplitudes}},
  \href{http://xxx.lanl.gov/abs/1308.1697}{{\tt arXiv:1308.1697}}.

\bibitem{Bern:2006ew}
Z.~Bern, M.~Czakon, L.~J. Dixon, D.~A. Kosower, and V.~A. Smirnov, {\it {The
  Four-Loop Planar Amplitude and Cusp Anomalous Dimension in Maximally
  Supersymmetric Yang-Mills Theory}},  {\em Phys.Rev.} {\bf D75} (2007) 085010,
  [\href{http://xxx.lanl.gov/abs/hep-th/0610248}{{\tt hep-th/0610248}}].

\bibitem{Bern:2007ct}
Z.~Bern, J.~Carrasco, H.~Johansson, and D.~Kosower, {\it {Maximally
  supersymmetric planar Yang-Mills amplitudes at five loops}},  {\em Phys.Rev.}
  {\bf D76} (2007) 125020, [\href{http://xxx.lanl.gov/abs/0705.1864}{{\tt
  arXiv:0705.1864}}].

\bibitem{ArkaniHamed:2010kv}
N.~Arkani-Hamed, J.~L. Bourjaily, F.~Cachazo, S.~Caron-Huot, and J.~Trnka, {\it
  {The All-Loop Integrand For Scattering Amplitudes in Planar N=4 SYM}},  {\em
  JHEP} {\bf 1101} (2011) 041, [\href{http://xxx.lanl.gov/abs/1008.2958}{{\tt
  arXiv:1008.2958}}].

\bibitem{ArkaniHamed:2010gh}
N.~Arkani-Hamed, J.~L. Bourjaily, F.~Cachazo, and J.~Trnka, {\it {Local
  Integrals for Planar Scattering Amplitudes}},  {\em JHEP} {\bf 1206} (2012)
  125, [\href{http://xxx.lanl.gov/abs/1012.6032}{{\tt arXiv:1012.6032}}].

\bibitem{Dixon:2011nj}
L.~J. Dixon, J.~M. Drummond, and J.~M. Henn, {\it {Analytic result for the
  two-loop six-point NMHV amplitude in N=4 super Yang-Mills theory}},  {\em
  JHEP} {\bf 1201} (2012) 024, [\href{http://xxx.lanl.gov/abs/1111.1704}{{\tt
  arXiv:1111.1704}}].

\bibitem{Bourjaily:2011hi}
J.~L. Bourjaily, A.~DiRe, A.~Shaikh, M.~Spradlin, and A.~Volovich, {\it {The
  Soft-Collinear Bootstrap: N=4 Yang-Mills Amplitudes at Six and Seven Loops}},
   {\em JHEP} {\bf 1203} (2012) 032,
  [\href{http://xxx.lanl.gov/abs/1112.6432}{{\tt arXiv:1112.6432}}].

\bibitem{Dixon:2013eka}
L.~J. Dixon, J.~M. Drummond, M.~von Hippel, and J.~Pennington, {\it {Hexagon
  functions and the three-loop remainder function}},  {\em JHEP} {\bf 1312}
  (2013) 049, [\href{http://xxx.lanl.gov/abs/1308.2276}{{\tt
  arXiv:1308.2276}}].

\bibitem{Dixon:2014xca}
L.~J. Dixon, J.~M. Drummond, C.~Duhr, M.~von Hippel, and J.~Pennington, {\it
  {Bootstrapping six-gluon scattering in planar N=4 super-Yang-Mills theory}},
  {\em PoS} {\bf LL2014} (2014) 077,
  [\href{http://xxx.lanl.gov/abs/1407.4724}{{\tt arXiv:1407.4724}}].

\bibitem{Drummond:2009fd}
J.~M. Drummond, J.~M. Henn, and J.~Plefka, {\it {Yangian symmetry of scattering
  amplitudes in N=4 super Yang-Mills theory}},  {\em JHEP} {\bf 0905} (2009)
  046, [\href{http://xxx.lanl.gov/abs/0902.2987}{{\tt arXiv:0902.2987}}].

\bibitem{Drummond:2006rz}
J.~Drummond, J.~Henn, V.~Smirnov, and E.~Sokatchev, {\it {Magic identities for
  conformal four-point integrals}},  {\em JHEP} {\bf 0701} (2007) 064,
  [\href{http://xxx.lanl.gov/abs/hep-th/0607160}{{\tt hep-th/0607160}}].

\bibitem{Alday:2007hr}
L.~F. Alday and J.~M. Maldacena, {\it {Gluon scattering amplitudes at strong
  coupling}},  {\em JHEP} {\bf 0706} (2007) 064,
  [\href{http://xxx.lanl.gov/abs/0705.0303}{{\tt arXiv:0705.0303}}].

\bibitem{Beisert:2003yb}
N.~Beisert and M.~Staudacher, {\it {The N=4 SYM integrable super spin chain}},
  {\em Nucl.Phys.} {\bf B670} (2003) 439--463,
  [\href{http://xxx.lanl.gov/abs/hep-th/0307042}{{\tt hep-th/0307042}}].

\bibitem{Beisert:2006ez}
N.~Beisert, B.~Eden, and M.~Staudacher, {\it {Transcendentality and Crossing}},
   {\em J.Stat.Mech.} {\bf 0701} (2007) P01021,
  [\href{http://xxx.lanl.gov/abs/hep-th/0610251}{{\tt hep-th/0610251}}].

\bibitem{ArkaniHamed:2012nw}
N.~Arkani-Hamed, J.~L. Bourjaily, F.~Cachazo, A.~B. Goncharov, A.~Postnikov,
  and J.~Trnka, {\it {Scattering Amplitudes and the Positive Grassmannian}},
  \href{http://xxx.lanl.gov/abs/1212.5605}{{\tt arXiv:1212.5605}}.

\bibitem{ArkaniHamed:2009dn}
N.~Arkani-Hamed, F.~Cachazo, C.~Cheung, and J.~Kaplan, {\it {A Duality For The
  S Matrix}},  {\em JHEP} {\bf 1003} (2010) 020,
  [\href{http://xxx.lanl.gov/abs/0907.5418}{{\tt arXiv:0907.5418}}].

\bibitem{Mason:2009qx}
L.~Mason and D.~Skinner, {\it {Dual Superconformal Invariance, Momentum
  Twistors and Grassmannians}},  {\em JHEP} {\bf 0911} (2009) 045,
  [\href{http://xxx.lanl.gov/abs/0909.0250}{{\tt arXiv:0909.0250}}].

\bibitem{ArkaniHamed:2009vw}
N.~Arkani-Hamed, F.~Cachazo, and C.~Cheung, {\it {The Grassmannian Origin Of
  Dual Superconformal Invariance}},  {\em JHEP} {\bf 1003} (2010) 036,
  [\href{http://xxx.lanl.gov/abs/0909.0483}{{\tt arXiv:0909.0483}}].

\bibitem{Kaplan:2009mh}
J.~Kaplan, {\it {Unraveling L(n,k): Grassmannian Kinematics}},  {\em JHEP} {\bf
  1003} (2010) 025, [\href{http://xxx.lanl.gov/abs/0912.0957}{{\tt
  arXiv:0912.0957}}].

\bibitem{ArkaniHamed:2009dg}
N.~Arkani-Hamed, J.~Bourjaily, F.~Cachazo, and J.~Trnka, {\it {Unification of
  Residues and Grassmannian Dualities}},  {\em JHEP} {\bf 1101} (2011) 049,
  [\href{http://xxx.lanl.gov/abs/0912.4912}{{\tt arXiv:0912.4912}}].

\bibitem{Bourjaily:2012gy}
J.~L. Bourjaily, {\it {Positroids, Plabic Graphs, and Scattering Amplitudes in
  Mathematica}},  \href{http://xxx.lanl.gov/abs/1212.6974}{{\tt
  arXiv:1212.6974}}.

\bibitem{Amariti:2013ija}
A.~Amariti and D.~Forcella, {\it {Scattering Amplitudes and Toric Geometry}},
  {\em JHEP} {\bf 1309} (2013) 133,
  [\href{http://xxx.lanl.gov/abs/1305.5252}{{\tt arXiv:1305.5252}}].

\bibitem{Franco:2013nwa}
S.~Franco, D.~Galloni, and A.~Mariotti, {\it {The Geometry of On-Shell
  Diagrams}},  {\em JHEP} {\bf 1408} (2014) 038,
  [\href{http://xxx.lanl.gov/abs/1310.3820}{{\tt arXiv:1310.3820}}].

\bibitem{Du:2014jwa}
P.~Du, G.~Chen, and Y.-K.~E. Cheung, {\it {Permutation relations of generalized
  Yangian Invariants, unitarity cuts, and scattering amplitudes}},  {\em JHEP}
  {\bf 1409} (2014) 115, [\href{http://xxx.lanl.gov/abs/1401.6610}{{\tt
  arXiv:1401.6610}}].

\bibitem{Franco:2014nca}
S.~Franco, D.~Galloni, and A.~Mariotti, {\it {Bipartite Field Theories, Cluster
  Algebras and the Grassmannian}},  {\em J.Phys.} {\bf A47} (2014), no.~47
  474004, [\href{http://xxx.lanl.gov/abs/1404.3752}{{\tt arXiv:1404.3752}}].

\bibitem{Olson:2014pfa}
T.~M. Olson, {\it {Orientations of BCFW Charts on the Grassmannian}},  {\em
  JHEP} {\bf 08} (2015) 120, [\href{http://xxx.lanl.gov/abs/1411.6363}{{\tt
  arXiv:1411.6363}}].

\bibitem{Bork:2015fla}
L.~V. Bork and A.~I. Onishchenko, {\it {On soft theorems and form factors in $
  \mathcal{N}=4 $ SYM theory}},  {\em JHEP} {\bf 12} (2015) 030,
  [\href{http://xxx.lanl.gov/abs/1506.0755}{{\tt arXiv:1506.0755}}].

\bibitem{Frassek:2015rka}
R.~Frassek, D.~Meidinger, D.~Nandan, and M.~Wilhelm, {\it {On-shell Diagrams,
  Gra{\ss}mannians and Integrability for Form Factors}},
  \href{http://xxx.lanl.gov/abs/1506.0819}{{\tt arXiv:1506.0819}}.

\bibitem{Benincasa:2015zna}
P.~Benincasa, {\it {On-shell diagrammatics and the perturbative structure of
  planar gauge theories}},  \href{http://xxx.lanl.gov/abs/1510.0364}{{\tt
  arXiv:1510.0364}}.

\bibitem{Ferro:2012xw}
L.~Ferro, T.~Łukowski, C.~Meneghelli, J.~Plefka, and M.~Staudacher, {\it
  {Harmonic R-matrices for Scattering Amplitudes and Spectral Regularization}},
   {\em Phys.Rev.Lett.} {\bf 110} (2013), no.~12 121602,
  [\href{http://xxx.lanl.gov/abs/1212.0850}{{\tt arXiv:1212.0850}}].

\bibitem{Ferro:2013dga}
L.~Ferro, T.~Łukowski, C.~Meneghelli, J.~Plefka, and M.~Staudacher, {\it
  {Spectral Parameters for Scattering Amplitudes in N=4 Super Yang-Mills
  Theory}},  {\em JHEP} {\bf 1401} (2014) 094,
  [\href{http://xxx.lanl.gov/abs/1308.3494}{{\tt arXiv:1308.3494}}].

\bibitem{Beisert:2014qba}
N.~Beisert, J.~Broedel, and M.~Rosso, {\it {On Yangian-invariant regularization
  of deformed on-shell diagrams in $\mathcal{N}=4$ super-Yang-Mills theory}},
  {\em J.Phys.} {\bf A47} (2014) 365402,
  [\href{http://xxx.lanl.gov/abs/1401.7274}{{\tt arXiv:1401.7274}}].

\bibitem{Broedel:2014hca}
J.~Broedel, M.~de~Leeuw, and M.~Rosso, {\it {Deformed one-loop amplitudes in $
  \mathcal{N}=4 $ super-Yang-Mills theory}},  {\em JHEP} {\bf 1411} (2014) 091,
  [\href{http://xxx.lanl.gov/abs/1406.4024}{{\tt arXiv:1406.4024}}].

\bibitem{Bargheer:2014mxa}
T.~Bargheer, Y.-t. Huang, F.~Loebbert, and M.~Yamazaki, {\it {Integrable
  Amplitude Deformations for N=4 Super Yang-Mills and ABJM Theory}},  {\em
  Phys.Rev.} {\bf D91} (2015), no.~2 026004,
  [\href{http://xxx.lanl.gov/abs/1407.4449}{{\tt arXiv:1407.4449}}].

\bibitem{Ferro:2014gca}
L.~Ferro, T.~Łukowski, and M.~Staudacher, {\it {$\mathcal N=4$ scattering
  amplitudes and the deformed Graßmannian}},  {\em Nucl. Phys.} {\bf B889}
  (2014) 192--206, [\href{http://xxx.lanl.gov/abs/1407.6736}{{\tt
  arXiv:1407.6736}}].

\bibitem{Bern:1997nh}
Z.~Bern, J.~Rozowsky, and B.~Yan, {\it {Two loop four gluon amplitudes in N=4
  superYang-Mills}},  {\em Phys.Lett.} {\bf B401} (1997) 273--282,
  [\href{http://xxx.lanl.gov/abs/hep-ph/9702424}{{\tt hep-ph/9702424}}].

\bibitem{Bern:2007hh}
Z.~Bern, J.~Carrasco, L.~J. Dixon, H.~Johansson, D.~Kosower, et~al., {\it
  {Three-Loop Superfiniteness of N=8 Supergravity}},  {\em Phys.Rev.Lett.} {\bf
  98} (2007) 161303, [\href{http://xxx.lanl.gov/abs/hep-th/0702112}{{\tt
  hep-th/0702112}}].

\bibitem{Bern:2010tq}
Z.~Bern, J.~Carrasco, L.~J. Dixon, H.~Johansson, and R.~Roiban, {\it {The
  Complete Four-Loop Four-Point Amplitude in N=4 Super-Yang-Mills Theory}},
  {\em Phys.Rev.} {\bf D82} (2010) 125040,
  [\href{http://xxx.lanl.gov/abs/1008.3327}{{\tt arXiv:1008.3327}}].

\bibitem{Carrasco:2011mn}
J.~J. Carrasco and H.~Johansson, {\it {Five-Point Amplitudes in N=4
  Super-Yang-Mills Theory and N=8 Supergravity}},  {\em Phys.Rev.} {\bf D85}
  (2012) 025006, [\href{http://xxx.lanl.gov/abs/1106.4711}{{\tt
  arXiv:1106.4711}}].

\bibitem{Bern:2012uc}
Z.~Bern, J.~Carrasco, H.~Johansson, and R.~Roiban, {\it {The Five-Loop
  Four-Point Amplitude of N=4 super-Yang-Mills Theory}},  {\em Phys.Rev.Lett.}
  {\bf 109} (2012) 241602, [\href{http://xxx.lanl.gov/abs/1207.6666}{{\tt
  arXiv:1207.6666}}].

\bibitem{Arkani-Hamed:2014via}
N.~Arkani-Hamed, J.~L. Bourjaily, F.~Cachazo, and J.~Trnka, {\it {Singularity
  Structure of Maximally Supersymmetric Scattering Amplitudes}},  {\em
  Phys.Rev.Lett.} {\bf 113} (2014), no.~26 261603,
  [\href{http://xxx.lanl.gov/abs/1410.0354}{{\tt arXiv:1410.0354}}].

\bibitem{Arkani-Hamed:2014bca}
N.~Arkani-Hamed, J.~L. Bourjaily, F.~Cachazo, A.~Postnikov, and J.~Trnka, {\it
  {On-Shell Structures of MHV Amplitudes Beyond the Planar Limit}},  {\em JHEP}
  {\bf 06} (2015) 179, [\href{http://xxx.lanl.gov/abs/1412.8475}{{\tt
  arXiv:1412.8475}}].

\bibitem{Bern:2014kca}
Z.~Bern, E.~Herrmann, S.~Litsey, J.~Stankowicz, and J.~Trnka, {\it {Logarithmic
  Singularities and Maximally Supersymmetric Amplitudes}},  {\em JHEP} {\bf 06}
  (2015) 202, [\href{http://xxx.lanl.gov/abs/1412.8584}{{\tt
  arXiv:1412.8584}}].

\bibitem{Franco:2015rma}
S.~Franco, D.~Galloni, B.~Penante, and C.~Wen, {\it {Non-Planar On-Shell
  Diagrams}},  {\em JHEP} {\bf 06} (2015) 199,
  [\href{http://xxx.lanl.gov/abs/1502.0203}{{\tt arXiv:1502.0203}}].

\bibitem{Chen:2015bnt}
B.~Chen, G.~Chen, Y.-K.~E. Cheung, R.~Xie, and Y.~Xin, {\it {Top-forms of
  Leading Singularities in Nonplanar Multi-loop Amplitudes}},
  \href{http://xxx.lanl.gov/abs/1507.0321}{{\tt arXiv:1507.0321}}.

\bibitem{Badger:2015lda}
S.~Badger, G.~Mogull, A.~Ochirov, and D.~O’Connell, {\it {A Complete
  Two-Loop, Five-Gluon Helicity Amplitude in Yang-Mills Theory}},  {\em JHEP}
  {\bf 10} (2015) 064, [\href{http://xxx.lanl.gov/abs/1507.0879}{{\tt
  arXiv:1507.0879}}].

\bibitem{Bern:2015ple}
Z.~Bern, E.~Herrmann, S.~Litsey, J.~Stankowicz, and J.~Trnka, {\it {Evidence
  for a Nonplanar Amplituhedron}},
  \href{http://xxx.lanl.gov/abs/1512.0859}{{\tt arXiv:1512.0859}}.

\bibitem{Arkani-Hamed:2013jha}
N.~Arkani-Hamed and J.~Trnka, {\it {The Amplituhedron}},  {\em JHEP} {\bf 1410}
  (2014) 30, [\href{http://xxx.lanl.gov/abs/1312.2007}{{\tt arXiv:1312.2007}}].

\bibitem{Arkani-Hamed:2013kca}
N.~Arkani-Hamed and J.~Trnka, {\it {Into the Amplituhedron}},  {\em JHEP} {\bf
  1412} (2014) 182, [\href{http://xxx.lanl.gov/abs/1312.7878}{{\tt
  arXiv:1312.7878}}].

\bibitem{Enciso:2014cta}
M.~Enciso, {\it {Volumes of Polytopes Without Triangulations}},
  \href{http://xxx.lanl.gov/abs/1408.0932}{{\tt arXiv:1408.0932}}.

\bibitem{Bai:2014cna}
Y.~Bai and S.~He, {\it {The Amplituhedron from Momentum Twistor Diagrams}},
  {\em JHEP} {\bf 1502} (2015) 065,
  [\href{http://xxx.lanl.gov/abs/1408.2459}{{\tt arXiv:1408.2459}}].

\bibitem{Franco:2014csa}
S.~Franco, D.~Galloni, A.~Mariotti, and J.~Trnka, {\it {Anatomy of the
  Amplituhedron}},  {\em JHEP} {\bf 1503} (2015) 128,
  [\href{http://xxx.lanl.gov/abs/1408.3410}{{\tt arXiv:1408.3410}}].

\bibitem{Lam:2014jda}
T.~Lam, {\it {Amplituhedron cells and Stanley symmetric functions}},
  \href{http://xxx.lanl.gov/abs/1408.5531}{{\tt arXiv:1408.5531}}.

\bibitem{Arkani-Hamed:2014dca}
N.~Arkani-Hamed, A.~Hodges, and J.~Trnka, {\it {Positive Amplitudes In The
  Amplituhedron}},  {\em JHEP} {\bf 08} (2015) 030,
  [\href{http://xxx.lanl.gov/abs/1412.8478}{{\tt arXiv:1412.8478}}].

\bibitem{Bai:2015qoa}
Y.~Bai, S.~He, and T.~Lam, {\it {The Amplituhedron and the One-loop
  Grassmannian Measure}},  \href{http://xxx.lanl.gov/abs/1510.0355}{{\tt
  arXiv:1510.0355}}.

\bibitem{Ferro:2015grk}
L.~Ferro, T.~Lukowski, A.~Orta, and M.~Parisi, {\it {Towards the Amplituhedron
  Volume}},  \href{http://xxx.lanl.gov/abs/1512.0495}{{\tt arXiv:1512.0495}}.

\bibitem{Elvang:2014fja}
H.~Elvang, Y.-t. Huang, C.~Keeler, T.~Lam, T.~M. Olson, S.~B. Roland, and D.~E.
  Speyer, {\it {Grassmannians for scattering amplitudes in 4d $\mathcal{N}=4$
  SYM and 3d ABJM}},  {\em JHEP} {\bf 12} (2014) 181,
  [\href{http://xxx.lanl.gov/abs/1410.0621}{{\tt arXiv:1410.0621}}].

\bibitem{2005arXiv09129W}
L.~K. {Williams}, {\it {Shelling totally nonnegative flag varieties}},  {\em
  J.~Reine~Angew.~Math.} {\bf 609} (2007) 001,
  [\href{http://xxx.lanl.gov/abs/math/0509129}{{\tt math/0509129}}].

\bibitem{polymake}
E.~Gawrilow and M.~Joswig, {\it polymake: a framework for analyzing convex
  polytopes},  in {\em Polytopes Combinatorics and Computation} (G.~Kalai and
  G.~Ziegler, eds.), vol.~29 of {\em DMV Seminar}, pp.~43--73.
\newblock Birkhuser Basel, 2000.

\end{thebibliography}\endgroup
\end{document}